\documentclass[10pt, a4paper, spanish]{report}

\usepackage[spanish]{babel}
\usepackage{amsmath, amsfonts, amsthm, amssymb}
\usepackage{latexsym}
\usepackage{graphicx}
\usepackage{amscd, epsfig}
\usepackage[latin1]{inputenc}

\newcommand{\paruno} { \setlength{\parskip}{0.25 cm}  }
\newcommand{\parcero} { \setlength{\parskip}{0 cm}  }

\newcommand{\R} {\mathbb R}

\newcommand{\calA} {\mathcal A}
\newcommand{\calC} {\mathcal C}
\newcommand{\calD} {\mathcal D}
\newcommand{\calE} {\mathcal E}
\newcommand{\calG} {\mathcal G}
\newcommand{\calL} {\mathcal L}
\newcommand{\calM} {\mathcal M}
\newcommand{\calN} {\mathcal N}

\newcommand{\calT} {\mathcal T}
\newcommand{\calV} {\mathcal V}
\newcommand{\calW} {\mathcal W}

\newcommand{\Ode} {\mathcal O}
\newcommand{\dd } {\partial }

\newcommand{\ff} {\mathbf f}
\newcommand{\bh} {\mathbf h}
\newcommand{\bt} {\mathbf t}
\newcommand{\uu} {\mathbf u}
\newcommand{\vv} {\mathbf v}
\newcommand{\w} {\mathbf w}
\newcommand{\x} {\mathbf x}
\newcommand{\y} {\mathbf y}

\newcommand{\bA} {\mathbf A}
\newcommand{\bB} {\mathbf B}
\newcommand{\bI} {\mathbf I}

\newcommand{\bM} {\mathbf M}
\newcommand{\bR} {\mathbf R}
\newcommand{\bU} {\mathbf U}
\newcommand{\bZ} {\mathbf Z}

\newcommand{\bcero} {\mathbf 0}

\newtheorem{propo}{Proposici\'on}  
\newtheorem{coro}[propo]{Corolario}
\newtheorem{teo}[propo]{Teorema}
\newtheorem{lema}[propo]{Lema}

\theoremstyle{definition}
\newtheorem{defin}[propo]{Definici\'on}
\newtheorem{ejem}[propo]{Ejemplo}

\theoremstyle{remark}
\newtheorem{obs}[propo]{Observaci\'on}

\paruno

\begin{document}


\thispagestyle{empty}

\begin {center}

\includegraphics[scale=.3]{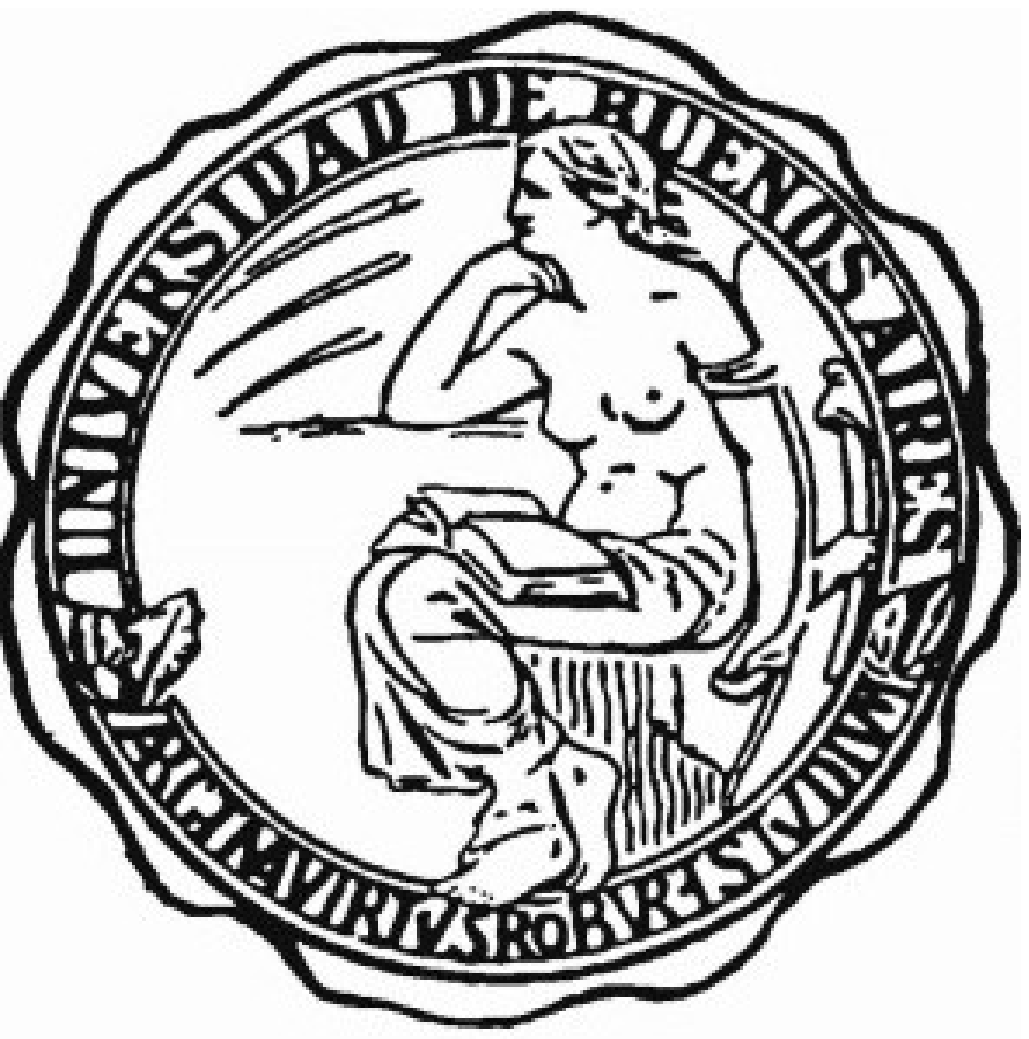}

\medskip
\textbf{UNIVERSIDAD DE BUENOS AIRES}

\smallskip

\textbf{Facultad de Ciencias Exactas y Naturales}

\smallskip

\textbf{Departamento de Matem\'atica}

\vspace{3.5cm}

\textbf{\large Tesis de Licenciatura}

\vspace{1.5cm}

\textbf{\large Aplicaci\'on de las Redes Neuronales al Reconocimiento de Sistemas Operativos}

\vspace{1.5cm}

\textbf{Carlos Sarraute}

\end {center}

\vspace{1.5cm}

\noindent \textbf{Director:} \ Mat\'{\i}as Gra\~na

\vspace{3cm}

\rightline{6/12/2007}


%
%
%

\newpage{\ }
\thispagestyle{empty}
\begin{abstract}
En este trabajo vamos a presentar algunas familias de redes neuronales,
las redes de perceptrones multi-capas, y algunos de los algoritmos
que se usan para entrenarlas (esperamos que con suficientes detalles
y precisi\'on como para satisfacer un p\'ublico matem\'atico).
Luego veremos como usarlas para resolver un problema que surge
del campo de la seguridad inform\'atica:
la detecci\'on a distancia del Sistema Operativo
(una de las etapas de recolecci\'on de informaci\'on, que forman
parte de la metodolog{\'\i}a de pentesting).
Este es el aporte de este trabajo: es una aplicaci\'on 
de t\'ecnicas cl\'asicas de Inteligencia Artificial a un problema de
clasificaci\'on que brind\'o mejores resultados que las 
t\'ecnicas cl\'asicas usadas para resolverlo.

\bigskip
\centering
{\bf Agradecimientos}

A Javier Burroni, Matias Graña, Pablo Groisman y Enrique Segura.

\end{abstract}

\tableofcontents


\chapter{Introducción}

\section{Inteligencia Artificial y analogías biológicas}

En este trabajo vamos a presentar algunas familias de redes neuronales,
y luego veremos cómo usarlas para resolver un problema que surge
del campo de la seguridad informática.

Las redes neuronales fueron creadas dentro del movimiento de investigación en
 Inteligencia Artificial (IA).
La inteligencia artificial surge (como campo) des-pués de la segunda guerra mundial, 
en particular se considera como fecha de nacimiento la 
conferencia de Dartmouth (1956), organizada por 
John McCarthy, Marvin Minsky, Nathaniel Rochester y Claude Shannon. 
En esa época tenían mucha fe en poder 
realizar con la ayuda de máquinas tareas cotidianas para un ser humano:
 algunas de las tareas estudiadas fueron el reconocimiento de patrones visuales,
reconocimiento de texto escrito a máquina o a mano,
 reconocimiento de voz, reconocimiento de rostros, 
el uso del lenguaje, la planificación de tareas, el diagnóstico médico...
En un artículo de 1950, ``Computing machinery and intelligence",
Alan Turing se pregunta si las máquinas pueden pensar.
Como aproximación a la respuesta, propone una prueba (el test de Turing)
para determinar si una máquina puede simular 
una conversación humana: un usuario dialogando mediante un canal escrito
debe determinar si su interlocutor es un ser humano o una máquina.
 
Para realizar estas tareas fueron creadas distintas clases de herramientas:
sistemas expertos, sistemas de inferencia lógica, sistemas de planificación...
 y la que nos interesa aquí: las Redes Neuronales Artificiales (RNA).
 Como el nombre lo indica, la inspiración para la
creación de las RNA fueron los conocimientos que las neurociencias iban adquiriendo
sobre el funcionamiento del cerebro humano. 
Si bien en la presentación que haremos de las redes neuronales (a partir del capítulo 2)
no haremos ninguna referencia a esta inspiración biológica, 
aprovechemos esta introducción para decir unas palabras.

En el cerebro humano hay aproximadamente cien mil millones 
de neuronas ($10^{11}$ células nerviosas). 
Cada neurona tiene un cuerpo celular (soma), que tiene un núcleo celular. 
A partir del cuerpo de la neurona se ramifican una cantidad de fibras 
llamadas dendritas y una única fibra larga llamada axón. 
El axón mide en general 100 veces más que el diámetro del cuerpo 
de la célula (aproximadamente 1 cm). Una neurona se conecta con 
entre 100 y 100.000 neuronas formando una conjunto de 
conexionas sinápticas. 
Se estima que hay unas $10^{14}$ sinapsis en el cerebro de un adulto.
Las señales se propagan entre neuronas 
mediante una reacción electroquímica: 
a nivel de la sinapsis, la neurona que genera la señal emite unos 
neurotransmisores que activan los receptores de la neurona que recibe la señal. 
\begin{figure}
\centering
\includegraphics[width=12cm]{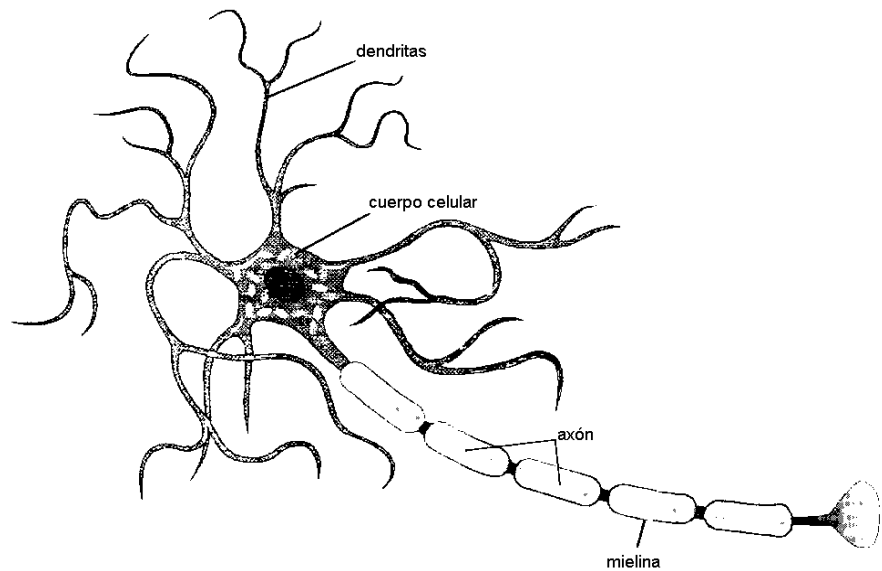}
\caption{Estructura básica de una neurona}
\end{figure}

En 1943, McCulloch y Pitts propusieron un modelo matemático sencillo 
del funcionamiento de las neuronas: la neurona dispara un potencial de acción 
cuando la combinación lineal de sus entradas supera un umbral. 
Este modelo se usa como unidad de procesamiento 
en redes neuronales artificiales, bajo el nombre de perceptron. 
Algunos de los primeros trabajos en IA se basaban en estas redes neuronales artificiales.
Otros nombres para este campo es computación neuronal, procesamiento 
distribuido paralelo y conexionismo.
En 1951 Marvin Minsky fue el primero en desarrollar una red neuronal en hardware.
En 1957 Frank Rosenblatt inventó el perceptrón moderno,
y demostró el teorema de convergencia del perceptrón.
Sin embargo después de esta primer época de entusiasmo,
el libro de Minsky y Papert donde 
analizan los límites de los perceptrones, marca el fin de los primeros esfuerzos de investigación en redes neuronales.
En los años 1980 aparecen los primeros sistemas expertos que resuelven problemas a escala industrial, y la Inteligencia Artificial se convierte en una industria.
Renace el conexionismo, a partir de 1986 se produce el impulso más fuerte,
cuando por lo menos cuatro grupos distintos reinventan en forma independiente
el algoritmo de retro-propagación, mencionado por primera vez por Bryson y Ho en 1969.

\section{Detección remota de Sistemas Operativos}

Uno de los conceptos unificadores de la teoría de la inteligencia artificial 
es el de agente racional: un agente que percibe su entorno, 
persiste durante un período de tiempo prolongado, 
se adapta a los cambios y es capaz de alcanzar diferentes objetivos.

Hemos propuesto en \cite{ataques} un modelo de ataques informáticos basado 
en el concepto de agentes: el atacante es modelado como un agente racional, 
que tiene una base de conocimiento del entorno y es capaz de ejecutar 
acciones para ampliar su conocimiento del entorno 
(lo que se llama Information Gathering) y de instalar nuevos agentes para 
proseguir con el ataque. 
Estas acciones tienen requerimientos y en caso de ejecución exitosa, 
le proporcionan al agente un resultado en la forma de información sobre 
el entorno o de modificación del entorno.

Como aplicación de las redes neuronales, veremos como usarlas
para resolver un problema del campo de la seguridad informática:
la detección a distancia del Sistema Operativo
(SO, en inglés Operating System = OS),
también llamada ``OS Fingerprinting" (toma de huellas del SO).
Es una etapa crucial de un test de intrusión (penetration test),
dado que el atacante necesita conocer el SO del objetivo
para elegir los exploits que va a usar.
Dicho de otra manera, es una de la acciones de Information Gathering 
dentro de nuestro modelo del atacante.
La detección de SO se realiza sniffeando paquetes de la red
en forma pasiva, 
y enviando en forma activa paquetes test al sistema objetivo,
para estudiar variaciones específicas en las respuestas
que revelen su sistema operativo.

Las primeras implementaciones de detección de SO
estaban basadas en el análisis de diferencias entre las implementaciones
de la pila TCP/IP.
La generación siguiente utilizó datos del nivel de aplicaciones,
tales como los puntos finales (endpoints) DCE RPC.
Aunque el análisis se hacía sobre nuevas fuentes de información,
la forma de interpretar esta información seguía siendo una
variación del algoritmo de ``best fit'' que consiste en buscar
el punto más cercano.
Esta estrategia tiene varios defectos:
no funciona en las situaciones no estándar,
y no permite extraer los elementos clave que identifican
en forma única a un sistema operativo.
Pensamos que el próximo paso es trabajar sobre las técnicas
usadas para analizar los datos.

Nuestro nuevo enfoque al problema se basa en el 
análisis de la composición de la información relevada
durante el proceso de identificación del sistema operativo
para descubrir los elementos clave y sus relaciones.
Para implementar este enfoque desarrollamos herramientas
que usan redes neuronales y técnicas de análisis estadístico.
Estas herramientas fueron integradas en un software
comercial llamado Core Impact.
Este software es un framework para automatizar el proceso de penetration testing
(test de intrusión), que consiste en atacar una red con las herramientas 
que usaría un atacante real, para evaluar la seguridad de la red y de las
medidas de protección (sistemas de detección y prevención de intrusiones).
Para comprometer maquinas de la red estudiada y acceder información sensible,
el auditor hace uso de vulnerabilidades en las maquinas objetivos, 
para tratar de explotar esas vulnerabilidades y tomar control de la maquina.

\chapter{Redes Neuronales}

\section{Conceptos básicos}\label{conceptos_basicos}

\begin{defin} 
Un {\em grafo dirigido} es un par $\calG = (\calV,\calE)$ 
donde $\calV$ es un conjunto finito de puntos
llamados nodos o vértices,
y los elementos de $\calE$ son pares ordenados de elementos de $\calV$,
llamados flechas o arcos.
\end{defin}

\begin{defin}\label{def:red}
Una {\em red neuronal} $\calN = ( \calV, \calE, \calW, \calT )$ es una estructura de procesamiento 
de información que se puede representar por un grafo dirigido $\calG = (\calV,\calE)$, 
con las siguientes características:
\begin{itemize}
\item{Los nodos del grafo (elementos de $\calV$) se llaman {\em unidades de procesamiento}, 
{\em elementos de procesamiento} o simplemente {\em neuronas}
(esta última palabra se considera que tiene un ``cool factor").
}
\item{
Las flechas del grafo (elementos de $\calE$) se llaman {\em conexiones}, 
representan un camino que conduce una señal en una única dirección.
}
\item{
Cada flecha $(i,j) \in \calE$ tiene un valor asociado $w_{j i}$ que representa 
la intensidad de la conexión, llamado {\em peso sináptico}
o simplemente {\em peso}. Notamos $\calW$ al conjunto de los pesos.
} 
\item{
Cada unidad de procesamiento $k \in \calV$ genera una señal de salida
a partir de las señales de entrada (las señales que recibe), 
aplicando una función $f_k$ llamada {\em función de transferencia}. 
Notamos $\calT$ al conjunto de funciones de transferencia.
}
\item{
Una unidad de procesamiento puede recibir cualquier cantidad de conexiones entrantes.
Una unidad de procesamiento puede tener varias conexiones salientes, 
pero todas transmiten la misma señal.
}
\item{
La red neuronal recibe un vector de entrada ${\mathbf x}$, de dimensión $d$.
La red neuronal genera un vector de salida ${\mathbf y}$, de dimensión $c$.
}
\item{
Tipos de señales: la señal puede ser un entero, un número real o un número complejo. 
Para nosotros las señales serán siempre números reales (desde el punto de vista teórico, 
o sea números de coma flotante en las implementaciones). 
}
\end{itemize}
\end{defin}

\begin{obs}
La presentación de redes neuronales que hacemos es parcial, solo cubre una porción de las distintas
redes neuronales que existen. Quedan fuera de esta presentación las
redes RBF (Radial basis function), 
las redes auto-organizadas de Kohonen (Kohonen self-organizing network),
las redes recurrentes (Hopfield network, Boltzmann machine),
las memorias asociativas y otros tipos de redes.
No es nuestra intención dar un panorama sobre redes neuronales, 
sino presentar las redes que usamos para la clasificación de sistemas operativos.
\end{obs}

\begin{obs}
Cuando la red neuronal ya está entrenada, los pesos están fijos y 
se puede pensar la red como una función $\calN : \R ^ {d} \rightarrow \R ^ {c}$, 
con una estructura particular,
tal que ${\mathbf y} = \calN (\x)$. 
\end{obs}

\begin{defin}
Las neuronas (o unidades de procesamiento) que transmiten a la red 
las señales de entrada
se llaman {\em neuronas de entrada}, 
las que transmiten las señales de salida 
hacia el ``exterior'' de la red se llaman {\em neuronas de salida}, 
todas las otras se llaman {\em neuronas escondidas}.
\end{defin}

\begin{defin}
Sea $\calG = (\calV,\calE)$ un grafo dirigido. 
Dados $v, w \in \calV$, un {\em camino dirigido} entre $v$ y $w$ es una sucesión
de flechas $e_1, \ldots, e_n$ que conectan $v$ y $w$.
Vale decir que existen nodos $v_1, \ldots, v_{n-1}$ tales que
\begin{align}
e_1 = (v, v_1), \ldots, e_k = (v_{k-1}, v_k), \ldots, e_n = ( v_{n-1}, w)
\end{align}
Si $v=w$ y los nodos $v, v_1, \ldots, v_{n-1}$ son todos distintos diremos que 
el camino es un {\em ciclo dirigido}. 
\end{defin}

\begin{defin}\label{def:pred}
Sea $\calG = ( \calV, \calE )$ un grafo dirigido. Dado un nodo $k \in \calV$
el conjunto de predecesores inmediatos, notado $\calE k$, es el conjunto
\begin{align}
\calE k = \{ i \in \calV : (i,k) \in \calE \}
\end{align}
El conjunto de sucesores inmediatos, notado $k \calE$, es el conjunto
\begin{align}
k \calE = \{ i \in \calV : (k,i) \in \calE \}
\end{align}
\end{defin}

\begin{defin}
Una red neuronal se dice que es de {\em alimentación hacia adelante}
(feed-forward) si el grafo $\calG = (\calV, \calE)$ que representa a la red no contiene
ciclos dirigidos. 
\end{defin}

\begin{obs}
En tal caso, se pueden 
asignar números a los nodos de $\calV$ de manera tal que 
las flechas vayan siempre de menor a mayor:
\begin{align}
(i, j) \in \calE  \Rightarrow  i < j
\end{align}
\end{obs}
\begin{proof}
Describimos una forma de hacerlo. Iniciar un contador $c=1$.
Elegir un nodo $v$ del grafo. Si $\calE v \neq \emptyset$,
elegir un nuevo nodo en $\calE v$. Repetir hasta encontrar
un nodo $w$ tal que $\calE w = \emptyset$. 
Siempre vamos a encontrar uno porque el grafo tiene finitos nodos,
y no podemos pasar nunca dos veces por el mismo nodo (significaría encontrar
un ciclo dirigido).

Asignarle a ese nodo $w$ el número $c$, incrementar $c$, borrar el nodo $w$ del grafo
y volver a hacer lo mismo con el grafo reducido. Al disminuir la cantidad de nodos en cada paso,
estamos seguros de que el algoritmo termina en $\# \calV$ pasos.
Si existe una conexión desde el nodo $i$ al nodo $j$ (donde $i,j$ representan los números
asignados por este algoritmo), al momento de asignarle número al nodo $j$, el nodo $i$ ya había sido
borrado del grafo, y por lo tanto $i < j$.
\end{proof}

\begin{defin}\label{def:capas}
Las redes neuronales usadas en la práctica 
tienen sus unidades de procesamiento distribuidas en capas.
Vale decir que existe una partición del conjunto $\calV$
en subconjuntos $C_0, \ldots, C_m$ llamados {\em capas } (layers)
tales que no hay conexiones entre unidades de la misma capa
\begin{align}
i \in C_a  \,\wedge\, j \in C_a \Rightarrow (i,j) \not\in \calE
\end{align}
y las conexions son entre una capa y la siguiente
\begin{align}
i \in C_a \,\wedge\, j \in C_b \,\wedge\,  (i,j) \in \calE \Rightarrow b = a+1.
\end{align}
La capa $C_0$ corresponde a las neuronas de entrada
(y se llama lógicamente {\em capa de entrada}), 
la capa $C_m$ a las neuronas de salida ({\em capa de salida}),
las otras capas a las neuronas escondidas ({\em capas escondidas}).
En este caso, se dice que la red tiene $m$ capas (no se cuenta
la capa de entrada, en la cual no se efectúa ningún procesamiento).
\end{defin}

\section{Redes con una capa}

Veamos en primer lugar las redes en las cuales
hay una sola capa de neuronas.
Las neuronas de entrada se conectan con las neuronas de salida
según flechas con pesos adaptativos (adaptive weights).

\subsection{Funciones discriminantes lineales}

\begin{defin}
Consideremos el problema de clasificar un conjunto de datos $X \subset \R^d$ en dos clases
($X$ es simplemente un conjunto de puntos en $\R^d$).
Llamamos {\em función discriminante} a una función $y : X \rightarrow \R$ 
tal que $\x \in X$ se asigna a la clase $\mathcal{C}_1$ si $y(\x) > 0$ 
y a la clase $\mathcal{C}_2$ si $y(\x) < 0$.

La forma más simple que puede tomar la función discriminante es
\begin{align}\label{eq_fdl}
y (\x) \; = \; \sum_{ i = 1 }^{ d } { w_i  x_i }  + w_0  \; = \; \w ^T  \x  +  w_0 
\end{align}

El vector $\w$ de dimensión $d$ se llama vector de pesos, 
el parámetro $w_0$ (que cumple $w_0 < 0$) se llama el {\em sesgo} (bias), aunque tiene un significado
un poco distinto del sesgo estadístico.
A veces $-w_0$ se llama el {\em umbral} (threshold), en referencia al modelo 
biológico de McCulloch y Pitts \cite{culloch}: 
cuando la suma ponderada de las señales que recibe la neurona supera el umbral, 
se dispara una señal por el axón de la neurona.
\end{defin}

\begin{figure}
\centering
\includegraphics[width=8cm]{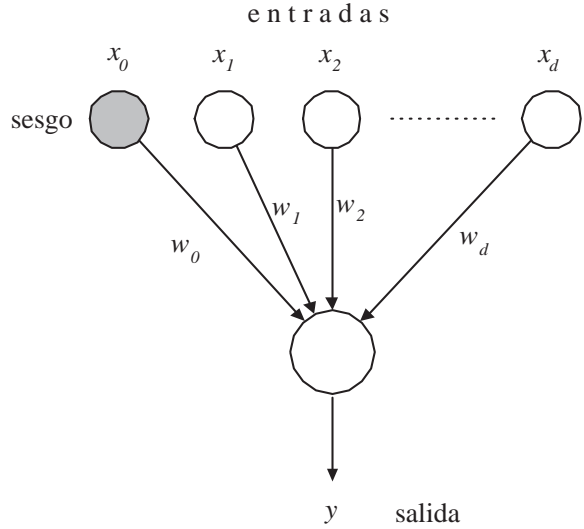}
\caption{Representación de una función discriminante lineal como 
unidad de una red neuronal.
El sesgo $w_0$ se puede considerar que corresponde a una entrada
constante $x_0 = +1$.}
\end{figure}

La interpretación geométrica de la expresión \eqref{eq_fdl} en el espacio $\R ^ d$
es simple: la frontera de decisión $y(\x) = 0$ es un hiperplano (de dimensión $d-1$)
que separa el espacio en dos partes, ortogonal al vector $\w$.
La distancia del hiperplano al origen es:
\begin{align}
dist =  - \frac { \w ^T \x} { ||\w|| }  = - \frac { w_0 } { || \w || }
\end{align}

Los discriminantes lineales se pueden extender a varias clases
${\mathcal C}_1 \ldots {\mathcal C}_c$
introduciendo una función discriminante $y_k$ para cada clase ${\mathcal C}_k$
de la forma
\begin{align}
y_k (\x) = \w_k^T \x  +   w_{k 0}
\end{align}

Dado un punto $\x$, será asignado a la clase ${\calC}_k$ 
si $y_k(\x) > y_j(\x) $ para todo $ j \neq k$.
La frontera de decisión entre las clases $\calC_k$ y $\calC_j$
viene dada por $y_k(\x) = y_j(\x)$ que corresponde al hiperplano
\begin{align}
(\w_k - \w_j)^T \x + (w_{k 0} - w_{j 0} ) = 0
\end{align}

En la representación de la red neuronal como un grafo, 
los pesos $w_{k i}$ corresponden a la conexión entre 
la entrada $x_i$ y la unidad de procesamiento $k$.
Resulta conveniente agregar un nodo de entrada con valor fijo $x_0 = 1$
para el sesgo, de manera que las funciones discriminantes se pueden escribir
\begin{align}
y_k (\x) = \sum_{i = 0}^{d} w_{k i} x_i
\end{align}

\begin{figure}
\centering
\includegraphics[width=10cm]{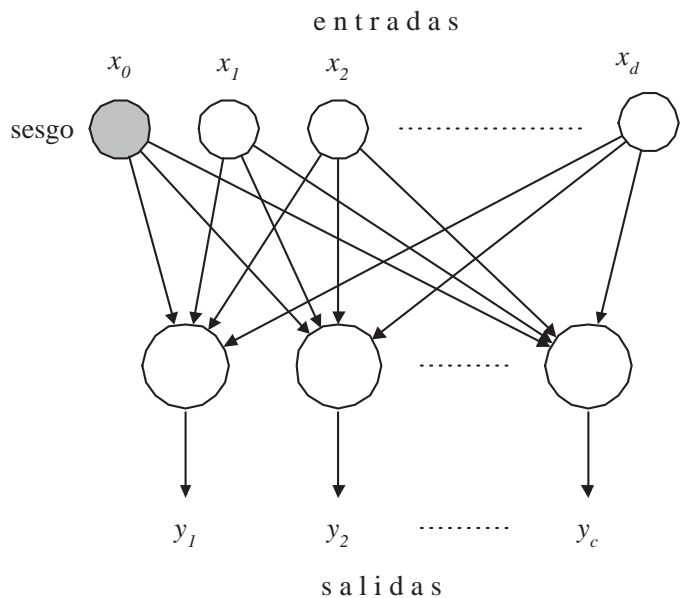}
\caption{Representación de funciones discriminantes lineales $y_k(\x)$
como diagrama de una red neuronal.}
\end{figure}

\subsection{Función de activación}

\begin{defin}
Una forma de generalizar las funciones discriminantes lineales 
es aplicar a la suma una función no lineal $g$, 
llamada {\em función de activación}.
En el caso de la clasificación en dos clases queda
\begin{align}
y = g( \w ^T \x + w_0 )
\end{align}
En general se usa una función monótona. Por lo tanto
la expresión anterior se puede seguir considerando una 
función discriminante lineal, dado que las fronteras de decisión 
siguen siendo lineales.
Para esta unidad de procesamiento, la función de transferencia
es el resultado de componer la función discriminante lineal con la 
función de activación.
\end{defin}

\begin{ejem}
Una elección clásica es la función de activación  logística sigmoidal
\begin{align}\label{eq:sigmoidal}
g(a) = \frac { 1 } { 1 + e^{-a} }
\end{align}
El término sigmoidal quiere decir con forma de S, 
esta función transforma
el intervalo $( - \infty, + \infty )$ en el intervalo $(0,1)$ 
y se puede aproximar por una función lineal 
cuando $|a|$ es pequeño.
\end{ejem}

\begin{ejem}
Otra elección similar es la tangente hiperbólica
\begin{align}\label{eq:tanh}
\tanh(a) = \frac { e^a - e^{-a} } { e^a + e^{-a} }
\end{align}
que difiere de la función logística sigmoidal $g$ solo por unas transformaciones lineales
\begin{align*}
2 \; g(2a) - 1 &= \frac {2} {1 + e^{-2a} } - 1 \\
&= \frac {1 - e^{-2a} } { 1 + e^{-2a} }  = \tanh(a)
\end{align*}
Vale decir que una red neuronal que use la función de activación \eqref{eq:tanh}
es equivalente a una que use \eqref{eq:sigmoidal} con distintos valores para los pesos
y sesgos. 
\end{ejem}

\begin{ejem}
Otra función de activación introducida por McCulloch y Pitts (1943)
para modelar el comportamiento de una neurona en un 
sistema nervioso biológico
es la función escalonada de Heaviside
(también llamada función umbral)
\begin{align} \label{eq:umbral}
g(a) = \left\{ \begin{array}{l l}
 0 & \textrm { si $a < 0$ } \\
 1 & \textrm { si $a \geq 0$ }
\end{array} \right.  
\end{align}
donde $ a = \w ^T \x + w_0 $.
Las entradas $x_i$ representan el nivel de actividad de las otras neuronas,
las pesos $w_i$ la fuerza de las conexiones sinápticas entre las neuronas,
$w_0$ un umbral a partir del cual la neurona dispara un nuevo potencial 
de acción. Este modelo inspirado en la biología se usó para el 
reconocimiento estadístico de patrones.
Rosenblatt (1962) estudió redes de unidades con función de activación
escalonada, que llamó {\em perceptrones}. También fueron estudiadas
por Widrow y Hoff (1960), que las llamaron {\em adalines}.
\end{ejem}

\begin{ejem}
Usando la función de activación umbral $g$, podemos representar funciones
booleanas básicas.
Esta fue una de las motivaciones de McCulloch y Pitts para el 
diseño de unidades individuales.
\begin{align*}
 \mbox{ AND} ( x_1, x_2 ) &= g ( x_1 + x_2 - 1,5 )  \\
 \mbox{ OR } ( x_1, x_2 ) &= g ( x_1 + x_2 - 0,5 ) \\
 \mbox{ NOT } ( x_1 ) &= g( - x_1 + 0,5 )
\end{align*}
Esto quiere decir que usando estas unidades podemos construir una red neuronal
que calcule cualquier función booleana de las entradas.
\end{ejem}

\subsection{Separabilidad lineal}

\begin{figure}
\centering
\includegraphics[width=12.5 cm]{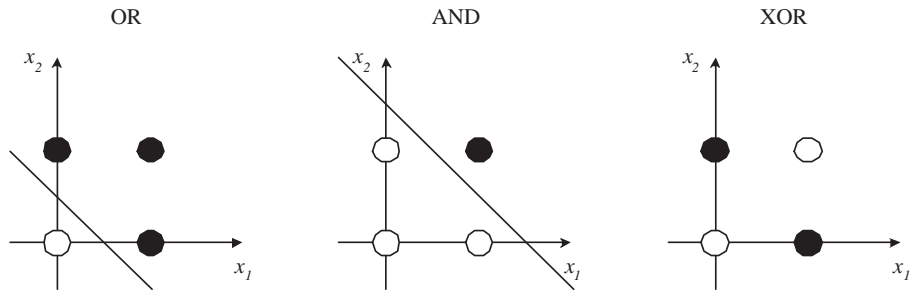}
\caption{Las funciones lógicas OR y AND son linealmente separables.
En cambio la función XOR no es linealmente separable.}
\label{fig:xor}
\end{figure}

Las funciones discriminantes que hemos visto tienen
tienen fronteras de decisión lineales (en general, un hiperplano en un
espacio de dimensión $d$).
Esto es una restricción importante de los sistemas con una sola capa
y es la motivación principal para considerar sistemas con varias
capas de neuronas.
 
\begin{defin}
Consideremos el problema de clasificación en dos clases $\calC_1$ y $\calC_2$.
Si se puede encontrar un hiperplano que separe los puntos de $\calC_1$ y 
$\calC_2$, entonces los puntos se dicen {\em linealmente separables}.
\end{defin}

\begin{obs}
En los años 60, Marvin Minsky y Seymour Papert iniciaron una campaña
para desacreditar la investigación en redes neuronales 
(y promover su propia visión de la inteligencia artificial).
Los argumentos de esta campaña fueron publicados en el libro
{\em Perceptrons} en 1969.
La crítica fundamental de Minsky y Papert era que el perceptron no sirve
para modelar la función XOR en dos dimensiones, dado que las clases
$\calC_1 = \{ (0,0), (1,1) \} $ y $\calC_2 = \{ (0,1), (1,0) \}$
no son linealmente separables
(ver figura \ref{fig:xor}).
Esto nos motiva a usar redes neuronales
con por lo menos dos capas para modelar problemas no lineales.  
Sin embargo redes con una sola capa siguen siendo
importantes en la práctica dado que son muy rápidas
de entrenar.
\end{obs}

\begin{obs}
Si no pedimos que la función de activación sea monótona,
tomando 
\begin{align}
g(a) = \frac {1} { (1 + e^{-a}) \cdot (1 + e^{a - 10} )  } - \frac{1}{2}
\end{align}
obtenemos que las clases $y(\x) > 0$ e $y(\x) < 0$ separan los valores de XOR
(ver figura \ref{fig:xor2}).

\begin{figure}
\centering
\includegraphics[width=4 cm]{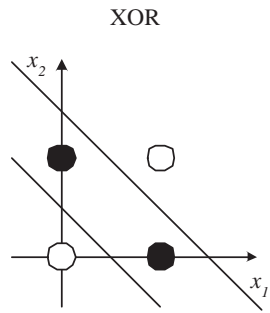}
\caption{Solución con una función de activación no monótona.}
\label{fig:xor2}
\end{figure}

\end{obs}

\subsection{El Perceptron de Rosenblatt}

Rosenblatt estudió redes de neuronas de una capa, 
usando la función de activación umbral \eqref{eq:umbral}. 
En 1962 publicó 
``Principios de Neurodinámica: Perceptrones y la
teoría de los mecanismos cerebrales'' \cite{rosenblatt}.
Construyó implementaciones en hardware de estas redes de perceptrones,
cuyo mecanismo de entrenamiento veremos a continuación.
Las usó para problemas de clasificación, usando como entrada imágenes
blanco y negro de letras y formas simples.

Al mismo tiempo que Rosenblatt desarrollaba los perceptrones,
Widrow y Hoff trabajaban en sistemas similares que llamaron
{\em adalines}, cuyo nombre viene de ADAptive LINear Elements,
y se refiere a una unidad de procesamiento con función de activación no lineal
muy parecida al perceptron \cite{widrow}.

Ya hablamos de las limitaciones de una red con una sola capa de pesos:
solo puede discriminar regiones linealmente separables.
Para mejorar la capacidad del perceptron, Rosenblatt usó una capa de 
elementos de procesamiento fijos para transformar los datos de entrada.
Estos eran habitualmente unos cables fijos conectados a un conjunto aleatorio
de los pixeles de entrada, y emitían su salida usando una función de 
activación \eqref{eq:umbral}. 
Notaremos $f_j$ a estos elementos de procesamiento fijos. 
Como ya es costumbre, agregamos una función $f_0$
cuya salida es siempre 1, con su correspondiente sesgo $w_0$.
Por lo tanto la salida del perceptron está dada por
\begin{align}
y = g \left ( \sum_{j=0}^{M} w_j  f_j (\x) \right ) = g ( \w ^T  \bf f )
\end{align}
La función de activación usada es de la forma (versión antisimétrica)
\begin{align}
g(a) = \left\{ \begin{array}{l l}
 -1 & \textrm { si $a < 0$ } \\
 +1 & \textrm { si $a \geq 0$ }
\end{array} \right.  
\end{align}

\subsection{El criterio del Perceptron}

Para entrenar la red, se necesita una función de error que se pueda minimizar fácilmente.
Consideramos aquí una función continua y lineal a trozos llamada el 
criterio del perceptron. Cuando un vector de entrada $\x ^n$ es presentado a los sensores 
de la red, genera un vector de activaciones $\ff ^n$ en la primer capa de elementos fijos.
A cada vector de entrada $\x^n$ asociamos un valor esperado $t^n$ que
vale $t^n = +1$ si la entrada pertenece a la clase $\calC_1$ 
y vale $t^n = -1$ si la entrada pertenece a la clase $\calC_2$. 
Queremos que se cumpla $\w^T \ff^n > 0$ para los vectores de la clase $\calC_1$
y $\w^T \ff^n < 0$ para los vectores de la clase $\calC_2$.
Podemos simplificar estas condiciones, pidiendo que para todos los vectores de entrada
\begin{align}
\w ^T (\ff^n t^n) > 0
\end{align}
Esto sugiere minimizar la función de error siguiente, conocida como criterio del perceptron
\begin{align}
E^{perc} ( \w ) = \sum_{\ff^n \in \calM} - \w^T (\ff^n t^n)
\end{align}
donde $\calM$ es el conjunto de todos los vectores $\ff^n$ que fueron mal clasificados
con el vector de pesos actual $\w$. Para los vectores clasificados erroneamente
$\w ^T (\ff^n t^n) < 0$ luego $E^{perc}$ es una suma de términos positivos
y vale 0 si todos los vectores están correctamente clasificados.

\section{Redes con varias capas}
\label{sec:red_varias_capas}

\begin{figure}
\centering
\includegraphics[width=12 cm]{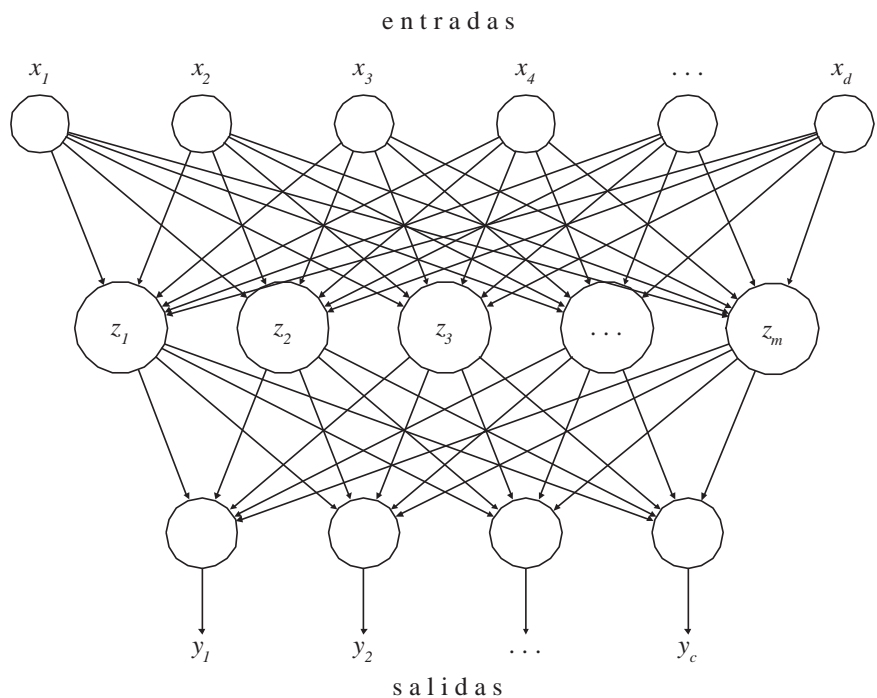}
\caption{Red neuronal de perceptrones con dos capas}
\end{figure}

En la sección anterior vimos redes donde las neuronas de salida
se conectan directamente con las neuronas de entrada, que solo funcionan
para clasificar conjuntos linealmente separables.
Consideramos ahora redes con unidades ocultas organizadas en capas, 
donde cada unidad recibe entradas únicamente de las unidades
de la capa que la precede inmediatamente.

Cuando definimos los conceptos básicos de redes neuronales (en la sección \ref{conceptos_basicos})
dijimos que la salida de una unidad de procesamiento $k \in \calV$ era una 
función $f_k$ (llamada función de transferencia)
de las entradas de la unidad y de los pesos de las conexiones con unidades anteriores.
En la práctica, la función de transferencia consiste
en una función discriminante lineal seguida por una función de activación $g_k$:
\begin{align}\label{eq:gk}
y = f_k( \x ) = g_k \left( \sum_{i \in \calE k}w_{k i} \; x_i \right)
\end{align}
Llamando $\calA = \{ g_k : k \in \calV \} $ al conjunto de las funciones de activación,
podemos describir una red neuronal como una tupla $\calN = (\calV, \calE, \calW, \calA) $
en lugar de $\calN = ( \calV, \calE, \calW, \calT )$:
precisamos las funciones de activación en lugar de las funciones de transferencia.
De ahora en adelante todas las redes neuronales tendrán esa forma.

\subsection{Notación para las neuronas}

Consideremos una red neuronal de $m$ capas, cuyas unidades están dispuestas 
en capas $C_0, \ldots, C_m$.
Hay (por lo menos) dos opciones para indexar las unidades de la red:
numerarlas en forma secuencial, ó elegir indices que reflejen la disposición en capas.
Siguiendo la segunda opción, indicamos con $(i, j)$ a la neurona que ocupa
el $j$-ésimo lugar dentro de la capa $i$. Notemos $z^{(i)}_{j}$ al valor de activación
(señal de salida) de esa neurona.
Como casos particulares tenemos los valores de entrada de la red $\x = (x_1, \ldots, x_d)$
\begin{align}
x_i = z^{(0)}_{ i} \mbox{ para } 1 \leq i \leq d
\end{align}
y los valores de salida de la red $\y = (y_1, \ldots, y_c)$
\begin{align}
y_k = z^{(m)}_{k} \mbox{ para } 1 \leq k \leq c
\end{align}
Las conexiones son siempre entre capas sucesivas, por ejemplo desde la neurona 
$(i-1, j)$ hacia la neurona $(i, k)$.
De acuerdo con la definición \ref{def:red}, el peso de esa conexión debería notarse
$w_{ (i,k), (i-1,j) }$, podemos aliviar un poco esta notación dejando $w^{(i)}_{kj}$.

Supongamos (caso típico) que las neuronas de una capa
están conectadas con todas las neuronas de la capa siguiente.
Llamemos $d_i$ a la cantidad de neuronas en la capa $C_i$ 
(en particular la dimensión de entrada $d_0 = d$ y 
de salida $d_m = c$).
La fórmula \eqref{eq:gk} se escribe entonces
\begin{align}
z^{(i)}_{k} = g^{(i)}_{k} \left ( \sum_{j=1}^{d_{i-1}} w^{(i)}_{k j} \; z^{(i-1)}_{j} \right )
\end{align}

Veamos la otra opción: numerar las neuronas en forma arbitraria
 (por ejemplo empezando por las neuronas de entrada y terminando por las de salida).
Como la red tiene topología de alimentación hacia adelante,
podemos hacerlo de manera que $(i,j) \in \calE \Rightarrow i < j$.
El valor de activación de $i \in \calV$ se nota $z_i$ y el peso de la conexión 
desde $i$ hacia $j$ se nota $w_{j i}$ (ver definición \ref{def:red}).

Usando la notación introducida en la definición \ref{def:pred},
la fórmula anterior se escribe como 
\begin{align}\label{eq:zk2}
z_k = g_k \left( \sum_{ i \in \calE k }  w_{k i} \; z_i \right )
\end{align}
La ventaja de esta notación es que se generaliza en forma inmediata 
a una red neuronal con topología arbitraria 
(donde no se pide que las neuronas estén distribuidas en capas).

\begin{ejem}
Veamos un último ejemplo de la fórmula de propagación hacia adelante.
El caso más común de redes con unidades ocultas consiste en una red con una única capa oculta
(según la definición \ref{def:capas}, una red con dos capas, aunque algunos autores
contando la capa de entrada $C_0$ la llaman red de tres capas).
Para precisar ideas, supongamos que se usan solo 2 funciones de activación diferentes.
La fórmula para una red con dos capas, que usa
la función de activación $g$ en la primer capa y $\hat{g}$ en la segunda es
\begin{align}
y_k = \hat{g} \left( \sum_{j=0}^{d_1} w^{(2)}_{k j} \;  g \left( \sum_{i=0}^{d} w^{(1)}_{j i} \; x_i \right)  \right)
\end{align}
\end{ejem}

\subsection{Expresividad: teorema de Kolmogorov}

En esta sección vamos a abordar la pregunta acerca de la expresividad
de las redes de perceptrones con varias capas.
Una pregunta importante (por lo menos desde el punto de vista teórico) es: 
cuales son las funciones que pueden representarse como una red neuronal?

Una respuesta proviene de un teorema de Andrei Kolmogorov de 1957 [Kol].
El origen de este teorema se remonta al año 1900,
cuando David Hilbert publicó su famosa lista de 23 problemas para los
matemáticos del siglo XX.
El problema 13 de Hilbert abordaba el problema de representar
funciones de varias variables como superposición de funciones de menos variables
(una superposición es una función de funciones).
Conjeturó que existen funciones de 3 variables que no se pueden representar
como superposición de funciones de 2 variables.
La conjetura fue refutada por Vladimir Arnold (1957).
Sin embargo, Kolmogorov probó un resultado mucho más general y sorprendente.
Sea $I$ el intervalo unitario $I = [ 0, 1 ]$.

\begin{teo}
(Kolmogorov, 1957).
Dada una dimensión $d \geq 2$,
existen funciones reales continuas $h_{ki} (x) $ definidas en el intervalo $I$
tales que:
para toda función real continua $f(x_1, \ldots, x_d)$ definida en el hipercubo $I^d$,
existen funciones reales continuas $g_k$ tales que
\begin{align}
f(x_1, \ldots, x_d) = \sum_{k=1}^{2d+1} g_k \left(  \sum_{i=1}^{d} h_{ki} (x_i)   \right)
\end{align}
\end{teo}

Este teorema fue extendido por George Lorentz en 1966 y luego en 1976 
(ver ``Sobre el problema 13 de Hilbert'' [Lor2]).

\begin{teo}\label{teo:lorentz}
Dada una dimensión $d \geq 2$,
existen constantes $\{\lambda_1, \ldots, \lambda_d \}$ con
\begin{align}
\forall i \; \lambda_i > 0 \; \mbox{ y } \;
 \sum_{i=1}^{d} \lambda_i \leq 1
\end{align}y funciones reales $\{ h_1, \ldots, h_{2d+1} \}$ continuas estrictamente crecientes 
$h_k : I \rightarrow I$, con la propiedad siguiente:\\
Para toda $f \in C(I^d)$, existe $g \in C(I)$ tal que
\begin{align}\label{eq:lorentz}
f(x_1, \ldots, x_d) = \sum_{k=1}^{2d+1} g \left( \sum_{i=1}^{d}  \lambda_i \; h_k (x_i)   \right) 
\end{align}
\end{teo}
En otras palabras dice que toda función continua de varias variables se puede
representar como superposición de funciones continuas de una variable.
Es bastante sorprendente que los coeficientes $\lambda_j$ 
y las funciones $h_k$ están fijas (no dependen de la función $f$ que queremos representar).
Ahora la expresión \eqref{eq:lorentz} se puede transformar en forma bastante directa en
una red neuronal, para obtener el teorema siguiente.

\begin{teo} (Teorema de Existencia de Redes Neuronales)
Dada una función continua $f : I^d \rightarrow \R$,
existe una red neuronal de 3 capas que implementa exactamente $f$.
En esta red, la capa de entrada tiene $d$ neuronas, 
la primer capa oculta $d \cdot (2d+1)$ neuronas,
la segunda capa oculta $2d+1$ neuronas,
y la capa de salida tiene 1 neurona.
\end{teo}

\begin{proof}
De acuerdo con el teorema \ref{teo:lorentz}, la función $f$ se puede representar
como \eqref{eq:lorentz}. Modelemos esa expresión como red neuronal.
La capa de entrada tiene las variables $x_1, \ldots, x_d$.
En la primer capa escondida, ponemos $d$ grupos con $2d+1$ neuronas cada uno,
cada grupo tiene neuronas con funciones de activación $h_1, \ldots, h_{2d+1}$
conectadas con una única neurona de entrada, generando 
el valor de activación
\begin{align}
z^{(1)}_{k i} = h_k (x_i)
\end{align}
En la segunda capa escondida, cada neurona está conectada con $d$ neuronas
de la capa anterior, con pesos $\lambda_i$, y función de activación $g$.
Genera el valor
\begin{align}
z^{(2)}_{k} = g \left( \sum_{i=1}^{d} \lambda_i \; z^{(1)}_{k i} \right)
\end{align}
En la neurona de salida, tomando la identidad como función de activación,
calculamos la suma de los términos anteriores para obtener
\begin{align}
y = \sum_{k=1}^{2d+1} z^{(2)}_{k}
 =  \sum_{k=1}^{2d+1} g \left( \sum_{i=1}^{d}  \lambda_i \; h_k (x_i)   \right)
\end{align}
\end{proof}

\begin{coro}
Dada una función continua $\ff : I^d \rightarrow \R^c$,
existe una red neuronal con 3 capas que implementa exactamente $\ff$.
En esta red, la capa de entrada tiene $d$ neuronas, 
la primer capa oculta $d \cdot (2d+1)$ neuronas,
la segunda capa oculta $2d+1$ neuronas,
y la capa de salida tiene $c$ neuronas.
\end{coro}

\begin{proof}
Para cada componente $f_m$ ($1 \leq m \leq c$), existe una función de activación $g_m$.
Simplemente extendemos la red anterior, para que en cada neurona de salida
\begin{align}
y_m = \sum_{k=1}^{2d+1} g_m \left( \sum_{i=1}^{d}  \lambda_i \; h_k (x_i)   \right)
\end{align}
\end{proof}

\begin{obs}
El teorema de Kolmogorov es notable desde un punto de vista teórico:
nos dice que cualquier función continua se puede representar como red neuronal
de simplemente 3 capas, y que aproximar funciones con redes neuronales tiene
sentido.
Desde el punto de vista práctico, el teorema nos garantiza la existencia de las
funciones $g$ y $h_k$, pero no nos da ninguna técnica constructiva para
encontrarlas. Por otra parte, solo se puede garantizar que esas funciones sean continuas.
Si pedimos que las funciones $h_k$ sean diferenciables, el teorema deja de ser cierto.

Además la función de activación $g$ depende de la función $f$ a implementar.
Esto es al revés del funcionamiento normal de las redes neuronales:
las funciones de activación están fijas, y son los pesos los que varían.
Otra característica de las redes neuronales (que no se cumple acá)
es que sirven para modelar
una función $f$ de la cual solo conocemos el valor en algunos puntos 
(el conjunto de datos $\calD$).
\end{obs}

\chapter{Entrenamiento y Preprocesamiento}\label{chap:entrenamiento}

\section{Aprendizaje de observaciones}

Hemos visto en el capítulo anterior la estructura de las redes de perceptrones.
Ahora abordamos la tarea de lograr que la red ``aprenda'' a resolver un problema.
Las redes neuronales se pueden usar para problemas de regresión (calcular una función
continua de las entradas) o problemas de clasificación (calcular una función de valores discretos: la pertenencia a una clase).

Suponemos que existe una función $\y = \ff (\x)$ con $\ff : \R^d \rightarrow \R^c$
y vamos a usar una red neuronal para aproximar la función $\ff$.
La realidad es un poco más compleja, dado que en general solo podemos
observar $\hat{\y} = \bR( \ff(\x) )$ donde $\bR$ es una función de ruido (estocástica).

\begin{defin}
El aprendizaje se dice {\em supervisado} cuando el objetivo es aprender una función a partir
de ejemplos de sus entradas y salidas. El entrenamiento de las redes neuronales
es un caso de aprendizaje supervisado.
\end{defin}

\begin{defin}
El conjunto de datos $\calD$ es un conjunto de patrones de la forma $(\x^n, \bt^n)$ 
donde $\x^n \in \R^d$ es una entrada y $\bt^n \in \R^c$ es la salida esperada,
vale decir que $\bt^n = \ff(\x^n)$ si $\ff$ es la función a modelar.
Llamamos $N$ a la cantidad de patrones, de manera que
\begin{align}
\calD = \{ (\x^n, \bt^n ) : 1 \leq n \leq N \}
\end{align}
\end{defin}

\begin{defin}\label{nosupervisado}
El aprendizaje se dice {\em no supervisado} cuando el objetivo es aprender a partir de
patrones de entrada para los que no se especifican los valores de salida.
Veremos técnicas no supervisadas de preprocesamiento del conjunto de datos.
\end{defin}

La forma de representar la información aprendida es mediante 
la modificación de los pesos sinápticos.
Durante el entrenamiento, consideramos la red neuronal como una función 
que depende también de los pesos $w_1, \ldots, w_M$ :
\begin{align}
\y = \calN (\x, \w) \; \textrm{  con  } \; \calN : \R ^ d \times \R ^ M \rightarrow \R ^ {c}   
\end{align}

\begin{obs}
El algoritmo para aprendizaje supervisado recibe como entrada el valor correcto $\bt^n$
para determinados valores $\x^n$ de una función desconocida $\ff$ y debe aproximar la función.
Esta es una tarea de inferencia inductiva (o inducción): dada la colección de ejemplos $\calD$,
devolver una función $\bh$ (llamada hipótesis) que aproxime a $\ff$.
Una hipótesis estará {\em bien generalizada} si puede predecir ejemplos que no conoce.
Por lo tanto, hay que evaluar el desempeño de la red con un conjunto de datos independiente de $\calD$.
\end{obs}

\begin{obs}
Cuando el conjunto de hipótesis posibles es muy grande, ese grado de libertad puede
llevar a encontrar ``regularidades'' poco significativas. Este fenómeno se llama {\em sobreajuste}:
la red memoriza los datos, pero no los generaliza.
Se puede producir si la estructura de la red es demasiado grande (con muchas capas o muchas neuronas
en las capas escondidas).
Es como hacer mínimos cuadrados con polinomios de grado alto: se ajustan mejor a los datos
de entrenamiento, pero producen una función con mayores oscilaciones,
y por lo tanto una representación más pobre de la función $\ff$ que queremos aproximar. 
\end{obs}

\section{Algoritmo de retro-propagación}
\label{sec:retropropagacion}

El problema del entrenamiento consiste entonces en aprender una función a partir de un 
conjunto de datos $\calD$.
El aprendizaje consiste en modificar los pesos $\w$, dicho de otra manera,
en moverse en el espacio de pesos hasta encontrar un valor óptimo.
Para evaluar como nos está yendo, necesitamos una función de error.
Un ejemplo importante es la suma de diferencias al cuadrado
\begin{align}
E = \frac{1}{2} \sum_{n=1}^{N} || \calN (\x^n, \w) - \bt^n || ^ 2
\end{align}
Veremos esta función de error más en detalle en la sección \ref{sec:error}.
Por ahora consideramos una función de error genérica $E$,
que vamos a minimizar respecto de los pesos.

Consideramos funciones de activación $g$ diferenciables
(por ejemplo, la función logística sigmoidal o la tangente hiperbólica), de manera que
la salida es una función diferenciable respecto de las variables 
de entrada y respecto de los pesos:
\begin{align}\label{eq:gk}
y = g \left( \sum_{i \in \calE k}w_{k i} \; x_i \right)
\end{align}
Por lo tanto, la función de error $E$
se puede pensar como una función de los pesos.
Se pueden calcular derivadas del error respecto de los pesos,
y usar esas derivadas para encontrar pesos que minimicen 
la función de error.

Los algoritmos de entrenamiento son procesos iterativos,
donde los pesos se van ajustando en una serie de pasos.
Podemos descomponer cada paso en dos etapas:
\begin{itemize}
\item{Evaluar las derivadas de la función de error respecto de los pesos.
El algoritmo de retro propagación (back propagation) cumple esta función.
}
\item{Usar estas derivadas para ajustar los pesos. Una de las técnicas para
hacer esto es el descenso del gradiente (gradient descent).
}
\end{itemize}

\subsection{Evaluación de las derivadas de la función de error}

En esta sección vamos a derivar el algoritmo de retro-propagación 
siguiendo Bishop \cite{bishop}.
El algoritmo de retro-propagación fue publicado por primera vez por Bryson y Ho en 1969.
Sin embargo, no se enteró todo el mundo, y se volvió a descubrir
varias veces en forma independiente, por ejemplo
por Werbos en 1974 y Parker en 1985. 

Consideramos una red neuronal $\calN = (\calV, \calE, \calW, \calA) $ con una topología de alimentación hacia
adelante (feed-forward) arbitraria, 
funciones de activación no lineales diferenciables arbitrarias $\calA = \{ g_k : k \in \calV \} $, 
y una función de error diferenciable $E$ arbitraria.

Sea $\calG = (\calV,\calE)$ el grafo asociado a la topología de la red, cada unidad
de procesamiento $k \in \calV$ calcula una suma ponderada de sus entradas
\begin{align}\label{eq:ak}
a_k = \sum_{i \in \calE k} w_{k i} \; z_i
\end{align}
donde la suma se toma sobre todas las unidades conectadas con la unidad $k$,
o sea tales que $(i,k) \in \calE$, y $w_{k i}$ es el peso asociado a esa conexión.
El valor $z_i$ es el valor de activación de la unidad $i$,
con esta notación incluimos los valores de entrada de la red
(usualmente notados $x_i$) y los valores de salida de la red
(usualmente $y_i$).
También unificamos los pesos y los sesgos, considerando que los
sesgos son pesos que corresponden a neuronas con valor de activación constantemente 
igual a +1.
A esta suma ponderada se le aplica una función de activación $g_k$
para obtener
\begin{align}\label{eq:zk}
z_k = g_k (a_k) = g_k  \left( \sum_{i \in \calE k} w_{k i} \; z_i \right)
\end{align}

Dado un patrón $(\x, \vv)$ fijo en el conjunto de datos $\calD$
(un vector de entrada $\x$ con su salida esperada ${\mathbf v}$),
consideramos el error como una función diferenciable respecto de las
variables de salida de la red
\begin{align}
E = E(y_1, \ldots, y_c)
\end{align}
Como las $y_k$ dependen en forma diferenciable de los pesos, 
el error también.

\begin{defin}
Dado el vector de entrada $\x$, calculamos los valores de
activación de las neuronas escondidas y de salida
aplicando sucesivamente la fórmula \eqref{eq:zk}.
Este proceso se llama {\em propagación hacia adelante} (forward propagation),
dado que hacemos fluir la información siguiendo las flechas del grafo.
\end{defin}

\begin{lema}\label{obs:odew}
La complejidad del algoritmo de propagación hacia adelante es $ \Ode ( \# \calW ) $
\end{lema}
\begin{proof}
Como el grafo de la red neuronal no tiene ciclos dirigidos, 
podemos numerar los elementos de $\calV$ de manera que
$ (i,j) \in \calE \Rightarrow i < j$.
Si recorremos las neuronas siguiendo el orden numérico,
cuando aplicamos la fórmula \eqref{eq:zk} para calcular el valor de $z_k$,
los términos $z_i$ que aparecen en la suma ya han sido calculados
(dado que $i \in \calE k \Rightarrow i < k$).
Vale decir que el algoritmo está bien definido.

Cuando la cantidad de neuronas y de conexiones crece, 
la cantidad de conexiones (igual a la cantidad de pesos) 
tiende a ser mucho más grande que la cantidad de neuronas.
En el algoritmo de propagación hacia adelante,
cada peso interviene en una multiplicación y una suma
(despreciamos las evaluaciones de las funciones de activación, 
dado que son tantas como neuronas en la red),
por lo tanto el costo computacional es $\Ode (\# \calW)$.
\end{proof}

\begin{propo}\label{teo_backpropagation}
Dados una red neuronal $\calN = (\calV, \calE, \calW, \calA) $ 
con una topología de alimentación hacia adelante arbitraria,
un patrón $(\x, \vv) \in \calD$, 
y una función de error diferenciable $E$ arbitraria;
existe un algoritmo eficiente que permite calcular las derivadas parciales de 
$E$ respecto de los pesos $w_{k i}$
a partir de las derivadas parciales de $E$ respecto de la variables de salida $y_k$,
cuya complejidad es $\Ode (\# \calW)$.
\end{propo}

\begin{proof}
Antes de empezar, precisemos qué consideramos como datos del problema:
la topología de la red neuronal $\calG = (\calV, \calE)$ es fija,
las funciones de activación $g_k$ son fijas, son derivables y también conocemos
sus derivadas $g_k'$. También es dato el patrón $(\x, \vv) \in \calD$.
En cambio, los pesos $w_{k i}$ son variables, y sus variaciones provocan cambios
en los $z_k$ y $a_k$.
 
Para deducir las fórmulas de este algoritmo, vamos a hacer un cambio de
variables. Consideramos como nuevas variables las $a_k$ definidas por \eqref{eq:ak},
y vamos a evaluar las derivadas parciales de $E$ respecto de $a_k$.
Como juegan un rol importante les ponemos un nombre:
\begin{align}
\delta_k = \frac {\dd E} {\dd a_k}
\end{align}
 
Para las neuronas de salida, volvemos a la notación habitual, $y_k = g_k(a_k)$
(con $y_k$ en lugar de $z_k$).
Cuando $a_k$ varía, el único valor que cambia es el de $y_k$.
Usando la regla de la cadena obtenemos la expresión
\begin{align}\label{eq:deltaksalida}
\delta_k \;=\; \frac {\dd E} {\dd a_k} 
 \;=\; \frac {\dd E} {\dd y_k} \; \frac {\dd y_k} {\dd a_k}
 \;=\; \frac {\dd E} {\dd y_k} \; g_k'(a_k)
\end{align}
donde las $ {\dd E} / {\dd y_k}$ se pueden calcular usando 
la fórmula del error.

Para las neuronas escondidas, es un poco más complicado.
La fórmula \eqref{eq:ak} escrita en función de 
las variables $a_k$ toma la forma
\begin{align}
a_m = \sum_{k \in \calE m} w_{m k} \; g_k(a_k)
\end{align}
Una variación en $a_k$ provoca variaciones en las $a_m$
de las unidades $m$ tales que $(k,m) \in \calE$
(tales que hay una conexión desde $k$ hacia $m$, ver la figura \ref{retroprop}).
Usamos de vuelta la regla de la cadena:
\begin{align}\label{eq:deltak}
\delta_k \;=\; \frac {\dd E} {\dd a_k} 
 \;=\; \sum_{m \in k \calE} \, \frac {\dd E} {\dd a_m} \; \frac {\dd a_m} {\dd a_k}
\end{align}
En efecto, una variación en $a_k$ solo puede provocar una variación en $E$
a través de las variables $a_m$ consideradas.

\begin{figure}
\centering
\includegraphics[width=10 cm]{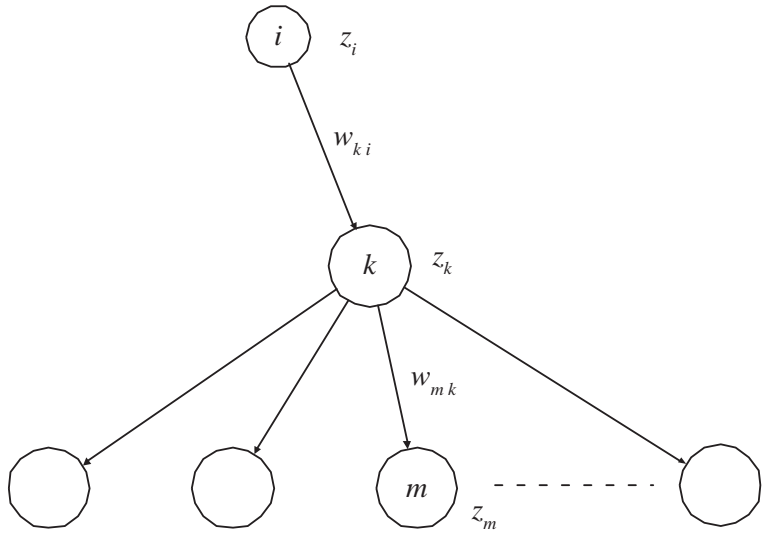}
\caption{Esquema de la regla de la cadena}
\label{retroprop}
\end{figure}

Ahora $a_m$ es una suma donde el término $w_{m k} g_k(a_k)$
aparece una única vez (para $k \in \calE m$)
\begin{align}
a_m \;=\; \ldots + w_{m k} \; g_k(a_k) + \ldots
\end{align}
Luego
\begin{align}
\frac {\dd a_m} {\dd a_k} = w_{m k} \; g_k'(a_k)
\end{align} 
Reemplazando en \eqref{eq:deltak},
obtenemos la fórmula de retro-propagación (back-propagation)
\begin{align}\label{eq:formretro}
\delta_k = g_k'(a_k) \sum_{m \in k \calE} w_{m k} \; \delta_m
\end{align}
que nos dice que podemos calcular un valor $\delta_k$ de una neurona escondida
a partir de los valores $\delta_m$ de las neuronas situadas inmediatamente debajo
($m \in k \calE$).
Partiendo de las neuronas de salida y yendo hacia atrás, 
podemos calcular todos los $\delta$, cualquiera sea la topología de la red.
En efecto, podemos numerar las neuronas de $\calV$ de manera que
$(i,j) \in \calE \Rightarrow i<j$, luego $m \in k \calE \Rightarrow k < m$.
Al recorrer las neuronas por orden decreciente, nos aseguramos
que los valores $\delta_m$ que aparecen en la suma de \eqref{eq:formretro} 
ya hayan sido evaluados en el momento de calcular $\delta_k$.

Evaluemos ahora la derivada de $E$ respecto del peso $w_{k i}$.
$E$ depende de $w_{k i}$ a través de la suma $a_k$.
Usando la regla de la cadena
\begin{align}
\frac {\dd E} {\dd w_{k i} } = \frac {\dd E} {\dd a_k} \; 
  \frac {\dd a_k} {\dd w_{k i} }
= \delta_k \; \frac {\dd a_k} {\dd w_{k i} }
\end{align}
De acuerdo con \eqref{eq:ak} tenemos
\begin{align}
 \frac {\dd a_k} {\dd w_{k i} } = z_i
\end{align}
Luego
\begin{align}\label{eq:parciales}
\frac {\dd E} {\dd w_{k i} } = \delta_k \; z_i
\end{align}
Vale decir que calculando los $\delta_k$ para todas las neuronas 
escondidas y de salida, podemos calcular las derivadas parciales de $E$
respecto de los pesos.

Calculemos la complejidad. 
Como vimos en la observación \ref{obs:odew},
cuando la cantidad de neuronas y de conexiones crece, 
la cantidad de conexiones tiende a ser mucho más grande que la cantidad de neuronas
(por lo cual despreciamos las cuentas que se hacen una sola vez por neurona).
En este algoritmo, cada peso interviene en una multiplicación y una suma
de manera que el costo es $\Ode (\# \calW)$.
\end{proof}

\begin{defin}
El algoritmo de la demostración anterior se llama algoritmo de 
{\em retro-propagación del error} (error back-propagation).
Los valores $\delta_k$ calculados para cada $k \in \calV$ se llaman ``errores''
(son estimaciones del error cometido).
Resumamos aquí los pasos del algoritmo:
\begin{enumerate}
\parcero
\item{Aplicar un vector de entrada $\x$ a la red y propagar hacia adelante
usando \eqref{eq:zk} para obtener los valores de activación de todas
las neuronas escondidas y de salida.}
\item{Evaluar los $\delta_k$ para las neuronas de salida aplicando 
\eqref{eq:deltaksalida}.}
\item{Propagar hacia atrás usando \eqref{eq:deltak} para obtener
los $\delta_k$ para las neuronas escondidas.}
\item{Las derivadas parciales se calculan con \eqref{eq:parciales}.}
\end{enumerate}
\end{defin}

\begin{obs}
Esto nos da las derivadas del error para un patrón del juego de datos
$\calD = \{ \x^n, \vv^n \} $. Cambiando un poco de notaciones, 
si llamamos $E^n$ el error para el patrón $n$, y el error total
\begin{align}
E = \sum_{n} E^n
\end{align}
entonces 
\begin{align}
\frac {\dd E} {\dd w_{k i}} = \sum_{n} \frac {\dd E^n} {\dd w_{k i}}
\end{align}
En la demostración consideramos el caso en que cada neurona $k \in \calV$ puede tener
una función de activación $g_k$ diferente. En la práctica,
la mayoría de las redes neuronales usan una misma función de activación $g$
para todas las neuronas.
\end{obs}

\section{ Funciones de error}\label{sec:error}

En los problemas de clasificación (que son los que nos interesan)
el objetivo es modelar las probabilidades a posteriori de pertenencia a las clases.
En la salida de la red neuronal dedicamos una dimensión (una neurona)
a cada clase, de manera que el valor de activación de la neurona
aproxime la probabilidad de pertenencia a esa clase. 

La función de error suma de cuadrados es una función clásica para problemas de 
clasificación. Veremos en esta sección que es una elección razonable,
además de dar buenos resultados en la práctica.

Como ya mencionamos, el objetivo principal durante el entrenamiento de una red neuronal
no es memorizar el conjunto de datos, sino modelar el generador
de datos subyacente. 
Las observaciones compiladas en el conjunto $ \calD = \{ (\x^n, \bt^n) \} $ son valores
que corresponden a una función ideal a la cual se le suma un ruido estocástico.
Por lo tanto, la descripción más completa del generador viene dada por la
función densidad de probabilidad $ p(\x, \bt) $ en el espacio
conjunto de entradas - salidas esperadas.

Podemos descomponer la densidad conjunta en el producto
de la densidad condicional de las salidas por la densidad
incondicional de las entradas:
\begin{align}
p(\x,\bt) = p(\bt | \x) \; p(\x)
\end{align}
donde $p(\bt | \x)$ representa la densidad de $\bt$ dado que $\x$
toma un cierto valor, y $p(\x)$ representa la densidad incondicional
de $\x$ dada por
\begin{align}
p(\x) = \int p(\bt, \x) d\bt . 
\end{align}
Como queremos hacer predicciones sobre $\bt$ dados nuevos
 valores de $\x$, es la probabilidad condicional $p(\bt | \x)$
la que queremos modelar.

Vamos a motivar el uso de la suma de cuadrados como función de error 
usando el principio
de máxima verosimilitud (maximum likelihood).
Para un conjunto de datos de entrenamiento $\{ \x^n, \bt^n \}$
la verosimilitud se puede escribir como
\begin{align}
\calL &= \prod_{n} p( \x^n, \bt^n)  \\
 &= \prod_{n} p( \bt^n | \x^n ) \; p( \x^n ) 
\end{align}
donde asumimos que cada punto $(\x^n, \bt^n)$ del conjunto de datos
es extraído en forma independiente siguiendo la misma distribución,
y podemos por lo tanto multiplicar las probabilidades.
En lugar de maximizar la verosimilitud, es más fácil 
(como el logaritmo es monótono, es equivalente)
 minimizar el logaritmo negativo
de la verosimilitud. Luego vamos a minimizar
\begin{align}
E = - \ln \calL = - \sum_{n} \ln p(\bt^n | \x^n) - \sum_{n} \ln p(\x^n)
\end{align}
 donde decimos que $E$ es una {\em función de error}.
Veremos que una red neuronal se puede ver como una herramienta para 
modelar la densidad condicional $p(\bt | \x)$.
El segundo término de la fórmula anterior no depende de los parámetros de
la red neuronal, representa una constante que se puede eliminar.
Obtenemos como función de error
\begin{align}\label{eq:error}
E = - \sum_{n} \ln p(\bt^n | \x^n ) 
\end{align}
Tiene la forma de una suma sobre patrones de un término de error que
depende de cada patrón por separado, esto se debe al hecho de que suponemos
que los datos son independientes.

\subsection{La suma de cuadrados como función de error}

Consideremos el caso de $c$ variables de salida $t_1, \ldots, t_c$,
supongamos que las distribuciones de las distintas variables son 
independientes de manera que podemos escribir
\begin{align}\label{eq:producto}
p(\bt | \x ) = \prod_{k=1}^{c} p( t_k | \x )
\end{align}

Suponemos además que los datos objetivos (los datos de salida esperados)
siguen una distribución normal.
Más específicamente suponemos que la variable objetivo $t_k$
está dada por una función determinística $h_k(x)$
con un ruido agregado Gaussiano $\epsilon$ de manera que
\begin{align}\label{eq:ruido_e_k}
t_k = h_k(\x) + \epsilon_k
\end{align}
Suponemos que los errores $\epsilon_k$ tienen una distribución normal
con promedio cero, y desviación estándar $\sigma$ que no depende
de $\x$ ni de $k$.
Luego la distribución de $\epsilon_k$ está dada por
\begin{align}
p(\epsilon_k) = \frac {1}{ ( 2 \pi \sigma^2 ) ^ {1/2} } \exp \left( - \frac {\epsilon_k ^ 2} {2 \sigma^2} \right)
\end{align}
Ahora queremos modelar las funciones $h_k(\x)$ con una red neuronal con 
funciones de salida $y_k(\x, \w)$ donde $\w$ es el vector de pesos que determina
la configuración de la red neuronal. 
Reemplazando en la expresión \eqref{eq:ruido_e_k} la función desconocida $h_k(\x)$
por nuestro modelo $y_k(\x, \w)$ obtenemos que el ruido se expresa como
\begin{align}
\epsilon_k = t_k - y_k(\x, \w)
\end{align}
Por lo tanto la distribución de las variables objetivo está dada por
\begin{align}
p( t_k | \x ) = \frac {1}{ ( 2 \pi \sigma^2 ) ^ {1/2} } \exp \left( - \frac { ( y_k(\x, \w) - t_k ) ^ 2} {2 \sigma^2} \right)
\end{align}
Reemplazando en \eqref{eq:error} y \eqref{eq:producto},
obtenemos que 
\begin{align}
E &= - \sum_{n=1}^{N} \sum_{k=1}^{c} \left\{ \ln \left( \frac {1} { ( 2 \pi \sigma^2 ) ^ {1/2} } \right)
- \frac { ( y_k(\x^n, \w) - t_k^n ) ^ 2} {2 \sigma^2} \right\} \\
&= \frac {1 } {2 \sigma^2} \sum_{n=1}^{N} \sum_{k=1}^{c} ( y_k(\x^n, \w) - t_k^n ) ^ 2 
+ \frac{N c}{2} \ln ( 2 \pi \sigma^2 )
\end{align}
Durante el entrenamiento de la red, buscamos minimizar el error modificando los pesos $\w$.
Vale decir que el segundo término de la suma anterior, y el factor $1 / \sigma^2$ 
son constantes (no dependen de $\w$) y se pueden omitir.
Finalmente obtuvimos para el error la familiar suma de cuadrados
\begin{align}
E &= \frac {1 } {2 } \sum_{n=1}^{N} \sum_{k=1}^{c} ( y_k(\x^n, \w) - t_k^n ) ^ 2 \\
&= \frac{1}{2} \sum_{n=1}^{N} || \y(\x^n, \w) - \bt^n || ^ 2
\end{align}

Además de llegar a algo que ya conocíamos, el interés de este análisis
es formular un poco más precisamente el hecho de que los observaciones
son los valores de la función objetivo más un ruido.

\begin{obs}
Esto permite hacer inferencia, en particular permite:
\begin{itemize}
\item{Testear si el modelo es correcto.}
\item{Dar intervalos de confianza.}
\item{Hacer test de hipótesis.}
\item{Cuantificar al estimador.}
\end{itemize}
\end{obs}

\subsection{Estimando las probabilidades a posteriori}
\label{sec:a_posteriori}

Cuando usamos una red neuronal para resolver un problema de clasificación,
el enfoque más general y eficiente es usar la red para modelar
las probabilidades a posteriori de pertenencia a cada clase.
La forma típica de hacerlo es poner una neurona de salida para cada clase,
y hacer que el valor de activación de las neuronas de salida represente
la probabilidad a posteriori $p(\calC_k | \x)$ de que el vector de entrada $\x$
pertenezca a la clase $\calC_k$.
Esas probabilidades son luego usadas para llegar a una clasificación 
mediante un proceso de toma de decisión.

En la sección \ref{sec:salidas_redes} veremos los ejemplos de las redes 
que usamos para la detección de sistemas operativos.
Uno de los análisis consiste en clasificar la respuesta de una máquina
a los tests de Nmap en una de 6 clases $\calC_1, \ldots, \calC_6$ que 
corresponden a las familias de sistemas operativos 
Windows, Linux, Solaris, OpenBSD, FreeBSD, y NetBSD.

Con una red neuronal diseñada para aproximar probabilidades a posteriori 
podemos {\em minimizar la probabilidad de error en la clasificación}.
En efecto para minimizar la probabilidad de error,
cada vector de entrada $\x$ debe ser asignado a la clase que tiene la mayor
probabilidad a posteriori $p(\calC_k | \x)$.

También se pueden {\em compensar diferentes probabilidades a priori}.
Supongamos que queremos entrenar una red neuronal para distinguir
imágenes de tejido sano (clase $\calC_1$) e imágenes de tejido con tumor (clase $\calC_2$).
Este es un ejemplo sacado de \cite{bishop}.
Supongamos que estadísticas médicas nos indican que en la población general
$\hat{p}(\calC_1) = 0.99$ y $\hat{p}(\calC_2) = 0.01$. 
Sin embargo para entrenar la red,
necesitamos tener una cantidad razonable de datos de cada clase.
Podemos tomar una cantidad similar de imágenes de cada clase,
por ejemplo $p(\calC_1) = 0.5$ y $p(\calC_2) = 0.5$.
La red entrenada con este conjunto nos da las probabilidades $p(\calC_k | \x)$.
Como evaluar las probabilidades $\hat{p}(\calC_k | \x)$ correspondientes
a la población general?

De acuerdo con el teorema de Bayes
\begin{align}
p(\calC_k | \x ) &= \frac { p(\x | \calC_k ) \; p(\calC_k) } { p(\x) }  \\
\hat{p}(\calC_k | \x ) &= \frac { p(\x | \calC_k ) \; \hat{p}(\calC_k) } { \hat{p}(\x) } 
\end{align}
Suponemos que las probabilidades $p(\x | \calC_k )$ condicionadas  a la clase no cambian.
Entonces
\begin{align}
\hat{p}(\calC_k | \x) = p(\calC_k | \x) \frac{\hat{p}(\calC_k)}{p(\calC_k)} \frac{p(\x)}{\hat{p}(\x)}
\end{align}
Vale decir que hay que dividir las salidas de la red por las probabilidades a priori
del conjunto de datos, multiplicar por las nuevas probabilidades a priori
de la población real, y luego normalizar los resultados.
De esta manera se pueden compensar cambios en las probabilidades a priori sin 
necesidad de reentrenar la red.

\section{ Optimización de los parámetros }

En las secciones anteriores hemos formulado
 el problema del entrenamiento de la red como el de 
minimizar una función de error $E$.
Este error depende de los parámetros adaptativos de la red
(los pesos y los sesgos), los cuales pueden ser agrupados 
en un vector $\w = (w_1, \ldots, w_W)$ de tamaño $W$.

Hemos visto en la sección \ref{sec:retropropagacion}
que en las redes de perceptrones de varias capas, las derivadas
parciales de la función de error respecto de los parámetros de la red
se pueden calcular en forma eficiente usando el algoritmo
de retro-propagación. La información de este gradiente va a ser de vital 
importancia para encontrar algoritmos de entrenamiento suficientemente
rápidos como para usar en aplicaciones prácticas de gran escala.

El problema de minimizar una función de varias variables continua y diferenciable
ha sido ampliamente estudiado y muchos de los enfoques tradicionales se 
pueden aplicar directamente al entrenamiento de redes neuronales.
En general se usan algoritmos de búsqueda local, que funcionan guardando
un solo estado en memoria, y moviéndose a vecinos de ese estado
(la búsqueda no toma en cuenta el camino para llegar a ese estado).
El descenso del gradiente\footnote{Ver en la 
wikipedia \tt{http://en.wikipedia.org/wiki/Gradient\_descent } } (\emph{gradient descent}) es una búsqueda local voraz, donde cada paso
consiste en moverse al vecino que tiene el mejor valor de la función objetivo.
El problema de esta búsqueda es que no puede salir de un mínimo local, y tampoco
puede avanzar por una meseta (donde la función objetivo es plana).
Una variación es el descenso del gradiente estocástico, donde los movimientos
se hacen a un vecino elegido al azar, de manera que los vecinos que corresponden a mejores
valores tienen más probabilidad de ser elegidos.
Otra variación es el descenso del gradiente con reinicio aleatorio: partiendo
de un estado al azar obtenemos un óptimo local. Quedándonos con el mejor de los 
óptimos locales obtenidos, vamos aproximando (y eventualmente alcanzamos) el mínimo global. 
Otro ejemplo es la búsqueda de \emph{simulated annealing}\footnote{Ver en la 
wikipedia \tt{http://en.wikipedia.org/wiki/Simulated\_annealing}}
(traducido como temple simulado o recocido simulado), 
una especie de descenso estocástico
donde el factor aleatorio va disminuyendo a medida que pasa el tiempo (se va enfriando).
Veremos como usar el descenso del gradiente para el entrenamiento de una red neuronal,
 sus limitaciones y algunas modificaciones heurísticas que permiten mejorarlo.

\subsection{Superficies de error}

El problema que estudiamos en esta sección es el de encontrar
un vector de pesos $\w$ que minimice una función de error $E(\w)$.  
Podemos hacernos una representación geométrica del 
proceso de minimización de error pensando $E(\w)$ como una
{\em superficie de error} definida en un espacio de pesos.

En las redes que tienen más de una capa, la función de error
es una función no lineal de los pesos, y tendrá posiblemente
muchos puntos estacionarios que satisfacen
\begin{align}
\nabla E = 0
\end{align}
entre los cuales puede haber puntos silla, mínimos y máximos locales,
y mínimos globales.

Dado que la función de error es no lineal, no existe en general una
fórmula cerrada para encontrar los mínimos.
Consideraremos algoritmos que involucran una búsqueda a través del espacio de pesos
y que consisten en una serie de pasos de la forma
\begin{align}
\w^{\tau + 1} = \w^{\tau} + \Delta \w^{\tau}
\end{align}
donde cada $\tau$ representa una iteración del algoritmo,
y el incremento al vector de pesos $\Delta \w^{\tau}$ depende del algoritmo elegido.

\subsection{Simetrías en el espacio de pesos}

Consideremos una red neuronal como las que hemos descripto
en la sección \ref{sec:red_varias_capas}. 
Supongamos para fijar ideas una red de 2 capas con $M$ unidades
en la capa escondida, con conectividad completa entre capas, y que use la
función de activación $tanh$ dada por \eqref{eq:tanh}.

Si cambiamos el signo de todos los pesos que llegan
a una neurona escondida determinada, se invierte el signo de la salida de la función 
de activación de esa neurona (dado que la función $tanh$ es impar).
Esto se puede compensar cambiando los signos de todos los pesos que salen 
de esa neurona escondida, de manera que la función representada por la red neuronal
no se modifica.
De esta manera encontramos dos conjuntos de pesos que generan la misma función.
Para $M$ neuronas escondidas, tenemos $M$ de estas simetrías de cambio de signo,
y por lo tanto $2^M$ vectores de pesos equivalentes.

De manera similar, consideremos dos neuronas escondidas: podemos intercambiar todos los 
pesos que entran y salen de estas dos neuronas, y obtener una red equivalente
(aunque el vector de pesos es diferente).
Para $M$ neuronas escondidas, hay $M!$ formas de reordenarlas, luego tenemos
$M!$ vectores de pesos equivalentes por esta simetría de intercambio.
Por lo tanto, el espacio de pesos de la red tiene un factor de simetría de $M! \; 2^M$.
 Para redes con más de una capa escondida, el factor total de simetrías
es el producto
\begin{align}
\prod_{k=1}^{m} M_k ! \; 2^{M_k}
\end{align}
donde $m$ es la cantidad de capas escondidas, y $M_k$ la cantidad de neuronas en la capa $k$.
Resulta que este factor da cuenta de todas las simetrías del espacio de pesos, 
salvo algunas posibles simetrías accidentales 
(ver el artículo de Chen et alter: ``On the geometry of feedforward 
neural network error surfaces" \cite{chen}).

Esto quiere decir que cualquier mínimo local o global va a estar replicado
una gran cantidad de veces en el espacio de pesos. Esto responde la pregunta sobre
la eventual unicidad de un mínimo global por ejemplo ;-)
De todas maneras los algoritmos que veremos hacen una búsqueda local en cada paso
y no se ven afectados por los numerosos puntos equivalentes que hay en el espacio de pesos.

\subsection{El descenso del gradiente}

Uno de los algoritmos de entrenamiento más sencillos y estudiados
es el des-censo del gradiente 
({\em gradient descent}, también llamado {\em steepest descent}).
Tiene dos versiones: 
\begin{enumerate}
\item{por lotes ({\em batch})}
\item{secuencial, también llamada basada en patrones ({\em pattern based}).}
\end{enumerate}
En ambos casos se parte de un vector de pesos inicial $\w^0$
(generalmente elegido al azar).
Luego se actualiza iterativamente el vector de pesos, 
moviéndonos en cada paso $\tau$ una pequeña distancia en la dirección
de mayor decrecimiento del error, vale decir en la dirección contraria
al gradiente del error evaluado en $\w^\tau$.

\begin{defin}
En la versión de descenso del gradiente {\em por lotes}, se calcula el gradiente del error
sobre el conjunto de los datos de entrenamiento $\calD$,
para obtener el desplazamiento
\begin{align}
\Delta \w^{\tau} = - \lambda \; \nabla E | _{ \w ^ \tau }. 
\end{align}
Este es el ``verdadero" gradiente, y consiste en la suma de los gradientes causados
por cada patrón. 
Recordemos que $\calD$ es un conjunto de patrones de la forma $\calD = \{ (\x^n, \bt^n ) : 1 \leq n \leq N \}$.
Por lo tanto, en la versión por lotes, cada paso del entrenamiento requiere
pasar por todos los datos $\{ (\x^n, \bt^n ) : 1 \leq n \leq N \}$ antes de actualizar los pesos.
\end{defin}

\begin{defin}
En la versión de descenso del gradiente {\em secuencial}, se calcula el gradiente
del error para cada patrón $(\x^n, \bt^n )$ del conjunto de datos, y los pesos
se actualizan usando
\begin{align}
\Delta \w^{\tau} = - \lambda \; \nabla E^n | _{ \w ^ \tau }. 
\end{align}
Dicho de otra manera, el ``verdadero" gradiente es aproximado por el gradiente
de la función de error evaluado en un solo patrón.
Los pesos son actualizados en un monto proporcional a este gradiente aproximado.
Vale decir que en la versión secuencial, los pesos se actualizan para cada patrón en $\calD$.
En la sección \ref{sec:retropropagacion}, en particular con el Teorema \ref{teo_backpropagation},
vimos como calcular $\nabla E^n$ para un patrón $(\x^n, \bt^n ) \in \calD$.
Para conjuntos de datos grandes, el descenso del gradiente secuencial
es mucho más rápido que el descenso del gradiente por lotes, esta es la razón por la cual 
se lo prefiere.
\end{defin}

\begin{defin}
En la versión de descenso del gradiente secuencial, 
llamamos {\em generación} del entrenamiento
el hecho de iterar (actualizando los pesos $\w$) sobre los $N$ patrones del conjunto de datos $\calD$.
Los diferentes patrones se toman siempre en el mismo orden,
o se reordenan al azar al principio de cada generación.
En la práctica se recomienda reordenar los datos al azar, para minimizar la influencia
que el orden de los datos pueda tener sobre el entrenamiento de la red. 
\end{defin}

\begin{defin}
El parámetro $\lambda$ se llama {\em taza de aprendizaje}.
\end{defin}

Si la taza de aprendizaje es suficientemente pequeña, en la versión por lotes,
podemos esperar que el valor del error $E$ disminuya en cada iteración,
llegando eventualmente a un vector de pesos tal que $\nabla E = 0$.

En la versión secuencial, también podemos esperar una reducción del error
para $\lambda$ suficientemente pequeño, dado que la dirección promedio del
movimiento en el espacio de pesos aproxima la inversa del gradiente $- \nabla E$.
El descenso del gradiente secuencial es similar al procedimiento de
Robbins-Monro (ver \cite{robbins}).
Si la taza de aprendizaje $\lambda$ disminuye en cada paso del algoritmo
de acuerdo con los requerimientos del teorema de \cite{luo},
se puede garantizar la convergencia del algoritmo.
Por ejemplo se puede tomar $\lambda^{(\tau)} \approx 1 / \tau$ en el paso $\tau$,
aunque esta elección lleva a una convergencia muy lenta.
En la práctica se usa un valor constante de $\lambda$, o incluso una estrategia
de taza de aprendizaje adaptativo, donde $\lambda$ puede aumentar durante el 
entrenamiento (ver la sección \ref{sec:taza_adaptativa}).
Estas estrategias llevan a mejores resultados, aunque se pierde la garantía de
convergencia.

\subsection{Momentum}

Cuando la función de error tiene distintas curvaturas en diferentes 
direcciones, por ejemplo cuando la superficie de error tiene la forma 
de un valle alargado, el gradiente negativo $ - \nabla E$
no apunta hacia el mínimo de la función de error.
Imaginemos que apunta en una dirección casi ortogonal,
los pasos del descenso del gradiente van a dar como
resultado oscilaciones que van progresando muy lentamente 
hacia el mínimo del error.

\begin{defin}
Una forma de encarar este problema es agregar un término 
llamado {\em momentum}
 a la fórmula de descenso del gradiente.
Esto le agrega inercia al movimiento en el espacio de pesos,
y suaviza las oscilaciones que ocurrían en el escenario planteado.
La fórmula modificada es
\begin{align}\label{eq:momentum}
\Delta \w^{\tau} = - \lambda \; \nabla E | _{ \w ^ \tau } + \mu \; \Delta \w^{\tau - 1} 
\end{align}
donde $\mu$ es el parámetro de {\em momentum}.
\end{defin}

Para entender un poco el efecto del momentum, consideremos dos 
casos simples. Cuando el gradiente se mantiene constante, el efecto
del momentum es de acelerar el movimiento.
Cuando el gradiente va oscilando, las contribuciones sucesivas 
del término correspondiente al momentum tienden a cancelarse
(no se amplifican las oscilaciones).
Por otra parte, mirando la fórmula \eqref{eq:momentum},
está claro que $\mu$ tiene que tomar sus valores en el rango
$0 \leq \mu \leq 1$.
En la práctica observamos que la inclusión del momentum
produce una mejora en la velocidad de convergencia
del descenso del gradiente.

\subsection{Taza de aprendizaje adaptativa}

Otra técnica para mejorar la convergencia del entrenamiento,
también llamada técnica {\em bold driver} 
(ver los artículos de \cite{vogl} y \cite{battiti} sobre 
métodos para acelerar la convergencia del algoritmo de retropropagación).

Después de cada generación (vale decir después de haber
hecho los cálculos para todos los pares de entrada / salida),
volvemos a evaluar la función de error $E$.
Si el error es más grande,
es que nos pasamos de largo, vale decir que la taza de aprendizaje
es demasiado grande y tenemos que disminuirla.
Por el contrario, si el error es más pequeño,
quiere decir que estamos en la dirección correcta
y aumentamos la taza de aprendizaje para desplazarnos más rápido.

La taza de aprendizaje se actualiza de la manera siguiente:
\begin{align}
\lambda_{nuevo} = \left\{
\begin{array} {l l}
\rho \cdot \lambda_{previo} & \mbox{si } \Delta E < 0 \\
\sigma \cdot \lambda_{previo} & \mbox{si } \Delta E > 0 \\
\end{array} \right.
\end{align}
El parámetro $\rho$ se toma ligeramente mayor que 1 
(un valor típico es $\rho = 1.1$).
El parámetro $\sigma$ se toma significativamente menor que 1,
para que el algoritmo no siga moviéndose en una dirección equivocada
(un valor típico es $\sigma = 0.5$).

En la sección \ref{sec:taza_adaptativa} mostramos
los gráficos de la evolución del error en la aplicación a detección de SO,
comparando las dos estrategias: taza de aprendizaje fija
y taza de aprendizaje adaptativa.

\subsection{Cómo empezar y cómo terminar}

Cuando uno usa los algoritmos que hemos descripto, tiene que resolver una serie de asuntos prácticos.
En particular, sobre cómo iniciar y cómo terminar el algoritmo.

Todos los algoritmos que hemos descripto empiezan con pesos elegidos al azar. 
De esta manera al repetir el proceso de entrenamiento, obtenemos valores diferentes 
para el error final (y podemos quedarnos con el mejor).
Dicho de otra manera, caemos en distintos mínimos locales.
El tema de los pesos iniciales es muy importante y ha sido ampliamente discutido
por la comunidad, pero no queremos adentrarnos en ese tema en esta exposición.

Cuando se usan algoritmos de optimización no lineal, hay que elegir un criterio
respecto de cuando parar el proceso de entrenamiento. 
Algunas elecciones posibles son:
\begin{enumerate}
\parcero
\item{Parar después de una cantidad fija de iteraciones. 
El problema es que es difícil saber de antemano cuantas iteraciones
serían apropiadas, aunque se puede tener una idea haciendo tests preliminares.
Cuando varias redes son entrenadas, la cantidad de iteraciones necesarias 
para cada red puede ser diferente.
}
\item{Parar después de usar un cierto tiempo de CPU.
Nuevamente, se requieren tests preliminares para estimar la cantidad de tiempo
necesaria.
}
\item{Parar cuando la función de error cae debajo de un umbral determinado.
Esta opción es usada habitualmente junto con una limitación sobre la cantidad de
iteraciones o el tiempo de CPU, dado que el umbral para el error puede no ser alcanzado nunca.
Este es el criterio que usamos en la aplicación de redes neuronales al reconocimiento 
de sistemas operativos.
}
\item{Parar cuando el cambio relativo de la función de error cae debajo de cierto umbral.
Esto puede llevar a una terminación prematura si la función de error decrece lentamente
durante parte del proceso de entrenamiento.
}
\item{Parar cuando el error medido por un proceso de validación independiente
empieza a aumentar. Esta opción se usa como parte de un proceso de optimización 
general del rendimiento de una red.
}
\end{enumerate}

\section{ Análisis en Componentes Principales}\label{sec:acp}

En esta sección seguimos el libro de Bishop. El análisis en componentes principales
es una técnica para reducir las dimensiones de entrada de la red neuronal.

\begin{obs}
Las altas dimensiones son un problema.
Este fenómeno ha sido llamado por Richard Bellman ``the curse of dimensionality"\footnote{ver la 
wikipedia \tt{http://en.wikipedia.org/wiki/Curse\_of\_dimensionality}}
y es un tema recurrente en Inteligencia Artificial.
Por ejemplo si discretizamos los datos de entrada, y agrupamos los valores
para cada dimensión de entrada en $m$ subcategorías, entonces 
la cantidad total de categorías (hipercubos) es $m^d$, crece exponencialmente con la cantidad de dimensiones $d$.
Si queremos tener una muestra de observaciones representativa (que haya por lo menos
una observación por categoría) entonces la cantidad de observaciones también crece
exponencialmente con $d$.
Si la cantidad de observaciones es fija, al crecer $d$ aumenta exponencialmente la cantidad de categorías vacías,
los datos están muy esparsos.
Dicho de otra manera, cuando la cantidad de dimensiones aumenta,
todo queda lejos de todo.
Más concretamente, al reducir las dimensiones de entrada de una red neuronal,
disminuye el tiempo de entrenamiento de la red y mejora su rendimiento.
\end{obs}

\begin{defin}
Una {\em característica} (feature) es una combinación
de las variables de entrada.
Las variables de entrada se combinan en características
de manera que el conjunto de características tenga
una dimensión menor que el conjunto de variables de entrada.
\end{defin}

De acuerdo con la Definición \ref{nosupervisado}, dado un conjunto de datos $\calD = \{ \x^n, \vv^n \}$,
las técnicas de aprendizaje que solo usan los datos de entrada $\{ \x^n \}$, 
sin usar las salida esperadas $\{ \vv^n \}$
se llaman {\em aprendizaje no supervisado} (unsupervised learning).

Vamos a ver una técnica de aprendizaje no supervisado 
para extraer características de las variables,
o sea para reducir las dimensiones de la entrada de la red neuronal,
llamada \emph{análisis en componentes principales}.
Esta técnica consiste en proyectar 
vectores $\x = (x_1, \ldots, x_d)$
en un espacio de dimensión $p$ con $p < d$,
preservando la mayor cantidad de información posible.
La idea del análisis en componentes principales
es de maximizar la varianza (la mayor varianza de cualquier proyección en un 
subespacio de $p$ dimensiones viene de proyectar sobre los $p$ primeros
vectores de la nueva base) 
y minimizar la diferencia entre el vector proyectado y el vector original.

Para hacer esto vamos a construir una nueva base
$\{ \uu_1, \ldots, \uu_d \}$ de vectores ortonormales
(tales que $\uu_i^T \uu_j = \delta_{i j} $).
Llamando $(z_1, \ldots, z_d)$ a las coordenadas de $\x$en esta nueva base, tenemos
que
\begin{align}
\x = \sum_{i = 1}^{d} z_i \uu_i
\end{align}
donde los coeficientes $z_i$ se pueden calcular como
\begin{align}
z_i = \x^T \uu_i
\end{align}

\begin{propo}
Dada una base $\{ \uu_i \}$, nos vamos a quedar con $p$ de los coeficientes $z_i$, 
los otros serán reemplazados por constantes $b_i$
para obtener como aproximación al vector $\x$ un vector $\hat{\x}$
\begin{align}\label{eq:aprox}
\hat{\x} = \sum_{i = 1}^{p} z_i \uu_i  +   \sum_{i = p+1}^{d} b_i \uu_i
\end{align}
Para minimizar el error cometido con esta aproximación,
hay que elegir
\begin{align}
b_i = \frac{1}{N} \sum_{n=1}^{N} z_i^n .
\end{align}
\end{propo}

\begin{proof}
Considerando el conjunto de datos $\calD$, este contiene $N$ vectores
$\{ \x^1, \ldots, \x^N \} $. Queremos calcular los $b_i$ para que la
aproximación \eqref{eq:aprox} sea la mejor posible, haciendo el promedio
sobre el conjunto de datos.
El error para cada vector $\x^n$ es
\begin{align}
\x^n - \hat{\x}^n = \sum_{i = p+1}^{d} (z_i^n - b_i) \uu_i
\end{align}

Definimos como mejor aproximación la que minimiza
la suma de los errores al cuadrado sobre el conjunto de datos,
vale decir que queremos minimizar
\begin{align}\label{eq:acpep}
E_p = \frac{1}{2} \sum_{n = 1}^{N} || \x^n - \hat{\x}^n || ^ 2
 = \frac{1}{2} \sum_{n=1}^{N} \sum_{i = p+1}^{d} (z_i^n - b_i)^2
\end{align}
donde usamos que los vectores $\uu_i$ son ortonormales.
Igualando a cero la derivada parcial de $E_p$ respecto de $b_i$
obtenemos
\begin{align}
\frac {\dd E_p}{\dd b_i} = \sum_{n=1}^{N} \left( z_i^n - b_i \right) = 0
\end{align}
Por lo tanto 
\begin{align}\label{eq:acpbi}
b_i = \frac{1}{N} \sum_{n=1}^{N} z_i^n = \bar{\x}^T \uu_i
\end{align}
donde $\bar{\x}$ es el vector promedio de los datos
\begin{align}
\bar{\x} = \frac{1}{N} \sum_{n=1}^{N} \x^n
\end{align}
Vale decir que para minimizar la diferencia entre el proyectado y el real,
primero hacemos una translación al centro de gravedad
de los datos, y luego proyectamos.
\end{proof}

\begin{obs}
Usando \eqref{eq:acpbi}, la fórmula del error \eqref{eq:acpep}
se puede escribir
\begin{align}\label{eq:acpep2}
\nonumber E_p &= \frac{1}{2} \sum_{n=1}^{N} \sum_{i=p+1}^{d} (( \x^n - \bar{\x} )^T \uu_i )^2 \\
\nonumber &= \frac{1}{2} \sum_{i=p+1}^{d} \sum_{n=1}^{N}  \uu_i^T ( \x^n - \bar{\x} )  ( \x^n - \bar{\x} )^T \uu_i  \\
&= \frac{1}{2} \sum_{i=p+1}^{d} \uu_i^T \bZ \uu_i
\end{align}
donde $\bZ$ es la matriz de covarianza definida por
\begin{align}\label{eq:acpz}
\bZ = \sum_{n=1}^{N}  ( \x^n - \bar{\x} )  ( \x^n - \bar{\x} )^T
\end{align}
\end{obs}

\subsection{ Interludio: derivadas parciales de matrices}\label{sec:derivadas}

Como vamos a calcular derivadas parciales de expresiones que involucran 
matrices y vectores, conviene repasar algunos hechos básicos
para fijar la notación.
Notamos a la matriz $\bA$ que tiene como elemento $a_{ij}$ en 
el lugar $(i,j)$ (fila $i$ columna $j$)
\begin{align}
\bA = \left (a_{ij} \right )_{ij}
\end{align}
donde los $a_{ij}$ son funciones de varias variables.
La matriz que resulta de calcular la derivada parcial de 
cada elemento de $\bA$ respecto de una variable $x$ se nota
\begin{align}
\frac {\dd \bA} {\dd x} = \left ( \frac {\dd a_{ij}}{\dd x} \right )_{ij} 
\end{align}
Con esta notación, el producto de $\bA$ por $\bB$ se escribe
\begin{align}
\bA \bB = \left ( \sum_{k} a_{ik} b_{kj} \right ) _{ij}
\end{align}
y la derivada parcial del producto
\begin{align}
\nonumber \frac{ \dd (\bA \bB) } {\dd x} &= \left ( \sum_{k} \frac {\dd ( a_{ik} b_{kj}) } { \dd x} \right ) _{ij} \\
\nonumber &= \left ( \sum_{k} \frac {\dd  a_{ik}  } { \dd x}  b_{kj} + a_{ik} \frac {\dd  b_{kj}  } { \dd x}   \right ) _{ij}  \\
& = \frac{ \dd \bA } {\dd x} \bB + \bA \frac{ \dd \bB } {\dd x} 
\end{align}
En particular si $\uu$ y $\vv$ son vectores columna
\begin{align}\label{eq:acpduv}
\frac {\dd ( \uu^T \vv ) } { \dd x } = \left ( \frac{\dd \uu}{\dd x} \right ) ^ T \vv 
+ \uu^T \left ( \frac{\dd \vv}{\dd x} \right )   
\end{align}
Otra notación que usamos es 
\begin{align}\label{eq:acpdvec}
\frac{\dd a}{\dd \x} = \left ( \frac{\dd a}{\dd x_1}, \ldots, \frac{\dd a}{\dd x_d}  \right ) ^ T
\end{align}

\begin{obs}\label{derivada1}
Usando \eqref{eq:acpduv} y la notación \eqref{eq:acpdvec},
si $\uu$ y $\vv$ son dos vectores columna que representan variables independientes
(o sea tales que $ \dd \uu / \dd v_i = 0 $ y $ \dd \vv / \dd u_i = 0$ para todo $1 \leq i \leq d$),
entonces 
\begin{align}
\frac {\dd ( \uu^T \vv ) } { \dd \uu } = \vv \; \; \mbox{  y  } \; \; 
\frac {\dd ( \uu^T \vv ) } { \dd \vv } = \uu
\end{align}
\end{obs}

\begin{obs}\label{derivada2}
Nos interesa este caso: si $\dd \bA / \dd \uu = \mathbf{0}$ entonces
\begin{align}
\frac{\dd ( \uu^T \bA \uu )}{\dd \uu} = (\bA + \bA^T) \uu
\end{align}
\end{obs}

\begin{proof}
Derivando primero respecto de una variable
\begin{align}
\frac{ \dd ( \uu^T \bA \uu ) } {\dd x} 
= \left ( \frac{\dd \uu}{\dd x} \right ) ^ T \bA \uu
+ \uu^T \left( \frac{\dd \bA}{\dd x} \uu + \bA \frac{\dd \uu}{\dd x}  \right)
\end{align}
Si $\dd \bA / \dd x = 0$ entonces queda
\begin{align}
\frac{ \dd ( \uu^T \bA \uu ) } {\dd x} 
= \left ( \frac{\dd \uu}{\dd x} \right ) ^ T ( \bA + \bA^T ) \uu
\end{align}
Derivando respecto de $u_1, \ldots, u_d$ obtenemos el resultado.
\end{proof}

\subsection{ La transformación de Karhunen - Loéve }

\begin{propo}
Consideremos el problema de encontrar una base ortonormal tal que,
dado $p$ $(1 \leq p \leq d)$, 
el error $E_p$ sea mínimo.
\begin{align}\label{eq:acpep3}
E_p = \frac{1}{2} \sum_{i=p+1}^{d} \uu_i^T \bZ \uu_i
\end{align}
donde $\bZ$ es la matriz de covarianza (constante)
y las variables son los vectores $\uu_i$.
Ponemos como condición que los vectores sean ortonormales
\begin{align}\label{eq:acpdij}
\uu_i^T \uu_j - \delta_{i j} = 0
\end{align}
La solución es tomar como base $\{ \uu_i \}$ la base de autovectores de $\bZ$
y quedarse con los $d - p$ vectores que corresponden a los
$d - p$ autovalores más chicos de $\bZ$.
\end{propo}

\begin{proof}
Vamos a usar la técnica de los multiplicadores de Lagrange
para minimizar $E_p$:
agregamos un multiplicador $\mu_{i j}$ para cada condición \eqref{eq:acpdij}
para construir la función Lagrangiana
\begin{align}\label{eq:lagrangiano}
L = \frac{1}{2} \sum_{i=p+1}^{d} \uu_i^T \bZ \uu_i 
 - \frac{1}{2} \sum_{i=p+1}^{d} \sum_{j=p+1}^{d} \mu_{i j} (\uu_i^T \uu_j - \delta_{i j} )
\end{align}
El mínimo de $E_p$ sujeto a las condiciones \eqref{eq:acpdij}
se obtiene minimizando el Lagrangiano $L$ respecto de las variables $\uu_i$ y 
de los multiplicadores $\mu_{ij}$.
Como $\mu_{ij}$ y $\mu_{ji}$ corresponden a la misma condición,
pedimos además que $\mu_{ij} = \mu_{ji}$.

Equipados con las observaciones de la sección \ref{sec:derivadas},
estamos listos para calcular las derivadas del Lagrangiano \eqref{eq:lagrangiano}
respecto de las variables $\uu_k$. 
Llamemos $\bM$ a la matriz con elementos $\mu_{ij}$ 
(la matriz $\bM$ es simétrica dado que $\mu_{ij} = \mu_{ji}$).
Llamemos $\bU$ a la matriz cuyas columnas son los vectores $\uu_j$.
$\bM, \bU, \bZ$ son matrices cuadradas de $d \times d$.
Los índices $i,j,k$ toman sus valores en el rango
$p+1 \leq i,j,k \leq d$.
Notemos $Col_k(\bA)$ a la columna $k$ de la matriz $\bA$.
Volviendo al Lagrangiano tenemos por una parte (usando la observación \ref{derivada2})
\begin{align}
\nonumber \frac{\dd }{\dd \uu_k } \left( \sum_{i} \uu_i^T \bZ \uu_i \right) 
& = (\bZ + \bZ^T) \uu_k \\
& = Col_k ( (\bZ + \bZ^T) \bU )
\end{align}
y por otra parte (usando la observación \ref{derivada1})
\begin{align}
\nonumber \frac{\dd }{\dd \uu_k } \left( \sum_{i} \sum_{j} \mu_{ij} ( \uu_i^T \uu_j - \delta_{ij} ) \right) 
              & = \sum_{i} \sum_{j}  \mu_{ij} \; \frac{\dd }{\dd \uu_k } \left( \uu_i^T \uu_j  \right)  \\
\nonumber &= \sum_{j \neq k} \mu_{kj} \uu_j + \sum_{i \neq k} \mu_{ik} \uu_i + \mu_{kk} (2 \uu_k) \\
\nonumber & = \sum_{j} \mu_{kj} \uu_j + \sum_{i} \mu_{ik} \uu_i \\
&= Col_k( \bU \bM^T ) + Col_k( \bU \bM )
\end{align}
Por lo tanto, $\dd L / \dd \uu_k = 0$ implica
\begin{align}
Col_k ( (\bZ + \bZ^T) \bU ) - Col_k( \bU ( \bM + \bM^T ) ) = 0
\end{align}
Esto vale para todo $p+1 \leq k \leq d$.
Podemos tomar los subespacios en los cuales calculamos el error como sucesión creciente:
cada uno incluido en el siguiente (cuando $p$ decrece).
Luego podemos agregar condiciones similares a la anterior para $1 \leq k \leq p$,
para obtener la ecuación matricial
\begin{align}
(\bZ + \bZ^T) \bU - \bU ( \bM + \bM^T ) = 0
\end{align}
Las matrices $\bM$ y $\bZ$ son simétricas, por lo tanto la expresión anterior se reduce a 
la ecuación (con $\bU$ y $\bM$ incógnitas)
\begin{align}
\bZ  \bU =  \bU  \bM 
\end{align}
Usando que los vectores $\uu_k$ son ortonormales, $\bU ^ T \bU = \bI$
\begin{align}
\bU^T \bZ \bU = \bM
\end{align}
Ahora que lo tenemos escrito así, vemos que una solución es tomar
como $\bU$ la matriz formada por los autovectores de $\bZ$
y como $\bM$ la matriz diagonal, cuyos elementos son los autovalores $\lambda_i$.
De esta manera los vectores $\uu_i$ verifican
\begin{align}
\bZ \uu_i = \lambda_i \uu_i
\end{align}
El error \eqref{eq:acpep3} toma la forma
\begin{align}
E_p = \frac{1}{2} \sum_{i=p+1}^{d} \uu_i^T \lambda_i \uu_i 
= \frac{1}{2} \sum_{i=p+1}^{d}  \lambda_i 
\end{align}
Por lo tanto el error se minimiza eligiendo como vectores a descartar
los $d-p$ autovectores que corresponden a los $d-p$ autovalores
más chicos de $\bZ$.
\end{proof}

\begin{defin}
Cada uno de los vectores $\uu_i$ se llama {\em componente principal}.
Este proceso de reducción de dimensiones se llama
{\em transformación de Karhunen-Loéve} o {\em análisis en componentes principales}.
Resumamos el algoritmo:
\begin{itemize}
\item{Calcular el promedio $\bar{\x}$.
}
\item{Sustraer el promedio a cada vector: $\x^n - \bar\x$.
}
\item{Calcular la matriz de covarianza $\bZ = \sum_{n}  ( \x^n - \bar{\x} )  ( \x^n - \bar{\x} )^T$.
}
\item{Calcular los autovalores y autovectores de $\bZ$.
}
\item{Ordenar los autovectores por autovalor en forma decreciente.
}
\item{Elegir los $p$ primeros autovectores para proyectar los datos.
}
\end{itemize}
\end{defin}

\begin{obs}
El análisis en componentes principales es una técnica clásica de reducción de dimensiones,
muy usada en estadística.
Una interpretación alternativa es la siguiente:
\begin{itemize}
\item{Buscar una dirección que maximice la variabilidad de los datos (cuando son proyectados en esa 
dirección). Esto da la primera componente principal. }
\item{Luego se repite en el subespacio ortogonal a la primera componente principal, para obtener la 
segunda componente principal.}
\item{Así siguiendo se construye la base buscada.}
\end{itemize}
El análisis en componentes principales también sirve para ``visualizar" los datos (por ejemplo
proyectando en 2 dimensiones).
\end{obs}

\chapter{Reconocimiento de Sistemas Operativos}

Uno de los conceptos unificadores de la teoría de 
la Inteligencia Artificial es el de 
agente racional. Basándonos en esta idea 
propusimos en \cite{ataques} un modelo de ataques informáticos, donde juegan un rol central los
agentes que modelan el atacante, y las acciones que puede ejecutar. Estas son acciones probabilísticas, que requieren ciertas condiciones para su ejecución, y producen un resultado cuya probabilidad de éxito refleja la incertidumbre sobre las condiciones reales de ejecución y que no entran en el modelo.

El prototipo de un ataque informático sigue básicamente las etapas de un penetration test, la primera es denominada 
``information gathering" (recolección de información),
base que sirve luego para lanzar ``exploits" (acciones que explotan una vulnerabilidad y permiten instalar un agente y tomar control de una máquina).
La detección remota del sistema operativo de una máquina
forma parte de esta etapa de recolección de información.
Su importancia radica en que el agente atacante necesita
conocer detalles sobre el sistema operativo del objetivo para elegir que exploits va a usar.

La detección de SO se realiza 
enviando en forma activa paquetes test al sistema objetivo,
para estudiar variaciones específicas en las respuestas
que revelen su sistema operativo.
Estas variaciones se deben a diferencias entre las implementaciones
de los protocolos de comunicación de red,
fundamentalmente el protocolo TCP/IP.

\section{El protocolo IP}

Las máquinas interconectadas en redes locales o a través de Internet
se comunican usando distintos protocolos.
Vamos a ver los más importantes (los más usados): la familia TCP/IP.
Es necesario conocer algunos detalles de los protocolos 
para entender las diferencias entre las implementaciones.

El escenario de ataque más común supone un atacante
que se conecta con las máquinas objetivo a través de Internet.
La red Internet fue pensada con un diseño abierto,
suponiendo cierta relación de confianza o buena fe entre los interlocutores.
La información enviada por Internet escapa al control del que la mandó,
y actualmente la habilidad de un atacante de adquirir información
plantea riesgos y peligros.

El idioma oficial de Internet es el {\em Internet Protocol} (IP),
especificado en el RFC 791.
Este RFC\footnote{RFC = Request For Comments} fue escrito en 1981 por la 
Universidad de Southern California para el proyecto de red
de la DARPA\footnote{DARPA = Defense Advanced Research Projects Agency}
(ver \cite{rfc791} en la bibliografía).

Los paquetes IP corresponden al tercer nivel (nivel de red) del modelo 
OSI\footnote{OSI = Open System Interconnection}.

El {\em encabezado} (header) de un paquete contiene la información necesaria para 
entregar el paquete a su destinatario.
El resto del paquete se llama {\em carga útil} (payload) y contiene la información que 
le queremos hacer llegar al destinatario (así como encabezados de niveles superiores).

Las direcciones IP son números de 32 bits (en la versión 4 de IP).
Una mascara de red especifica un conjunto de direcciones IP.

\subsection{El encabezado IP (IP header)}

Estos son los campos que contiene el encabezado IP de cualquier paquete:
\begin{itemize}
\item{Versión del protocolo (4 bits). El valor estándar es 4, por IPv4.
La última versión del protocolo, IPv6, aún no está muy difundida.
}
\item{Longitud del encabezado (header length) (4 bits).
Normalmente la longitud del encabezado es de 20 bytes.
Esta es la longitud mínima de un encabezado IP (cuando no tiene opciones).
Vale decir que un paquete con este campo menor que 20 bytes debería 
ser descartado por una buena implementación
(este no es siempre el caso).
}
\item{Tipo de servicio (TOS = Type Of Service) (8 bits).
Este campo se divide en tres segmentos: 
la prioridad del paquete (3 bits), el método de ruteo deseado por el emisor (4 bits),
y un bit reservado. 
Esta especificación se basa en una relación de confianza y buena fe entre los interlocutores,
y por lo tanto es habitualmente ignorada en la Internet abierta.
La última parte es un bit reservado, que debería valer siempre 0
(de nuevo, no es siempre el caso).
}
\item{Longitud total (TL = Total Length) del paquete (16 bits).
Los paquetes IP circulan sobre protocolos de más bajo nivel, 
por ejemplo Ethernet (segundo nivel) cuyo MTU (Maximum Transmission Unit) es 1500 bytes.
Por lo tanto un paquete de longitud total mayor a 1500 bytes deberá ser fragmentado
en paquetes más chicos para poder circular por Ethernet.
}
\item{Identificador del cuarto nivel (8 bits). Especifica a qué protocolo de cuarto nivel (TCP, UDP, ICMP)
corresponde la carga útil del paquete.
}
\item{Dirección IP de origen (32 bits)
}
\item{Dirección IP de destino (32 bits)
}
\item{Tiempo de vida (TTL = Time To Live) (8 bits).
Para evitar bucles infinitos si algo anda mal en el medio de las comunicaciones,
este contador se decrementa cada vez que un paquete pasa por un servidor intermedio.
Cuando el contador llega a 0, el paquete se tira y el sistema manda 
un mensaje de error al emisor mediante un paquete ICMP.
El TTL máximo es 255, pero dependiendo del emisor el valor inicial puede ser distinto.

\begin{defin}
El método {\em traceroute} es una técnica para analizar la ruta hasta un sistema remoto.
Consiste en mandar paquetes con TTL crecientes (1, 2, 3, etc...)
y estudiar los paquetes ICMP recibidos. 
El origen de los paquetes ICMP permite dibujar un mapa de la ruta de los paquetes
desde el emisor hasta el destinatario.
\end{defin}

}
\item{Flags DF, MF y fragment offset (16 bits).
Cuando un paquete grande debe pasar por un sistema intermedio 
cuyo MTU es menor que el tamaño del paquete,
se fragmenta (divide) el paquete en pedazos.

Todos los fragmentos (salvo el último) tienen prendido el flag MF (More Fragments). 
Cuando un sistema recibe un paquete con MF prendido, asigna lugar en memoria 
para poder reensamblar el paquete original.
El fragment offset indica que lugar ocupaba ese fragmento en el paquete original
(este dato es necesario dado que el protocolo IP no garantiza el orden de los paquetes,
en particular no garantiza que los fragmentos lleguen en orden).

El flag DF (Don't Fragment) avisa a los sistemas que reciben el paquete
que no debe ser fragmentado.

\begin{defin}
El proceso de fragmentación y reensamblado de paquetes es bastante ineficiente.
Por lo tanto sería conveniente para el emisor poder averiguar cual es
el MTU más pequeño de una ruta para elegir el tamaño máximo de los paquetes.
El MTU más chico de una ruta se llama {\em PMTU} (Path Maximum Transmission Unit).
\end{defin}

\begin{defin}\label{pmtud}
El protocolo IP no ofrece una forma estándar de hacer esto, pero existe una técnica
llamada {\em PMTUD} (Path MTU Discovery) para evaluar el MTU
más chico de una ruta.
Consiste en mandar paquetes con el flag DF (Don't Fragment).
Cuando un sistema recibe un paquete con DF prendido y necesita fragmentarlo, 
manda un mensaje de error ICMP al emisor avisando que 
``requiere fragmentación, pero el flag DF está prendido''.
Esto significa que el tamaño del paquete es mayor que el PMTU.
Una búsqueda binaria permite evaluar exactamente el PMTU.
\end{defin}

}
\item{Número de identificación IP (IPID) (16 bits).
Permite identificar los fragmentos que corresponden a un mismo paquete original.
Cuando dos paquetes son fragmentados, sus fragmentos pueden llegar 
mezclados al receptor. El IPID asegura que en el momento de reensamblar los fragmentos,
se puedan distinguir los paquetes originales (sus fragmentos tienen mismo IPID).
Nos interesa el hecho de que distintos sistemas operativos tienen distintas
formas de generar los IPID.
}
\item{Checksum (16 bits).
Método de detección de errores. El checksum se vuelve a calcular en cada hop
(en cada salto por un servidor intermedio), dado que cambia el TTL (Time To Live). 
Es una simple suma, y por lo tanto no constituye  
un mecanismo de protección de integridad.
}
\end{itemize}

Con el protocolo IP no hay forma de verificar que el emisor del paquete es efectivamente quien dice ser.

\section{El protocolo TCP}

TCP quiere decir {\em Transmission Control Protocol} y está definido 
en el RFC 793.
Los protocolos TCP y UDP introducen el concepto de puertos (de origen y destino).
Un puerto sirve para mandar datos a una aplicación o servicio específico
dentro de un sistema multitareas (como cualquier máquina Windows, Linux, etc...)
Los puertos hacen que sea posible para un cliente hablar con varios procesos
al mismo tiempo. También permite que varias aplicaciones cliente
se conecten con el mismo servicio de un servidor, sin que se mezclen las comunicaciones.

\subsection{Encabezado TCP (TCP header)}

Estos son los campos que tiene el encabezado de un paquete TCP
(vienen a continuación de los campos del encabezado IP).

\begin{itemize}
\item{Puerto de origen (source port) (16 bits).
}
\item{Puerto de destino (destination port) (16 bits).
}
\item{Número de secuencia (sequence number) (32 bits).
Este valor asegura la integridad de la sesión, y permite que el 
emisor y el receptor se den cuenta cuando paquetes se perdieron en el medio.
El emisor va incrementando el número de secuencia con
la cantidad de bytes que manda, y espera que el receptor
le devuelva ese valor como número de confirmación.
}
\item{Número de confirmación (acknowledgement number) (32 bits).
Es el próximo número de secuencia que el sistema espera recibir.
En la sección \ref{handshake} detallamos el uso de estos dos números.
}
\item{Offset de datos (data offset) (4 bits). 
Indica donde termina el encabezado TCP y donde comienzan los datos
(la carga útil).
}
\item{Bits reservados (4 bits). 
Deberían valer 0 (como de costumbre, no es siempre el caso).
}
\item{Flags (8 bits).\\
    FIN : el emisor del paquete no tienen más datos que mandar. \\
    SYN :  sincronizar (synchronize) los números de secuencia. \\
    RST :  reiniciar (reset) la conexión. \\
    PSH :  función de ``entregar datos inmediatamente'' (push). \\
    ACK : el número de confirmación (acknowledgement) es significativo. \\ 
    URG : el campo ``urgent pointer'' es significativo.

  Dos bits, previamente reservados, fueron agregados a los flags del protocolo TCP en
2001 (ver RFC 3168): \\
    ECN : Explicit Congestion Notification Echo. \\
   CWR : Congestion Window Reduced. \\ 
}
\item{Tamaño de la ventana (window size) (16 bits).
Determina cuanta información se puede mandar sin esperar
un paquete de confirmación (acknowledgement).
Un valor más alto permite mandar más datos de una vez,
pero introduce una penalidad de performance
cuando se pierden datos y deben ser reenviados.
}
\item{Checksum (16 bits). Mecanismo de control de errores, 
similar al checksum IP (no provee integridad real de los datos).
}
\item{Urgent pointer (16 bits).
El campo es interpretado solo si el flag URG está prendido
(sino el valor de este campo es ignorado).
Con este flag el emisor le pide al destinatario que inserte 
este paquete en el flujo lógico en una posición anterior a la que 
ocuparía normalmente, debido a una situación ``urgente''.
Es raro que este mecanismo se use en comunicaciones normales.
}
\item{Opciones.
}
\end{itemize}

\subsection{Establecimiento de la conexión (three-way handshake)} \label{handshake}

Supongamos que el sistema A se quiere conectar al sistema B.
\begin{enumerate}
\item{
A manda un paquete con el flag SYN (synchronize) y 
un número de secuencia inicial $x_0$ (ISN = initial sequence number).
}
\item{
 B responde con un paquete con los flags SYN-ACK (acknowledgement)
un número de secuencia inicial $y_0$,
un número de confirmación $x_0 + 1$.
}
\item{
A manda un paquete con el flag ACK,
número de secuencia $x_0 + 1$,
número de confirmación $y_0 + 1$.
}
\end{enumerate}

A partir de ese momento, A y B se comunican intercambiando paquetes ACK,
incrementando sus números de secuencia para reflejar la cantidad
de bytes enviados, reenviando paquetes si necesario.
Por ejemplo, si A manda un paquete con número de secuencia $x$ y 
con $b$ bytes de información, 
B responde con un paquete con número de confirmación $x + b$
para confirmar que recibió bien los últimos $b$ bytes.

Cualquiera de los dos lados manda un paquete FIN para terminar la sesión.

\subsection{Problemas de seguridad relacionados con los ISN }\label{isnsec}

Después de establecer la conexión, los interlocutores A y B
pueden tener una confianza razonable de que se están comunicando
con la persona correcta.

\begin{defin}
Se llama {\em spoofing de IP }  a un ataque mediante el cual
el atacante A (cuya dirección es $IP_A$) le hace creer a 
un sistema B que se está
comunicando con un tercero cuya dirección $IP_C \neq IP_A$. 
\end{defin}

\begin{defin}
Una forma de realizar este ataque, denominada {\em blind spoofing}
(spoofing de IP ciego), consiste en mandar un paquete SYN a B
con una dirección de origen trucha $IP_C$,
no recibir la respuesta (que contiene el número de secuencia inicial $y_0$)
dado que le llega a $IP_C$,
adivinar $y_0$ y mandar un nuevo paquete con 
número de confirmación $y_0 + 1$.
\end{defin}

Dado que los números de secuencia son números de 32 bits, 
la tarea de adivinar $y_0$ puede parecer imposible.
Pero en la práctica, las características de los generadores
de ISN la vuelven muchas veces accesible. 

Algunos sistemas usan siempre el mismo ISN, lo que vuelve el ataque
trivial. Ejemplos: algunos hubs 3Com (usan 0x803),
 impresoras Apple LaserWriter (usan 0xC7001).

Otros sistemas usan ISN múltiplos de $64\,000$. 
Ejemplos: switches 3Com, AmigaOS, Apple Mac OS 7.5-9,
algunas versiones viejas de Unix.
De esta manera el 
espacio de ISN se reduce a $2^{32} / \; 64\,000 \approx 67\,108$,
que se puede explorar por fuerza bruta.

Otro caso atacable son incrementos aleatorios, pero conteniendo
solo 7 bits de aleatoriedad (por ejemplo Windows 98).
La forma de atacar estos generadores es establecer primero conexiones legítimas
para estudiar los ISN generados, y usar esa información 
para adivinar el próximo ISN.

Ver el capítulo 10 del libro de Zalewski \cite{zalewski} para un estudio
mucho más detallado de los generadores de ISN de los
principales sistemas operativos, o los artículos disponibles 
en su página web \cite{zalewski2} y \cite{zalewski3}.

\section{Otros protocolos}

\subsection{El protocolo UDP}

UDP quiere decir {\em User Datagram Protocol} y está definido 
en el RFC 768.
Agrega el mecanismo de puertos para la entrega de datos
a una aplicación determinada, 
tiene poco costo extra (sobre el protocolo IP)
pero no provee ningún mecanismo de confiabilidad
(el emisor no sabe si el destinatario recibió los paquetes).
Se usa para comunicaciones simples, que no requieren mantener
información de estado.

\subsection{El protocolo ICMP}

ICMP quiere decir {\em Internet Control Message Protocol} y está definido 
en el RFC 792.
Los paquetes ICMP se usan para mandar información de diagnóstico
y notificaciones originadas por los otros protocolos.
Por ejemplo, indicando que un paquete no pudo ser entregado,
que expiró durante el camino, que fue rechazado por algún motivo.
La carga útil de un paquete ICMP depende del mensaje,
pero típicamente es una cita del principio del paquete
que originó el error.

\section{Diferencias entre implementaciones}

\begin{itemize}

\item{La prueba FIN.
Cuando uno recibe un paquete FIN (o cualquier paquete
que no tenga SYN o ACK), el comportamiento correcto,
de acuerdo con el RFC 793, es no responder.
Sin embargo implementaciones de Windows, Cisco, HP/UX
e IRIX responden con un RST.
}
\item{El bit DF (Don't Fragment).
Algunos sistemas están usando el bit DF prendido en el encabezado IP de los paquetes que
mandan (una posible razón es que el sistema esté tratando de
descubrir el MTU más chico de la ruta, ver definición \ref{pmtud}). 
Esto los distingue de los sistemas que usan el bit DF apagado.
} 
\item{El tamaño de la ventana (TCP window size).
Este campo aporta bastante información sobre la familia del sistema operativo.
Por ejemplo, AIX es el único operativo que usa 0x3F25.
Windows 2000 usa 0x402E, el mismo valor que usan OpenBSD y FreeBSD.
}
\item{Número de confirmación (acknowledgement).
Llamemos $seq$ y $ack$ a los números de secuencia y confirmación.
Como vimos en la sección \ref{handshake}, la respuesta normal
es mandar
\begin{align}
ack = seq \mbox{ recibido } + \mbox{ cantidad de bytes recibidos }
\end{align}
Sin embargo esto varía cuando el paquete recibido contiene
una combinación de flags extraña. 

Por ejemplo, cuando mandamos un paquete con FIN, PSH, URG a
 un puerto TCP cerrado, la mayoría de las implementaciones
devuelven $ack = seq$ recibido. Pero Windows y algunas impresoras
devuelven $ack = seq + 1$. 
Otro ejemplo, cuando mandamos un paquete con SYN, FIN, URG, PSH
a un puerto abierto,
Windows responde a veces $seq$, a veces $seq+1$ y a veces un
valor que parece aleatorio. 
}
\item{Tipo de servicio (TOS = Type Of Service).
Este es un campo del encabezado IP que terminó siendo ignorado
en las comunicaciones normales por Internet.
Cuando un sistema manda un mensaje de error ICMP ``puerto inalcanzable'',
el valor del campo TOS es normalmente 0. Sin embargo algunas versiones
de Linux usan 0xC0.
}

\item{Muestreo de valores IPID.
El número de identificación IPID sirve para reconocer los diferentes fragmentos de un paquete IP.
Existen diferencias en la forma de generar los IPID (que notaremos temporalmente ID).
Al recibir una serie de paquetes de un sistema, podemos estudiar la forma en que los genera.
\begin{itemize}
\item{
El ID vale siempre siempre 0, por ejemplo en Linux o en sistemas que implementan el método PMTUD
y mandan todos sus paquetes con DF prendido.
}
\item{
El ID es constante, por ejemplo en algunos routers Cisco.
}
\item{
El ID es incremental (la diferencia entre dos IDs consecutivos es menor que 10).
}
\item{
El ID es incremental, pero como el sistema guarda este valor en {\em host byte order} en lugar de
{\em network byte order}, las diferencias entre IDs consecutivos resultan múltiplos de 256.
Es el caso de Windows 95 y Windows 98.
}
\item{Quedan los casos en que el ID es aleatorio, o se incrementa en forma aleatoria.
}
\end{itemize}
}

\item{Muestreo de valores TCP ISN.
Al conectarnos varias veces (a intervalos de tiempo regulares) a un sistema, 
podemos estudiar la forma en que genera los ISN (esto es lo que realiza el test TSeq de Nmap).
Estos valores son números de secuencia de 32 bits.
Calculamos un vector $dif$ con las diferencias entre ISN consecutivos que recibimos 
(una diferencia negativa ser considera como un {\em wrap around} de 32 bits, 
y se le suma $2^{32}$).

Como parte del análisis, calculamos el $gcd$ de los valores de $dif$.
Podemos distinguir varias clases de generadores:
\begin{itemize}
\item{
Los ISN son siempre iguales. Por ejemplo, algunos switchs 3Com,
impresoras Apple, camaras Axis, etc...
}
\item{
El $gcd$ es divisible por $64\,000$.
Por ejemplo switches 3Com, AmigaOS, Apple Mac OS 7.5-9, etc...
}
\item{
El $gcd$ es divisible por 800 (pero no por $64\,000$).
Por ejemplo IBM OS/2 v3 y v4. 
}
\item{
Los ISN dependen del tiempo.
}
\item{
Los ISN se incrementan en forma aleatoria, o son generados en forma aleatoria.
}
\end{itemize}
}

\end{itemize}

\section{Detección basada en las firmas de Nmap}

Vamos a usar las diferencias que existen entre las implementaciones de los protocolos
para distinguir familias de sistemas operativos.
Para detectar diferencias usamos los tests de Nmap, un proyecto open source que tiene una extensa
base de datos, alimentada por mucha gente a lo largo de varios años.

\subsection{Los tests de Nmap}

Nmap (el mapeador de redes) es una herramienta de código abierto para 
exploración de red y auditoría de seguridad. 
Diseñada para analizar rápidamente grandes redes, 
funciona también contra equipos individuales. 
Nmap utiliza paquetes IP crudos (raw) con formatos extraños para determinar 
qué equipos se encuentran disponibles en una red, 
qué servicios (nombre y versión de la aplicación) ofrecen, 
qué sistemas operativos (y sus versiones) corren, 
qué tipo de filtros de paquetes o cortafuegos se están utilizando, etc...

Uno de los aspectos más conocidos de Nmap es la detección del sistema operativo
 en base a la comprobación de huellas TCP/IP. 
Nmap envía una serie de paquetes TCP y UDP al sistema remoto y analiza las respuestas. 
Nmap compara los resultados de pruebas tales como
el análisis de ISN de TCP, el soporte de opciones TCP y su orden, 
el análisis de IPID y las comprobaciones de tamaño inicial de ventana, con su base de datos.
En la tabla siguiente describimos brevemente los paquetes enviados
por cada test, para más información ver \cite{fyodor} y \cite{fyodor2}.

\begin{center}\begin{tabular}{| l | l | l | l  |}
\hline
Test  & enviar paquete   &  a un puerto  &  con los flags \\
\hline
T1  &  TCP  &  TCP abierto &  SYN, ECN \\
T2  &  TCP  &  TCP abierto  &  sin flags \\
T3  &  TCP  &  TCP abierto &  URG, PSH, SYN, FIN \\
T4  &  TCP  &  TCP abierto &  ACK \\
T5  &  TCP  &  TCP cerrado &  SYN \\
T6  &  TCP  &  TCP cerrado &  ACK \\
T7  &  TCP  &  TCP cerrado &  URG, PSH, FIN \\
PU &  UDP  &  UDP cerrado &  \\
TSeq &  TCP * 6  & TCP abierto & SYN \\
\hline
\end{tabular}
\end{center}

Nuestro método usa la base de datos de Nmap.
Esta base consta de 1684 huellas de sistemas operativos.
Cada huella contiene una descripción del sistema operativo, 
una clasificación que indica el nombre del proveedor (por ejemplo Sun), 
el sistema operativo subyacente (por ejemplo, Solaris), 
la versión del SO (por ejemplo, 10) 
y el tipo de dispositivo (propósito general, router, conmutador, consola de videojuegos, etc.).
Sigue un conjunto de reglas que describen cómo 
una versión / edición específica de un sistema operativo
responde a los tests.
Por ejemplo:

\begin{verbatim}
# Linux 2.6.0-test5 x86
Fingerprint Linux 2.6.0-test5 x86
Class Linux | Linux | 2.6.X | general purpose
TSeq(Class=RI%gcd=<6%SI=<2D3CFA0&>73C6B%IPID=Z%TS=1000HZ)
T1(DF=Y%W=16A0%ACK=S++%Flags=AS%Ops=MNNTNW)
T2(Resp=Y%DF=Y%W=0%ACK=S%Flags=AR%Ops=)
T3(Resp=Y%DF=Y%W=16A0%ACK=S++%Flags=AS%Ops=MNNTNW)
T4(DF=Y%W=0%ACK=O%Flags=R%Ops=)
T5(DF=Y%W=0%ACK=S++%Flags=AR%Ops=)
T6(DF=Y%W=0%ACK=O%Flags=R%Ops=)
T7(DF=Y%W=0%ACK=S++%Flags=AR%Ops=)
PU(DF=N%TOS=C0%IPLEN=164%RIPTL=148%RID=E%RIPCK=E%UCK=E%ULEN=134%DAT=E)
\end{verbatim}

La base de Nmap contiene 1684 huellas, 
vale decir que unas 1684 versiones / ediciones de sistemas operativos
pueden teóricamente ser distinguidas por este método.

Nmap funciona comparando la respuesta de una máquina
con cada huella en la base de datos.
Un puntaje es asignado a cada huella, calculado simplemente
como la cantidad de reglas que coinciden dividido por
la cantidad de reglas consideradas
(en efecto, las huellas pueden tener cantidad distintas de reglas,
o algunos campos pueden faltar en la respuesta de una máquina,
en tal caso las reglas correspondientes no se toman en cuenta).
Vale decir que Nmap realiza un especie de ``best fit''
basado en una distancia de Hamming en la cual todos los campos
de la respuesta tienen el mismo peso.

Uno de los problemas que presenta este método es el siguiente:
los sistemas operativos raros (improbables)
que generan menos respuesta a los tests obtienen un mejor puntaje!
(las reglas que coinciden adquieren un mayor peso relativo).
Por ejemplo, nos ocurrió en las pruebas de laboratorio que Nmap detecte un 
OpenBSD 3.1 como un ``Foundry FastIron Edge Switch (load balancer) 2402",
o que detecte un Mandrake 7.2 (Linux kernel versión 2.1) 
como un ``ZyXel Prestige Broadband router".
La riqueza de la base de datos se convierte en una debilidad!

\subsection{Estructura de redes jerárquicas}

Acá entran en juego las redes neuronales:
si representamos simbólicamente el espacio de sistemas operativos
como un espacio de 568 dimensiones 
(veremos a continuación el porqué de este número),
las respuestas posibles de las diferentes versiones de los sistemas
incluidos en la base de datos forma un nube de puntos.
Esta gran nube de puntos está estructurada de una manera particular,
dado que las familias de sistemas operativos forman grupos
más o menos reconocibles.
El método de Nmap consiste, a partir de la respuesta de una máquina,
a buscar el punto más cercano
(según la distancia de Hamming ya mencionada).
Está claro que estudiar el conjunto de datos puede aportar mucha información adicional.

Nuestro enfoque consiste en primer lugar a filtrar los 
sistemas operativos que no nos interesan
(siempre según el punto de vista del atacante,
concretamente los sistemas para los cuales no tiene exploits).
En nuestra implementación, nos interesan las familias
Windows, Linux, Solaris, OpenBSD, NetBSD y FreeBSD.
Estas son las familias de sistemas operativos que llamamos ``relevantes",
porque son las familias para las cuales hay exploits de Core Impact disponibles
(es una selección arbitraria, pero este análisis se puede aplicar para cualquier selección 
de sistemas operativos).

A continuación usamos la estructura de las familias de sistemas operativos
para asignar la máquina a una de las 6 familias consideradas.
El resultado es un módulo que utiliza varias redes neuronales
organizadas en forma jerárquica:
\begin{enumerate}
\item{
primer paso, una red neuronal para decidir si el SO es relevante.
}
\item{
segundo paso, una red neuronal para decidir la familia del SO:
Windows, Linux, Solaris, OpenBSD, FreeBSD, NetBSD.}
\item{
en el caso de Windows, usamos un módulo basado 
en los puntos finales (endpoints) DCE RPC
para refinar la detección.
}
\item{
en el caso de Linux, realizamos un análisis condicionado
(con otra red neuronal) para decidir la versión del kernel.
}
\item{
en el caso de Solaris y los BSD,
realizamos un análisis condicionado para decidir la versión.
}
\end{enumerate}

Usamos una red neuronal distinta para cada análisis.
De esta manera construimos 5 redes neuronales,
y cada una requiere una topología y un entrenamiento especial.

\subsection{Entradas de la red neuronal}

La primer cuestión a resolver es:
como traducir la respuesta de una máquina
en entradas para la red neuronal?
Asignamos un conjunto de neuronas de entrada para cada test.
Estos son los detalles para los tests T1 ... T7:
\begin{itemize}
\item{ una neurona para el flag ACK.}
\item{ una neurona para cada respuesta: S, S++, O.}
\item{ una neurona para el flag DF.}
\item{ una neurona para la respuesta : yes/no.}
\item{ una neurona para el campo \emph{Flags}.}
\item{ una neurona para cada flag: FIN, SYN, RST, PSH, ACK, URG, ECN, CWR
 (en total 8 neuronas).}
\item{ 10 grupos de 6 neuronas cada uno para el campo \emph{Options}.
Activamos una sola neurona en cada grupo según la opción - 
EOL, MAXSEG, NOP, TIMESTAMP, WINDOW, ECHOED -
respetando el orden de aparición (o sea en total 60 neuronas para las opciones). }
\item{ una neurona para el campo $W$ (tamaño de la ventana)
que tiene como entrada un valor hexadecimal.}
\end{itemize}

Para los flags o las opciones, la entrada es 1 o -1 (presente o ausente).
Otras neuronas tienen una entrada numérica, como 
el campo $W$ (tamaño de la ventana),
el campo $GCD$ (máximo común divisor de los números de secuencia iniciales)
o los campos $SI$ y $VAL$ de las respuestas al test Tseq.
En el ejemplo de un Linux 2.6.0, la respuesta:
\begin{verbatim} 
T3(Resp=Y%DF=Y%W=16A0%ACK=S++%Flags=AS%Ops=MNNTNW)
\end{verbatim}
se transforma en :

\noindent
\begin{tabular} {| c | c | c | c | c |  c | c | c | c | c |  c | c | c | c | c |    }
\hline
ACK & S & S++ & O & DF & Yes & Flags & E & U & A & P & R & S & F & $\ldots$ \\
\hline
1     & -1 & 1     & -1 &  1  &   1   & 1        & -1    &  -1     &   1    &   -1    &  -1   &   1   &  -1   &  $\ldots$ \\
\hline
\end{tabular}

De esta manera obtenemos una capa de entrada con 568 dimensiones,
con cierta redundancia.
La redundancia nos permite tratar de manera flexible las respuestas desconocidas
pero introduce también problemas de rendimiento y convergencia!
Veremos en la sección siguiente como resolver este problema
reduciendo la cantidad de dimensiones.
Las redes neuronales que usamos son redes de perceptrones
con 2 capas. 
Por ejemplo, la primer red neuronal (el filtro de relevancia) contiene:
la capa de entrada $C_0$ 96 neuronas,
la capa escondida $C_1$ 20 neuronas,
la capa de salida $C_2$ 1 neurona.

\subsection{Salidas de las redes neuronales}
\label{sec:salidas_redes}

Estamos usando las redes neuronales para resolver un problema 
de clasificación.
Como vimos en la sección \ref{sec:a_posteriori},
la forma más general de hacerlo es dejar que las salidas de la red
neuronal modelen las probabilidades de pertenencia a cada clase.

En el primer análisis, hay dos clases:
$\calC_1$ los sistemas operativos relevantes, y $\calC_2$ los 
sistemas operativos que queremos descartar.
En este caso basta con una sola neurona de salida,
cuyo valor de activación modela directamente $p(\calC_1 | \x)$
la probabilidad de que una entrada $\x$ pertenezca a $\calC_1$.

En el segundo análisis, hay 6 clases:
las familias de sistemas operativos 
Windows, Linux, Solaris, OpenBSD, FreeBSD, y NetBSD.
Por lo tanto ponemos 6 neuronas de salida.
El valor de salida $y_1(\x)$ representa la probabilidad de que una
máquina que generó el conjunto de respuestas $\x$ a los tests de Nmap
pertenezca a la familia Windows. Los valores de salida $y_2(\x) \ldots y_6(\x)$
se interpretan de manera similar.

\subsection{Generación de los datos de entrenamiento}

Para entrenar la red neuronal necesitamos un 
juego de datos $\calD = \{ (\x^n, \vv^n) : 1 \leq n \leq N \}$ 
con entradas $\{ \x^n \}$ (respuestas de máquinas)
y las salidas correspondientes $\{ \vv^n \}$
(sistema operativo de la máquina).
Como la base de huellas contiene 1684 reglas,
para tener una máquina correspondiente a cada regla,
necesitar{\'\i}amos una población de 1684 máquinas
para entrenar la red.
No teníamos acceso a semejante población...
y escanear la Internet no era una opción!

La solución adoptada es de generar el juego de datos
mediante una simulación Monte Carlo.
Para cada regla, generamos entradas que corresponden 
a esa regla.
La cantidad de entradas depende de la distribución empírica
de SO basada en datos estadísticos.
Vale decir que la distribución de sistemas operativos
en el  juego de datos generados 
respeta la distribución empírica.
Por ejemplo, de acuerdo con 
W3 Schools\footnote{\tt{http://www.w3schools.com/browsers/browsers\_os.asp}},
la repartición de sistemas operativos es la siguiente (datos de Julio 2007):

\begin{center}
\begin{tabular}{ | l | r |}
\hline
Sistema operativo & Porcentaje \\
\hline
Windows XP & 74,6 \% \\
Windows 2000 & 6,0 \% \\
Windows 98  & 0,9 \% \\
Windows Vista  & 3,6 \% \\
Windows 2003  & 2,0 \% \\
Linux   & 3,4 \% \\
Mac OS  &  4,0 \%  \\
\hline
\end{tabular}
\end{center}

Para generar una entrada $ \x^n $ que corresponda a una maquina $\vv^n$, 
tenemos que especificar la respuesta
de una máquina a partir de la regla de Nmap correspondiente.
Las reglas de Nmap especifican los valores de cada campo de la respuesta
como una constante, un conjunto acotado de valores posibles para ese campo,
o un intervalo de valores posibles.
Cuando la regla especifica una constante, usamos ese valor
para generar la respuesta simulada.
Cuando la regla especifica opciones o un intervalo de valores,
elegimos un valor siguiendo una distribución uniforme (porque no sabemos cual
es la distribución real).

Esto funciona porque las reglas de Nmap especifican la mayor{\'\i}a de los valores
como constantes. Veamos el ejemplo de la regla correspondiente a un Linux (versión de
kernel 2.6.10):
\begin{verbatim}
# Linux kernel 2.6.10 X86 Slackware 10.0
Fingerprint Linux 2.6.10
Class Linux | Linux | 2.6.X | general purpose
TSeq(Class=RI%gcd=<6%SI=<1FB5BDE&>51299%IPID=Z%TS=1000HZ)
T1(DF=Y%W=7FFF%ACK=S++%Flags=AS%Ops=MNNTNW)
T2(Resp=Y%DF=Y%W=0%ACK=S%Flags=AR%Ops=)
T3(Resp=Y%DF=Y%W=0%ACK=O%Flags=AR%Ops=)
T4(DF=Y%W=0%ACK=O%Flags=R%Ops=)
T5(DF=Y%W=0%ACK=S++%Flags=AR%Ops=)
T6(DF=Y%W=0%ACK=O%Flags=R%Ops=)
T7(DF=Y%W=0%ACK=S++%Flags=AR%Ops=)
PU(DF=N%TOS=C0%IPLEN=164%RIPTL=148%RID=E%RIPCK=E%UCK=E%ULEN=134%DAT=E)
\end{verbatim}
Para el test T1, todos los campos son especificados son constantes, 
y podemos generar la respuesta de un Linux 2.6.10 Slackware 10.0 en forma unívoca.
Lo mismo ocurre para los tests T2 hasta T7 y el test PU.
Salvo para el test TSeq, todos los valores especificados son constantes.
Por lo tanto, los datos generados corresponden a la respuesta de un Linux real,
salvo para los campos \emph{gcd} y \emph{SI} del test TSeq.
Para esos dos campos, se genera una respuesta plausible (obedece a la regla),
aunque no necesariamente corresponda a una máquina real.

Veamos otro ejemplo:
\begin{verbatim}
# Windows 2000 Version 5.0 Build 2195 SP 4 X86
Fingerprint Microsoft Windows 2000 SP4
Class Microsoft | Windows | NT/2K/XP | general purpose
TSeq(Class=TR%gcd=<6%IPID=I)
T1(DF=Y%W=4204|FFAF%ACK=S++%Flags=AS%Ops=MNWNNT)
T2(Resp=Y%DF=N%W=0%ACK=S%Flags=AR%Ops=)
T3(Resp=Y%DF=Y%W=4204|FFAF%ACK=S++%Flags=AS%Ops=MNWNNT)
T4(DF=N%W=0%ACK=O%Flags=R%Ops=)
T5(DF=N%W=0%ACK=S++%Flags=AR%Ops=)
T6(DF=N%W=0%ACK=O%Flags=R%Ops=)
T7(DF=N%W=0%ACK=S++%Flags=AR%Ops=)
PU(DF=N%TOS=0%IPLEN=38%RIPTL=148%RID=E%RIPCK=E%UCK=E%ULEN=134%DAT=E)
\end{verbatim}
En este caso, los únicos dos campos que presentan variaciones son el \emph{gcd} del test TSeq
y el W del test T1 (que puede tomar dos valores, 4204 o FFAF).
Todos los otros campos son constantes (entre máquinas Windows 2000 version 5.0 build 2195 service pack 4).
Al hacer este análisis sobre el conjunto de reglas de Nmap, vemos que la mayoría
de los campos son especificados como constantes y permiten generar una respuesta en forma unívoca.

\section{Reducción de las dimensiones y entrenamiento de la redes}

Cuando diseñamos la topología de las redes neuronales,
fuimos generosos con las entradas:
568 dimensiones, con una importante redundancia.
Una consecuencia es que la convergencia del entrenamiento
es más lenta y menos confiable,
tanto más que el juego de datos es muy grande.
Por \emph{convergencia del entrenamiento}, nos referimos a la evolución 
de la función de error cuando los pesos de la red se van ajustando 
usando los algoritmos de retro-propagación y descenso del gradiente
que vimos en el Capitulo \ref{chap:entrenamiento}.
El entrenamiento de la red se considera terminado cuando la función de error
cae debajo de un umbral prefijado, lo cual requiere (en nuestros experimentos)
varios miles de generaciones y varias horas de cómputo.
La solución al problema de la convergencia del entrenamiento fue reducir la cantidad de dimensiones,
usando la reducción de la matriz de correlación y el análisis en componentes principales.
Este análisis nos permite también entender mejor 
los elementos importantes de los tests usados.

El primer paso es considerar cada dimensión de entrada como
una variable aleatoria
$ X_i \, ( 1 \leq i \leq 568) $.
Las dimensiones de entrada tienen órdenes de magnitud diferentes:
los flags toman valores 1 / -1, en cambio
el campo ISN (número de secuencia inicial) es un entero de 32 bits.
Evitamos que la red tenga que aprender a sumar correctamente
estas variables heterogéneas normalizando las variables aleatorias
(restando la media $\mu$ y dividiendo por el desvío estándar $\sigma$):
\begin{align}
\frac { X_i - \mu_i } { \sigma_i }
\end{align}

\subsection{Matriz de correlación}

Luego calculamos la matriz de correlación $\bR$,
cuyos elementos son:
\begin{align}
R_{i j} = \frac{ E[ (X_i - \mu_i ) (X_j - \mu_j) ] } { \sigma_i \; \sigma_j }
\end{align}
El símbolo $E$ designa aquí la esperanza matemática.
Dado que normalizamos las variables,
$\mu_i = 0$ y $\sigma_i = 1$ para todo $i$,
la matriz de correlación es simplemente
\begin{align}\label{eq:matrizr}
 R_{i j} = E[ X_i \,X_j ]
\end{align}

\begin{obs}\label{nrz}
Sea $\bR$ la matriz de correlación, $\bZ$ la matriz de covarianza,
y $N$ la cantidad de patrones en el conjunto de datos $\calD$.
Si las variables $X_i$ están normalizadas, entonces
\begin{align}
N \bR = \bZ
\end{align}
\end{obs}

\begin{proof}
Notemos $\x^1, \ldots, \x^N$ los vectores de entrada del conjunto de datos.
La variable aleatoria $X_k$ es una variable discreta, que 
toma los valores $x^1_k, \ldots, x^N_k$. Luego la esperanza de $X_k$ es
\begin{align}
E[ X_k ] = \frac {1} {N} \sum_{n=1}^{N} x^n_k.
\end{align}
De la misma manera la esperanza del producto $X_i \, X_j $ es
\begin{align}
E[ X_i \, X_j ] = \frac{1}{N} \sum_{n=1}^{N} x^n_i \, x^n_j
\end{align}
Por lo tanto la matriz $N \bR$ tiene en cada lugar $(i,j)$
la suma $\sum_{n=1}^{N} x^n_i \, x^n_j$, lo que notamos
\begin{align}
N \bR  &= \left( \sum_{n} x^n_i\, x^n_j \right)_{i j} \\
&= \sum_{n} \left( x^n_i\, x^n_j \right)_{i j} \\
&= \sum_{n} \x^n\, (\x^n)^T = \bZ
\end{align}
donde $\bZ$ es la matriz de covarianza definida en \eqref{eq:acpz} como
\begin{align}
\bZ = \sum_{n=1}^{N}  ( \x^n - \bar{\x} )  ( \x^n - \bar{\x} )^T.
\end{align}
En efecto, como las variables están normalizadas, 
cada promedio $\bar{x_k} = 0$ y luego el vector promedio $\bar{\x} = 0$.
Por lo tanto da lo mismo mirar dependencia lineal entre columnas de $\bR$ o de $\bZ$.
\end{proof}

\begin{propo}
La dependencia lineal entre columnas de $\bR$ indica
variables dependientes.
Si existen coeficientes $\alpha_1 \ldots \alpha_d$ tales que 
$\sum_{k=1}^{d} \alpha_k Col_k(\bR) = 0$ entonces
\begin{align}
\sum_{k=1}^{d} \alpha_k X_k = 0
\end{align} 
\end{propo}

\begin{proof}
Llamemos $\bA$ a la matriz de dimensión $d \times N$, que tiene en cada columna 
el vector de entrada $\x^n$
\begin{align}
\bA = \left( x^n_i \right) _{i \, n}
\end{align} 
de manera que el producto
\begin{align}
\bA \cdot \bA^T = (x^n_i)_{i \, n} \cdot (x^n_j)_{n  j } = \left(\sum_{n=1}^{N} x^n_i \, x^n_j \right)_{ i  j } = \bZ.
\end{align}
Supongamos que $\bZ$ tiene columnas linealmente dependientes, vale decir que
existe un vector $\vv$ tal que $\bZ \vv = 0$ entonces
\begin{align}
\bA \, \bA^T \vv = 0
\end{align}
Consideremos el vector $\w = \bA^T \vv$, es una combinación lineal de las columnas de $\bA^T$,
vale decir que está generado por las filas de $\bA$. Como $\bA \w = 0$, tenemos que $\w$ es 
ortogonal a todas las filas de $\bA$. Por lo tanto, $\w = 0$ y $\vv^T \bA = 0$:
obtuvimos una combinación lineal entre las filas de $\bA$.
La fila $i$ de la matriz $\bA$ tiene la pinta
\begin{align}
( x^1_i, x^2_i, \ldots, x^N_i )
\end{align}
representa todos los valores que toma la variable $X_i$. 
La combinación lineal entre columnas de $\bR$ se traduce directamente
en una combinación lineal entre los valores de las variables $X_i$.
\end{proof}

Cuando encontramos variables linealmente dependientes,
nos quedamos con una y eliminamos las otras,
dado que no aportan información adicional.
Las constantes tienen varianza nula y son eliminadas
por este análisis.

Veamos el resultado en el caso de los sistemas OpenBSD.
Reproducimos abajo extractos de huellas de dos
versiones distintas de OpenBSD,
donde los campos que sobreviven a la reducción de la
matriz de correlación están marcados con cursivas.

\noindent \texttt{Fingerprint OpenBSD 3.6 (i386) \\
Class OpenBSD | OpenBSD | 3.X | general purpose \\
T1(DF=N \% \emph{W=4000} \% ACK=S++ \% Flags=AS \% Ops=\emph{MN}WNNT) \\
T2(Resp=N) \\
T3(\emph{Resp=N}) \\
T4(DF=N \% \emph{W=0} \% ACK=O \%Flags=R \% Ops=)   \\
T5(DF=N \% W=0 \% ACK=S++ \% Flags=AR \% Ops=)  }

\noindent \texttt{Fingerprint OpenBSD 2.2 - 2.3 \\
Class OpenBSD | OpenBSD | 2.X | general purpose \\
T1(DF=N \% \emph{W=402E} \% ACK=S++ \% Flags=AS \% Ops=\emph{MN}WNNT) \\
T2(Resp=N) \\
T3(\emph{Resp=Y} \% DF=N \% \emph{W=402E} \% ACK=S++ \% Flags=AS \% Ops=\emph{MN}WNNT) \\
T4(DF=N \% \emph{W=4000} \% ACK=O \% Flags=R \% Ops=) \\
T5(DF=N \% W=0 \% ACK=S++ \% Flags=AR \% Ops=)
}

Por ejemplo, para el test T1, los únicos campos que varían 
son $W$ y las dos primeras opciones,
los otros son constantes en todas las versiones de OpenBSD.
Otro ejemplo, para el test T4 solo $W$ es susceptible de variar,
y el test T5 directamente no aporta ninguna información
sobre la versión de OpenBSD examinada.

\begin{table}
\begin{center}
\begin{tabular} { | c | c | l | }
\hline
Indice & Antiguo índice & Nombre del campo \\
\hline
0 &   20  &  T1 : TCP OPT 1 EOL \\
1 &   26  &  T1 : TCP OPT 2 EOL \\
2  &  29  &  T1 : TCP OPT 2 TIMESTAMP \\
3  &  74  &  T1 : W FIELD \\
4  &  75  &  T2 : ACK FIELD \\
5  &  149  &  T2 : W FIELD \\
6 &   150  &  T3 : ACK FIELD \\
7  &  170  &  T3 : TCP OPT 1 EOL \\
8  &  179  &  T3 : TCP OPT 2 TIMESTAMP \\
9  &  224  &  T3 : W FIELD \\
10  &  227  &  T4 : SEQ S \\
11  &  299  &  T4 : W FIELD \\
12  &  377  &  T6 : SEQ S \\
13  &  452  &  T7 :  SEQ S \\
14  &  525  &  TSeq : CLASS FIELD \\
15  &  526  &  TSeq : SEQ TD \\
16  &  528  &  TSeq : SEQ RI \\
17  &  529  &  TSeq : SEQ TR \\
18  &  532  &  TSeq : GCD FIELD \\
19  &  533  &  TSeq : IPID FIELD \\
20  &  535  &  TSeq : IPID SEQ BROKEN INCR \\
21  &  536  &  TSeq : IPID SEQ RPI \\
22  &  537  &  TSeq : IPID SEQ RD \\
23  &  540  &  TSeq : SI FIELD \\
24  &  543  &  TSeq : TS SEQ 2HZ \\
25  &  546  &  TSeq : TS SEQ UNSUPPORTED \\
26  &  555  &  PU : UCK RID RIPCK EQ \\
27  &  558  &  PU : UCK RID RIPCK ZERO \\
28  &  559  &  PU : UCK RID RIPCK EQ \\
29  &  560  &  PU : UCK RID RIPCK FAIL \\
30  &  563  &  PU : UCK RID RIPCK EQ \\
31  &  564  &  PU : UCK RID RIPCK FAIL \\
32  &  565  &  PU : RIPTL FIELD \\
33  &  566  &  PU : TOS FIELD \\
\hline
\end{tabular}
\caption{Campos que permiten reconocer los OpenBSD}
\label{tabla-openbsd}
\end{center}
\end{table}

La tabla \ref{tabla-openbsd} muestra la lista completa de los campos que
sirven a reconocer las diferentes versiones de OpenBSD.
Como vimos, el test T5 no aparece,
en cambio los tests Tseq y PU conservan muchas variables,
lo que muestra que estos dos tests son los que mejor 
discriminan diferentes versiones dentro de la población de OpenBSD.

\subsection{Análisis en Componentes Principales}

Una reducción suplementaria de la dimensión de los datos de entrada usa el 
{\em Análisis en Componentes Principales} (ACP),
que hemos estudiado en la sección \ref{sec:acp}.
La idea es calcular una nueva base $\{ \uu_1, \ldots, \uu_d \}$ del espacio de entrada,
de tal manera que el menor error de cualquier proyección
del juego de datos en un subespacio de $p$ dimensiones ($p < d$),
provenga de proyectar sobre los $p$ primeros vectores de esta base.
Para esto tenemos que calcular los autovalores y autovectores de $\bZ$.
Notemos que el análisis anterior se puede ver como un caso de ACP:
buscar combinaciones lineales entre columnas de $\bZ$ es lo mismo que
buscar autovectores de autovalor 0.

Como ya hemos normalizado las variables de entrada,
tenemos que $\bar\x = \bcero$
y por lo tanto la matriz de covarianza $\bZ$ definida en \eqref{eq:acpz} vale
\begin{align}
\bZ = \sum_{n=1}^{N} (\x^n - \bar\x ) (\x^n - \bar\x )^T
 = \sum_{n=1}^{N} \x^n  (\x^n )^T 
\end{align}
De acuerdo con la Observación \ref{nrz}, $\bR = \frac{1}{N} \bZ $, y para el análisis en componentes principales,
podemos usar los autovectores de $\bR$ en lugar de los de $\bZ$.
El algoritmo ACP, modificado para aprovechar las cuentas que ya hicimos, consiste en:
\begin{itemize}
\item{Calcular los autovectores $\{ \uu_i \}$ y autovalores $\{ \lambda_i \}$ de $\bR$.
}
\item{Ordenar los vectores por autovalor decreciente.
}
\item{Quedarse con los $p$ primeros vectores para proyectar los datos.
}
\end{itemize}

El parámetro $p$ es elegido para mantener
por lo menos 98\% de la varianza total.
O sea pedimos que
\begin{align}
\sum_{i=1}^{p} \lambda_i \geq 0,98 \cdot \sum_{i=1}^{d} \lambda_i
\end{align}
Dicho al revés, pedimos que el error $E_p$ definido en \eqref{eq:acpep3} cumpla
\begin{align}
E_p = \frac{1}{2} \sum_{i=p+1}^{d} \lambda_i \leq 0,02 \cdot \frac{1}{2} \sum_{i=1}^{d} \lambda_i
\end{align}

Después de realizar el ACP obtuvimos las topologías siguientes
para las redes neuronales
(el tamaño de la capa de entrada original era de 568 en todos los casos):

\noindent
\begin{tabular} { | l | c | c | c | c | }
\hline
Análisis &  Capa de entrada & Capa de entrada 
& Capa & Capa de\\
& (post reducción & (post ACP) & escondida & salida\\
& de matriz $\bR$) & & & \\
\hline
Relevancia & 204 & 96 & 20 & 1 \\
Familia de SO & 145 & 66 & 20 & 6 \\
Linux & 100 & 41 & 18 & 8 \\
Solaris & 55 & 26 & 7 & 5 \\
OpenBSD & 34 & 23 & 4 & 3 \\
\hline
\end{tabular}

La dimensión de la capa de salida está determinada por la cantidad de clases
que queremos distinguir en cada análisis:\\
$\cdot$ En el primer análisis, una sola neurona de salida (relevante / no relevante).\\
$\cdot$ En el segundo análisis, una neurona de salida para cada familia de SO.\\
$\cdot$ En el análisis de versiones de Linux, distinguimos 8 grupos de versiones.\\
$\cdot$ En el caso de Solaris, distinguimos 5 grupos de versiones. \\
$\cdot$ En el caso de OpenBSD,  distinguimos 3 grupos de versiones. 

La dimensión de la capa escondida es uno de los parámetros difíciles de determinar
al usar redes de perceptrones. Una cantidad demasiado pequeña puede
impedir que la red permita distinguir las clases de salida, una cantidad demasiado grande
puede volver el entrenamiento demasiado lento o errático, y provocar el fenómeno 
de sobreajuste (overfitting).
En nuestros experimentos, la cantidad de neuronas escondidas representa entre 15\% y 45\% de 
la cantidad de neuronas en la capa de entrada, y fue determinada por un proceso de prueba y error
(viendo qué topología daba mejores resultados). 

Para concluir el ejemplo de OpenBSD,
a partir de las 34 variables que sobrevivieron a la reducción de la matriz de correlación,
se puede construir una nueva base con 23 vectores.
Las coordenadas en esta base son las entradas de la red,
la capa escondida solo contiene 4 neuronas
y la capa de salida 3 neuronas (dado que distinguimos
3 grupos de versiones de OpenBSD).
Una vez que sabemos que una máquina corre OpenBSD,
el problema de reconocer la versión es mucho más simple
y acotado, y se puede resolver con una red neuronal
de tamaño pequeño (más eficiente y rápida).

\subsection{Taza de aprendizaje adaptativa}
\label{sec:taza_adaptativa}

Es una estrategia para acelerar la convergencia del entrenamiento.
La taza de aprendizaje es el parámetro $\lambda$ que
interviene en las fórmulas de aprendizaje por retropropagación.

Dada una salida esperada de la red, podemos calcular
una estimación del error cuadrático 
\begin{align}
\frac { \sum_{i=1}^{n} ( y_i - v_i ) ^ 2 } {n}
\end{align}
donde  $v_i$ son las salidas esperadas y $y_i$ son las salidas de la red.

Después de cada generación (vale decir después de haber
hecho los cálculos para todos los pares de entrada / salida),
si el error es más grande, disminuimos la taza de aprendizaje.
Por el contrario, si el error es más pequeño, entonces
aumentamos la taza de aprendizaje.
La idea es desplazarse más rápido cuando estamos en la dirección correcta.

\begin{figure}
\centering
\includegraphics[width=12cm]{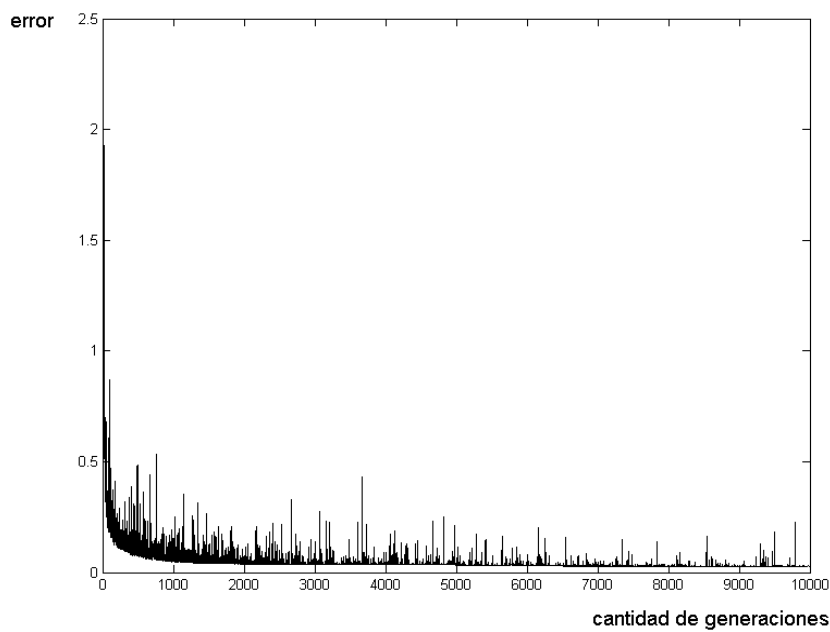}
\caption{Taza de aprendizaje fija}
\end{figure}
\begin{figure}
\centering
\includegraphics[width=12cm]{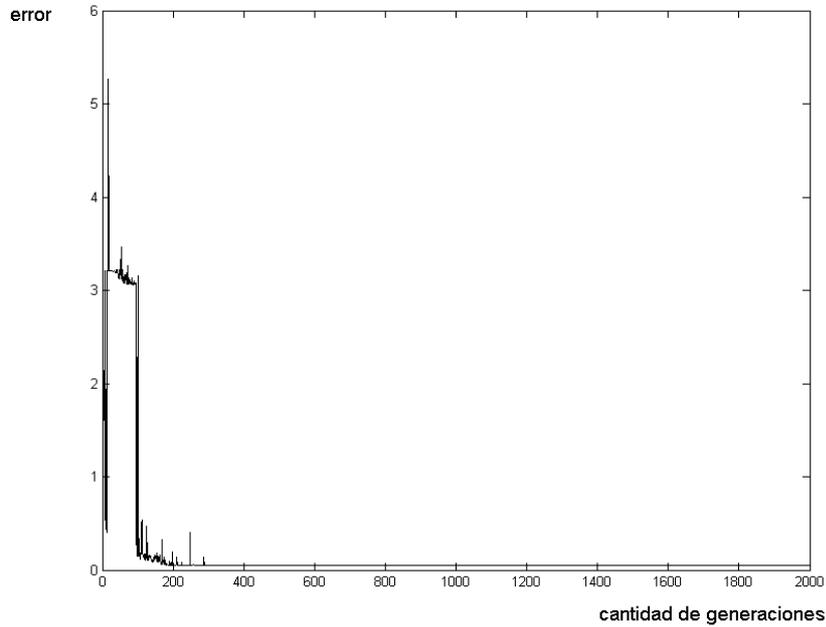}
\caption{Taza de aprendizaje adaptativa}
\end{figure}

Estos dos gráficos muestran la evolución del error cuadrático
promedio en función de la cantidad de generaciones para 
cada estrategia.
Cuando la taza de aprendizaje es fija,
el error disminuye y alcanza valores satisfactorios después
de 4000 o 5000 generaciones.
El error tiene una clara tendencia a disminuir,
pero con pequeños picos irregulares,
debidos a la naturaleza probabilística del entrenamiento de la red.

Utilizando una taza de aprendizaje adaptativa,
obtenemos al principio un comportamiento caótico,
con niveles de error más altos.
Pero una vez que el sistema encuentra la dirección correcta,
el error disminuye rápidamente para alcanzar un valor muy bajo
y constante después de 400 generaciones.
Estos resultados son claramente mejores y permiten 
acelerar el entrenamiento de las redes.
En nuestros experimentos, pasamos de tiempos de entrenamiento de 14 horas
a menos de una hora.

\subsection{Ejemplo de ejecución}

Reproducimos a continuación el resultado de ejecutar
el módulo contra una máquina Solaris 8.
El sistema correcto es reconocido con precisión.

\begin{verbatim}
Relevant / not relevant analysis
    0.99999999999999789    relevant 

Operating System analysis
    -0.99999999999999434   Linux 
    0.99999999921394744    Solaris 
    -0.99999999999998057   OpenBSD
    -0.99999964651426454 	 FreeBSD 
    -1.0000000000000000    NetBSD
    -1.0000000000000000    Windows

Solaris version analysis
    0.98172780325074482    Solaris 8 
    -0.99281382458335776   Solaris 9 
    -0.99357586906143880   Solaris 7 
    -0.99988378968003799   Solaris 2.X 
    -0.99999999977837983   Solaris 2.5.X 
\end{verbatim}

\section{DCE-RPC Endpoint mapper}

\subsection{Información provista por el servicio DCE-RPC}

DCE-RPC son las siglas de Distributed Computing Environment (DCE)
Remote Procedure Call (RPC). Es un sistema de llamada a procedimientos remotos
diseñado por la Open Software Foundation, y forma parte de su sistema de software
para computación distribuida (DCE).
Este sistema permite hacer funcionar software entre varias computadoras, 
como si estuviera corriendo en una sola.
Les permite a los programadores escribir software distribuido sin tener que 
preocuparse por el código de red subyacente.
La versión de Microsoft de DCE-RPC, también llamada MSRPC, 
fue usada para crear un modelo cliente/servidor en Windows NT
(y se siguió usando en las versiones siguientes de Windows).

En los sistemas Windows, el servicio 
DCE-RPC recibe conexiones enviadas al puerto TCP 135 de la maquina.
Enviando un pedido RPC, se puede determinar
qué servicios o programas están registrados en la base de datos 
del mapeador de endpoints RPC.
La respuesta incluye el UUID (Universal Unique IDentifier) 
de cada programa, el nombre anotado,
el protocolo utilizado,
la dirección de red a la cual el programa está ligado,
y el punto final del programa (endpoint).

El mecanismo de RPC fue diseñado para ser independiente del transporte:
diferentes protocolos pueden ser usados para transportar parámetros de 
un procedimiento remoto y resultados de una ejecución.
Cada protocolo de transporte tiene un identificador, 
en los sistemas Windows se usan habitualmente los protocolos 
siguientes\footnote{Los identificadores de transporte empiezan todos con las letras ``nca", por Network Computing Architecture}:\\
$\cdot$ {\tt ncacn\_ip\_tcp}: transporte por TCP/IP.\\
$\cdot$ {\tt ncadg\_ip\_udp}: transporte por UDP/IP.\\
$\cdot$ {\tt ncacn\_np}: transporte por \emph{named pipes}, usando SMB (Server Message Block).\\
$\cdot$ {\tt ncalrpc}: llamado RPC local.\\  
$\cdot$ {\tt ncacn\_http}: transporte por HTTP, usando IIS (Internet Information Services).

Un endpoint (punto final) es una entidad usada en el nivel de transporte para invocar remotamente un servicio RPC.
La naturaleza del endpoint depende del protocolo usado:\\
$\cdot$ {\tt ncacn\_ip\_tcp}: es un puerto TCP.\\
$\cdot$ {\tt ncadg\_ip\_udp}: es un puerto UDP.\\
$\cdot$ {\tt ncacn\_np}: es un \emph{named pipe}.\\
$\cdot$ {\tt ncalrpc}: es un puerto LPC (Local Procedure Call), usado para la comunicación
entre subsistemas internos en Windows NT.\\  
$\cdot$ {\tt ncacn\_http}: el puerto TCP 593.

Por ejemplo, en una maquina Windows 2000 edición profesional
service pack 0, el servicio RPC devuelve 8 endpoints que corresponden 
a 3 programas:
\begin{verbatim}
uuid="5A7B91F8-FF00-11D0-A9B2-00C04FB6E6FC"
annotation="Messenger Service"
 protocol="ncalrpc"      endpoint="ntsvcs"       id="msgsvc.1" 
 protocol="ncacn_np"     endpoint="\PIPE\ntsvcs" id="msgsvc.2" 
 protocol="ncacn_np"     endpoint="\PIPE\scerpc" id="msgsvc.3" 
 protocol="ncadg_ip_udp"                         id="msgsvc.4" 

uuid="1FF70682-0A51-30E8-076D-740BE8CEE98B"
 protocol="ncalrpc"      endpoint="LRPC"         id="mstask.1" 
 protocol="ncacn_ip_tcp"                         id="mstask.2" 

uuid="378E52B0-C0A9-11CF-822D-00AA0051E40F"
 protocol="ncalrpc"      endpoint="LRPC"         id="mstask.3" 
 protocol="ncacn_ip_tcp"                         id="mstask.4" 
\end{verbatim}

El interés de esta información reside en que
se pueden distinguir las versiones, ediciones y service packs
de Windows según la combinación de puntos finales
enviada por el servicio DCE-RPC.

\subsection{Nuevamente Redes Neuronales para analizar los endpoints}

Nuevamente nuestra idea es modelar la función que hace corresponder 
combinaciones de endpoints con versiones del sistema operativo
usando una red neuronal.

Usamos una red de perceptrones de 3 capas
(indicamos entre paréntesis la cantidad de neuronas en cada capa).
\begin{enumerate}
\item{capa de entrada (con 413 neuronas)
contiene una neurona por cada UUID
y una neurona por cada endpoint que corresponde a ese UUID.
Siguiendo con el ejemplo anterior, tenemos una neurona para el servicio
 Messenger y 4 neuronas para cada endpoint asociado a ese programa.
Esto nos permite responder con flexibilidad a la aparición de un endpoint desconocido:
retenemos de todas maneras la información del UUID principal.}
\item{capa de neuronas escondidas (con 42 neuronas),
donde cada neurona representa una combinación de neuronas de entrada.}
\item{capa de salida (con 25 neuronas),
contiene una neurona por cada versión y edición de Windows
(p.ej. Windows 2000 edición profesional),
y una neurona por cada versión y service pack de Windows
 (p.ej. Windows 2000 service pack 2). 
De esta manera la red puede distinguir la edición y el service pack de manera independiente:
los errores en una dimensión no afectan los errores en la otra dimensión.}
\end{enumerate}

\subsection{Resultados}


La tabla \ref{comparacion-dcerpc} muestra una comparación (de nuestro laboratorio)
entre el modulo DCE-RPC antiguo (que utiliza un algoritmo de ``best fit'')
y el nuevo modulo que utiliza una red neuronal para analizar la información.

\begin{table} [h]
\center
\begin{tabular} {| l | c | c | }
\hline
Resultado & Modulo DCE-RPC & DCE-RPC con \\
 & antiguo & redes neuronales \\
\hline
Concordancia perfecta & 6 & 7 \\
Concordancia parcial & 8 & 14 \\
Error & 7 & 0 \\
Sin respuesta & 2 & 2 \\
\hline
\end{tabular}
\caption{Comparación entre el antiguo y el nuevo módulo}
\label{comparacion-dcerpc}
\end{table}

Reproducimos a continuación el resultado de la ejecución del 
modulo contra una maquina Windows 2000 Edición Server SP1.
El sistema correcto es reconocido con precisión.

\begin{verbatim}
Neural Network Output (close to 1 is better):
Windows NT4: 4.87480503763e-005
Editions:
    Enterprise Server: 0.00972694324639
    Server: -0.00963500026763
Service Packs:
    6: 0.00559659167371
    6a: -0.00846224120952
Windows 2000: 0.996048928128
Editions:
    Server: 0.977780526016
    Professional: 0.00868998746624
    Advanced Server: -0.00564873813703
Service Packs:
    4: -0.00505441088081
    2: -0.00285674134367
    3: -0.0093665583402
    0: -0.00320117552666
    1: 0.921351036343
Windows 2003: 0.00302898647853
Editions:
    Web Edition: 0.00128127138728
    Enterprise Edition: 0.00771786077082
    Standard Edition: -0.0077145024893
Service Packs:
    0: 0.000853988551952
Windows XP: 0.00605168045887
Editions:
    Professional: 0.00115635710749
    Home: 0.000408057333416
Service Packs:
    2: -0.00160404945542
    0: 0.00216065240615
    1: 0.000759109188052
Setting OS to Windows 2000 Server sp1
Setting architecture: i386
\end{verbatim}

\section{Resultados experimentales}

En esta sección reproducimos numerosos ejemplos de ejecución del módulo
de reconocimiento de SO basado en los tests de Nmap, y en el caso
de máquinas Windows, de refinar la detección usando el DCE-RPC endpoint mapper.

\subsection{Windows 2000 Advanced Server SP 0}
\begin{verbatim}
Module "Neural Nmap OS Stack Fingerprinting" (v46560) started execution 
on Fri Aug 03 17:52:31 2007

Identifying host W2KADVSRV-SP0
 . Can't find a TCP open port in the container. A small Syn Scan will be launched
 . Found TCP open port 445.
 . Can't find a TCP closed port in the container. Using port 54305
 . Can't find a UDP closed port in the container. Using port 47666
T1(ACK=S++%Resp=Y%DF=Y%FLAGS=AS%W=64240%OPTIONS=MNWNNT)
T2(ACK=S%Resp=Y%DF=N%FLAGS=AR%W=0%OPTIONS=)
T3(ACK=S++%Resp=Y%DF=Y%FLAGS=AS%W=64240%OPTIONS=MNWNNT)
T4(ACK=O%Resp=Y%DF=N%FLAGS=R%W=0%OPTIONS=)
T5(ACK=S++%Resp=Y%DF=N%FLAGS=AR%W=0%OPTIONS=)
T6(ACK=O%Resp=Y%DF=N%FLAGS=R%W=0%OPTIONS=)
T7(ACK=S++%Resp=Y%DF=N%FLAGS=AR%W=0%OPTIONS=)
PU(RIPTL=328%RIPCK=E%DF=N%TOS=0%Resp=Y%IPLEN=56%DAT=E%RID=E%UCK=E%ULEN=308)
TSEQ(IPID=I%TS=0%SI=12491%CLASS=RI%GCD=1)

Neural Networks Output (close to 1 is better) for host : W2KADVSRV-SP0

Relevant analysis
Relevant: 0.999941687298

Operating System analysis
Linux: -1.0
Solaris: -0.999999999957
OpenBSD: -1.0
FreeBSD: -0.99999999989
NetBSD: -0.999999999997
Windows: 1.0

Setting (guessing) architecture: i386
Setting OS to windows unknown
--
Module finished execution after 4 secs.


Neural Network Output (close to 1 is better): 
2003: -0.999999749667
Editions:
	Enterprise Edition: -0.999999962087
	Web Edition: -0.999763009396
	Standard Edition: -0.999999955958
Service Packs:
	0: -0.999999765881
XP: -0.995926678826
Editions:
	Home: -0.999153319512
	Professional: -0.998379874255
Service Packs:
	1: -0.999440685451
	0: -0.999317709181
	2: -0.995951941804
2000: 0.999996512217
Editions:
	Professional: -0.99607795858
	Advanced Server: 0.999997142548
	Server: -0.999886282165
Service Packs:
	1: -0.427868026381
	0: 0.380428959868
	3: -0.999411870234
	2: -0.88105809509
	4: -0.999971974465
Vista: -0.999711954517
Editions:
	Enterprise Edition: -0.999992658054
	Home Premium: -0.99968153673
	Ultimate: -0.991802701272
	Business: -0.999894824162
	Home Basic: -0.999111052023
Service Packs:
	0: -0.999650521456
NT4: -0.999051377399
Editions:
	Enterprise Server: -0.999475956586
	Server: -0.999988987089
Service Packs:
	6a: -0.999813600039
	6: -0.999981293279
Setting OS to Windows 2000 Advanced Server sp0
 . Setting (guessing) architecture: i386
--
Module finished execution after 2 secs.
\end{verbatim}

\subsection{Windows 2000 Advanced Server SP 1}
\begin{verbatim}
Module "Neural Nmap OS Stack Fingerprinting" (v46560) started execution 
on Fri Aug 03 17:56:54 2007

Identifying host W2KADVSERV-SP1P
 . Can't find a TCP open port in the container. A small Syn Scan will be launched
 . Found TCP open port 445.
 . Can't find a TCP closed port in the container. Using port 49755
 . Can't find a UDP closed port in the container. Using port 41462
T1(ACK=S++%Resp=Y%DF=Y%FLAGS=AS%W=64240%OPTIONS=MNWNNT)
T2(ACK=S%Resp=Y%DF=N%FLAGS=AR%W=0%OPTIONS=)
T3(ACK=S++%Resp=Y%DF=Y%FLAGS=AS%W=64240%OPTIONS=MNWNNT)
T4(ACK=O%Resp=Y%DF=N%FLAGS=R%W=0%OPTIONS=)
T5(ACK=S++%Resp=Y%DF=N%FLAGS=AR%W=0%OPTIONS=)
T6(ACK=O%Resp=Y%DF=N%FLAGS=R%W=0%OPTIONS=)
T7(ACK=S++%Resp=Y%DF=N%FLAGS=AR%W=0%OPTIONS=)
PU(RIPTL=328%RIPCK=E%DF=N%TOS=0%Resp=Y%IPLEN=56%DAT=E%RID=E%UCK=E%ULEN=308)
TSEQ(IPID=I%TS=0%SI=7936%CLASS=RI%GCD=1)

Neural Networks Output (close to 1 is better) for host : W2KADVSERV-SP1P

Relevant analysis
Relevant: 0.999941686605

Operating System analysis
Linux: -1.0
Solaris: -0.999999999957
OpenBSD: -1.0
FreeBSD: -0.99999999989
NetBSD: -0.999999999997
Windows: 1.0

Setting (guessing) architecture: i386
Setting OS to windows unknown
--
Module finished execution after 3 secs.


Module "OS Detect by DCE-RPC Endpoint Mapper" (v44816) started execution 
on Fri Aug 03 18:00:06 2007

Neural Network Output (close to 1 is better): 
2003: -0.999999749667
Editions:
	Enterprise Edition: -0.999999962087
	Web Edition: -0.999763009396
	Standard Edition: -0.999999955958
Service Packs:
	0: -0.999999765881
XP: -0.995926678826
Editions:
	Home: -0.999153319512
	Professional: -0.998379874255
Service Packs:
	1: -0.999440685451
	0: -0.999317709181
	2: -0.995951941804
2000: 0.999996512217
Editions:
	Professional: -0.99607795858
	Advanced Server: 0.999997142548
	Server: -0.999886282165
Service Packs:
	1: -0.427868026381
	0: 0.380428959868
	3: -0.999411870234
	2: -0.88105809509
	4: -0.999971974465
Vista: -0.999711954517
Editions:
	Enterprise Edition: -0.999992658054
	Home Premium: -0.99968153673
	Ultimate: -0.991802701272
	Business: -0.999894824162
	Home Basic: -0.999111052023
Service Packs:
	0: -0.999650521456
NT4: -0.999051377399
Editions:
	Enterprise Server: -0.999475956586
	Server: -0.999988987089
Service Packs:
	6a: -0.999813600039
	6: -0.999981293279
Setting OS to Windows 2000 Advanced Server sp0
 . Setting (guessing) architecture: i386
--
Module finished execution after 2 secs.
\end{verbatim}

\subsection{Windows 2000 Advanced Server SP 3}
\begin{verbatim}
Module "Neural Nmap OS Stack Fingerprinting" (v46560) started execution 
on Fri Aug 03 18:03:11 2007

Identifying host W2KADVSRV-SP3
 . Can't find a TCP open port in the container. A small Syn Scan will be launched
 . Found TCP open port 445.
 . Can't find a TCP closed port in the container. Using port 56414
 . Can't find a UDP closed port in the container. Using port 40051
T1(ACK=S++%Resp=Y%DF=Y%FLAGS=AS%W=64240%OPTIONS=MNWNNT)
T2(ACK=S%Resp=Y%DF=N%FLAGS=AR%W=0%OPTIONS=)
T3(ACK=S++%Resp=Y%DF=Y%FLAGS=AS%W=64240%OPTIONS=MNWNNT)
T4(ACK=O%Resp=Y%DF=N%FLAGS=R%W=0%OPTIONS=)
T5(ACK=S++%Resp=Y%DF=N%FLAGS=AR%W=0%OPTIONS=)
T6(ACK=O%Resp=Y%DF=N%FLAGS=R%W=0%OPTIONS=)
T7(ACK=S++%Resp=Y%DF=N%FLAGS=AR%W=0%OPTIONS=)
PU(RIPTL=328%RIPCK=E%DF=N%TOS=0%Resp=Y%IPLEN=56%DAT=E%RID=E%UCK=E%ULEN=308)
TSEQ(IPID=I%TS=0%SI=10785%CLASS=RI%GCD=1)

Neural Networks Output (close to 1 is better) for host : W2KADVSRV-SP3

Relevant analysis
Relevant: 0.999941687039

Operating System analysis
Linux: -1.0
Solaris: -0.999999999957
OpenBSD: -1.0
FreeBSD: -0.99999999989
NetBSD: -0.999999999997
Windows: 1.0

Setting (guessing) architecture: i386
Setting OS to windows unknown
--
Module finished execution after 3 secs.


Module "OS Detect by DCE-RPC Endpoint Mapper" (v44816) started execution 
on Fri Aug 03 18:03:44 2007

Neural Network Output (close to 1 is better): 
2003: -0.999947255132
Editions:
	Enterprise Edition: -0.999982728984
	Web Edition: -0.99984844902
	Standard Edition: -0.999999877321
Service Packs:
	0: -0.999951724885
XP: -0.999988757916
Editions:
	Home: -0.999940293417
	Professional: -0.999988898767
Service Packs:
	1: -0.999987142258
	0: -0.999975151936
	2: -0.999297749564
2000: 0.999985538054
Editions:
	Professional: -0.999970218737
	Advanced Server: 0.999858892955
	Server: -0.997530987031
Service Packs:
	1: -0.999886704086
	0: -0.999999999998
	3: -0.672549039373
	2: -0.999999980088
	4: 0.674381703607
Vista: -0.999308071714
Editions:
	Enterprise Edition: -0.999997929828
	Home Premium: -0.999803939159
	Ultimate: -0.996110969976
	Business: -0.999919934419
	Home Basic: -0.999396429668
Service Packs:
	0: -0.999094772225
NT4: -0.999247289787
Editions:
	Enterprise Server: -0.999975675115
	Server: -0.999693970841
Service Packs:
	6a: -0.999605314625
	6: -0.999998540891
Setting OS to Windows 2000 Advanced Server sp4
 . Setting (guessing) architecture: i386
--
Module finished execution after 2 secs.
\end{verbatim}

\subsection{Windows 2000 Advanced Server SP 4}
\begin{verbatim}
Module "Neural Nmap OS Stack Fingerprinting" (v46560) started execution 
on Fri Aug 03 18:04:58 2007

Identifying host W2KADVSRV-SP4
 . Can't find a TCP open port in the container. A small Syn Scan will be launched
 . Found TCP open port 445.
 . Can't find a TCP closed port in the container. Using port 54482
 . Can't find a UDP closed port in the container. Using port 58537
T1(ACK=S++%Resp=Y%DF=Y%FLAGS=AS%W=64240%OPTIONS=MNWNNT)
T2(ACK=S%Resp=Y%DF=N%FLAGS=AR%W=0%OPTIONS=)
T3(ACK=S++%Resp=Y%DF=Y%FLAGS=AS%W=64240%OPTIONS=MNWNNT)
T4(ACK=O%Resp=Y%DF=N%FLAGS=R%W=0%OPTIONS=)
T5(ACK=S++%Resp=Y%DF=N%FLAGS=AR%W=0%OPTIONS=)
T6(ACK=O%Resp=Y%DF=N%FLAGS=R%W=0%OPTIONS=)
T7(ACK=S++%Resp=Y%DF=N%FLAGS=AR%W=0%OPTIONS=)
PU(RIPTL=328%RIPCK=E%DF=N%TOS=0%Resp=Y%IPLEN=56%DAT=E%RID=E%UCK=E%ULEN=308)
TSEQ(IPID=I%TS=0%SI=6068%CLASS=RI%GCD=1)

Neural Networks Output (close to 1 is better) for host : W2KADVSRV-SP4

Relevant analysis
Relevant: 0.999941686321

Operating System analysis
Linux: -1.0
Solaris: -0.999999999957
OpenBSD: -1.0
FreeBSD: -0.99999999989
NetBSD: -0.999999999997
Windows: 1.0

Setting (guessing) architecture: i386
Setting OS to windows unknown
--
Module finished execution after 3 secs.

Module "OS Detect by DCE-RPC Endpoint Mapper" (v44816) started execution 
on Fri Aug 03 18:05:21 2007

Neural Network Output (close to 1 is better): 
2003: -0.999947255132
Editions:
	Enterprise Edition: -0.999982728984
	Web Edition: -0.99984844902
	Standard Edition: -0.999999877321
Service Packs:
	0: -0.999951724885
XP: -0.999988757916
Editions:
	Home: -0.999940293417
	Professional: -0.999988898767
Service Packs:
	1: -0.999987142258
	0: -0.999975151936
	2: -0.999297749564
2000: 0.999985538054
Editions:
	Professional: -0.999970218737
	Advanced Server: 0.999858892955
	Server: -0.997530987031
Service Packs:
	1: -0.999886704086
	0: -0.999999999998
	3: -0.672549039373
	2: -0.999999980088
	4: 0.674381703607
Vista: -0.999308071714
Editions:
	Enterprise Edition: -0.999997929828
	Home Premium: -0.999803939159
	Ultimate: -0.996110969976
	Business: -0.999919934419
	Home Basic: -0.999396429668
Service Packs:
	0: -0.999094772225
NT4: -0.999247289787
Editions:
	Enterprise Server: -0.999975675115
	Server: -0.999693970841
Service Packs:
	6a: -0.999605314625
	6: -0.999998540891
Setting OS to Windows 2000 Advanced Server sp4
 . Setting (guessing) architecture: i386
--
Module finished execution after 1 secs.
\end{verbatim}

\subsection{Windows 2000 Professional SP 0}
\begin{verbatim}
Module "Neural Nmap OS Stack Fingerprinting" (v46560) started execution 
on Fri Aug 03 18:07:21 2007

Identifying host 2KPRO-SP0
 . Can't find a TCP open port in the container. A small Syn Scan will be launched
 . Found TCP open port 445.
 . Can't find a TCP closed port in the container. Using port 43224
 . Can't find a UDP closed port in the container. Using port 45951
T1(ACK=S++%Resp=Y%DF=Y%FLAGS=AS%W=64240%OPTIONS=MNWNNT)
T2(ACK=S%Resp=Y%DF=N%FLAGS=AR%W=0%OPTIONS=)
T3(ACK=S++%Resp=Y%DF=Y%FLAGS=AS%W=64240%OPTIONS=MNWNNT)
T4(ACK=O%Resp=Y%DF=N%FLAGS=R%W=0%OPTIONS=)
T5(ACK=S++%Resp=Y%DF=N%FLAGS=AR%W=0%OPTIONS=)
T6(ACK=O%Resp=Y%DF=N%FLAGS=R%W=0%OPTIONS=)
T7(ACK=S++%Resp=Y%DF=N%FLAGS=AR%W=0%OPTIONS=)
PU(RIPTL=328%RIPCK=E%DF=N%TOS=0%Resp=Y%IPLEN=56%DAT=E%RID=E%UCK=E%ULEN=308)
TSEQ(IPID=I%TS=0%SI=14571%CLASS=RI%GCD=1)

Neural Networks Output (close to 1 is better) for host : 2KPRO-SP0

Relevant analysis
Relevant: 0.999941687615

Operating System analysis
Linux: -1.0
Solaris: -0.999999999957
OpenBSD: -1.0
FreeBSD: -0.99999999989
NetBSD: -0.999999999997
Windows: 1.0

Setting (guessing) architecture: i386
Setting OS to windows unknown
--
Module finished execution after 3 secs.

Module "OS Detect by DCE-RPC Endpoint Mapper" (v44816) started execution 
on Fri Aug 03 18:07:53 2007

Neural Network Output (close to 1 is better): 
2003: -0.998262866887
Editions:
	Enterprise Edition: -0.999990150547
	Web Edition: -0.999963425494
	Standard Edition: -0.997768265695
Service Packs:
	0: -0.998472288517
XP: -0.999179877935
Editions:
	Home: -0.999417731381
	Professional: -0.998574408001
Service Packs:
	1: -0.999679181198
	0: -0.995553240272
	2: -0.999673575953
2000: 0.989100244713
Editions:
	Professional: 0.982645244795
	Advanced Server: -0.999696929425
	Server: -0.999954331719
Service Packs:
	1: -0.999999998573
	0: -0.999853407727
	3: -0.972853921662
	2: -0.957009794252
	4: -0.999999999996
Vista: -0.97961085952
Editions:
	Enterprise Edition: -0.997753326466
	Home Premium: -0.998188556643
	Ultimate: -0.982344378516
	Business: -0.999582288748
	Home Basic: -0.997915464289
Service Packs:
	0: -0.980773071549
NT4: -0.999461498513
Editions:
	Enterprise Server: -0.999783921503
	Server: -0.999967203157
Service Packs:
	6a: -0.999997408323
	6: -0.995010928976
Setting OS to Windows 2000 Professional
 . Setting (guessing) architecture: i386
--
Module finished execution after 1 secs.
\end{verbatim}

\subsection{Windows 2000 Professional SP 2}
\begin{verbatim}
Module "Neural Nmap OS Stack Fingerprinting" (v46560) started execution 
on Fri Aug 03 18:09:04 2007

Identifying host 2KPRO-SP2
 . Can't find a TCP open port in the container. A small Syn Scan will be launched
 . Found TCP open port 445.
 . Can't find a TCP closed port in the container. Using port 43197
 . Can't find a UDP closed port in the container. Using port 52544
T1(ACK=S++%Resp=Y%DF=Y%FLAGS=AS%W=64240%OPTIONS=MNWNNT)
T2(ACK=S%Resp=Y%DF=N%FLAGS=AR%W=0%OPTIONS=)
T3(ACK=S++%Resp=Y%DF=Y%FLAGS=AS%W=64240%OPTIONS=MNWNNT)
T4(ACK=O%Resp=Y%DF=N%FLAGS=R%W=0%OPTIONS=)
T5(ACK=S++%Resp=Y%DF=N%FLAGS=AR%W=0%OPTIONS=)
T6(ACK=O%Resp=Y%DF=N%FLAGS=R%W=0%OPTIONS=)
T7(ACK=S++%Resp=Y%DF=N%FLAGS=AR%W=0%OPTIONS=)
PU(RIPTL=328%RIPCK=E%DF=N%TOS=0%Resp=Y%IPLEN=56%DAT=E%RID=E%UCK=E%ULEN=308)
TSEQ(IPID=I%TS=0%SI=12824%CLASS=RI%GCD=1)

Neural Networks Output (close to 1 is better) for host : 2KPRO-SP2

Relevant analysis
Relevant: 0.999941687349

Operating System analysis
Linux: -1.0
Solaris: -0.999999999957
OpenBSD: -1.0
FreeBSD: -0.99999999989
NetBSD: -0.999999999997
Windows: 1.0

Setting (guessing) architecture: i386
Setting OS to windows unknown
--
Module finished execution after 3 secs.

Module "OS Detect by DCE-RPC Endpoint Mapper" (v44816) started execution 
on Fri Aug 03 18:09:37 2007

Neural Network Output (close to 1 is better): 
2003: -0.998262866887
Editions:
	Enterprise Edition: -0.999990150547
	Web Edition: -0.999963425494
	Standard Edition: -0.997768265695
Service Packs:
	0: -0.998472288517
XP: -0.999179877935
Editions:
	Home: -0.999417731381
	Professional: -0.998574408001
Service Packs:
	1: -0.999679181198
	0: -0.995553240272
	2: -0.999673575953
2000: 0.989100244713
Editions:
	Professional: 0.982645244795
	Advanced Server: -0.999696929425
	Server: -0.999954331719
Service Packs:
	1: -0.999999998573
	0: -0.999853407727
	3: -0.972853921662
	2: -0.957009794252
	4: -0.999999999996
Vista: -0.97961085952
Editions:
	Enterprise Edition: -0.997753326466
	Home Premium: -0.998188556643
	Ultimate: -0.982344378516
	Business: -0.999582288748
	Home Basic: -0.997915464289
Service Packs:
	0: -0.980773071549
NT4: -0.999461498513
Editions:
	Enterprise Server: -0.999783921503
	Server: -0.999967203157
Service Packs:
	6a: -0.999997408323
	6: -0.995010928976
Setting OS to Windows 2000 Professional
 . Setting (guessing) architecture: i386
--
Module finished execution after 1 secs.
\end{verbatim}

\subsection{Windows 2000 Professional SP 4}
\begin{verbatim}
Module "Neural Nmap OS Stack Fingerprinting" (v46560) started execution 
on Fri Aug 03 18:10:58 2007

Identifying host 2KPRO-SP4
 . Can't find a TCP open port in the container. A small Syn Scan will be launched
 . Found TCP open port 445.
 . Can't find a TCP closed port in the container. Using port 59149
 . Can't find a UDP closed port in the container. Using port 57127
T1(ACK=S++%Resp=Y%DF=Y%FLAGS=AS%W=64240%OPTIONS=MNWNNT)
T2(ACK=S%Resp=Y%DF=N%FLAGS=AR%W=0%OPTIONS=)
T3(ACK=S++%Resp=Y%DF=Y%FLAGS=AS%W=64240%OPTIONS=MNWNNT)
T4(ACK=O%Resp=Y%DF=N%FLAGS=R%W=0%OPTIONS=)
T5(ACK=S++%Resp=Y%DF=N%FLAGS=AR%W=0%OPTIONS=)
T6(ACK=O%Resp=Y%DF=N%FLAGS=R%W=0%OPTIONS=)
T7(ACK=S++%Resp=Y%DF=N%FLAGS=AR%W=0%OPTIONS=)
PU(RIPTL=328%RIPCK=E%DF=N%TOS=0%Resp=Y%IPLEN=56%DAT=E%RID=E%UCK=E%ULEN=308)
TSEQ(IPID=I%TS=0%SI=10107%CLASS=RI%GCD=1)

Neural Networks Output (close to 1 is better) for host : 2KPRO-SP4

Relevant analysis
Relevant: 0.999941686936

Operating System analysis
Linux: -1.0
Solaris: -0.999999999957
OpenBSD: -1.0
FreeBSD: -0.99999999989
NetBSD: -0.999999999997
Windows: 1.0

Setting (guessing) architecture: i386
Setting OS to windows unknown
--
Module finished execution after 3 secs.

Module "OS Detect by DCE-RPC Endpoint Mapper" (v44816) started execution 
on Fri Aug 03 18:16:06 2007

Neural Network Output (close to 1 is better): 
2003: -0.997547659226
Editions:
	Enterprise Edition: -0.990930851963
	Web Edition: -0.999597560204
	Standard Edition: -0.99999807532
Service Packs:
	0: -0.997757402622
XP: -0.994779776647
Editions:
	Home: -0.998557580864
	Professional: -0.996799534421
Service Packs:
	1: -0.999507541938
	0: -0.985142878974
	2: -0.999540693272
2000: 0.977796191204
Editions:
	Professional: 0.967703233132
	Advanced Server: -0.999676671726
	Server: -0.999978293796
Service Packs:
	1: -0.999999999691
	0: -0.985261423532
	3: -0.0312299372646
	2: -0.999999966164
	4: -0.99999999996
Vista: -0.989477503563
Editions:
	Enterprise Edition: -0.997084538343
	Home Premium: -0.998149546251
	Ultimate: -0.99369005782
	Business: -0.999364924897
	Home Basic: -0.997662707402
Service Packs:
	0: -0.988347647919
NT4: -0.999368373751
Editions:
	Enterprise Server: -0.985203586899
	Server: -0.999998438683
Service Packs:
	6a: -0.999997377969
	6: -0.994443401807
Setting OS to Windows 2000 Professional
 . Setting (guessing) architecture: i386
--
Module finished execution after 1 secs.
\end{verbatim}

\subsection{Windows XP Home SP 1}
\begin{verbatim}
Module "Neural Nmap OS Stack Fingerprinting" (v46560) started execution 
on Fri Aug 03 19:36:59 2007

Identifying host XPHSP1
 . Can't find a TCP open port in the container. A small Syn Scan will be launched
 . Found TCP open port 445.
 . Can't find a TCP closed port in the container. Using port 59926
 . Can't find a UDP closed port in the container. Using port 46541
T1(ACK=S++%Resp=Y%DF=Y%FLAGS=AS%W=16430%OPTIONS=MNWNNT)
T2(ACK=S%Resp=Y%DF=N%FLAGS=AR%W=0%OPTIONS=)
T3(ACK=S++%Resp=Y%DF=Y%FLAGS=AS%W=16430%OPTIONS=MNWNNT)
T4(ACK=O%Resp=Y%DF=N%FLAGS=R%W=0%OPTIONS=)
T5(ACK=S++%Resp=Y%DF=N%FLAGS=AR%W=0%OPTIONS=)
T6(ACK=O%Resp=Y%DF=N%FLAGS=R%W=0%OPTIONS=)
T7(ACK=S++%Resp=Y%DF=N%FLAGS=AR%W=0%OPTIONS=)
PU(RIPTL=32%RIPCK=E%DF=N%TOS=0%Resp=Y%IPLEN=56%DAT=E%RID=F%UCK=F%ULEN=12)
TSEQ(IPID=I%TS=0%SI=16864%CLASS=RI%GCD=1)

Neural Networks Output (close to 1 is better) for host : XPHSP1

Relevant analysis
Relevant: 0.999999999821

Operating System analysis
Linux: -1.0
Solaris: -0.999999999999
OpenBSD: -1.0
FreeBSD: -0.999820565463
NetBSD: -0.999999999999
Windows: 0.999999919035

Setting (guessing) architecture: i386
Setting OS to windows unknown
--
Module finished execution after 4 secs.

Module "OS Detect by DCE-RPC Endpoint Mapper" (v44816) started execution 
on Fri Aug 03 19:37:30 2007

Neural Network Output (close to 1 is better): 
2003: -0.97686628288
Editions:
	Enterprise Edition: -0.974276366215
	Web Edition: -0.997169984645
	Standard Edition: -0.98947240063
Service Packs:
	0: -0.976764636519
XP: 0.982033781694
Editions:
	Home: -0.782389513869
	Professional: 0.781715142886
Service Packs:
	1: -0.644004381716
	0: 0.643722767873
	2: -0.994851934287
2000: -0.986282751871
Editions:
	Professional: -0.981890661638
	Advanced Server: -0.986235902013
	Server: -0.99999958338
Service Packs:
	1: -0.999999996202
	0: -0.999994130333
	3: -0.996462185387
	2: -0.999998617596
	4: -0.999999997295
Vista: -0.999511628453
Editions:
	Enterprise Edition: -0.996902706404
	Home Premium: -0.998846032161
	Ultimate: -0.986313734364
	Business: -0.999727882801
	Home Basic: -0.997213662586
Service Packs:
	0: -0.999447953781
NT4: -0.995291293784
Editions:
	Enterprise Server: -0.992364589829
	Server: -0.999113094301
Service Packs:
	6a: -0.999991346331
	6: -0.984212301747
Setting OS to Windows XP Professional sp0
 . Setting (guessing) architecture: i386
--
Module finished execution after 1 secs.
\end{verbatim}

\subsection{Windows XP Professional SP 0}
\begin{verbatim}
Module "Neural Nmap OS Stack Fingerprinting" (v46560) started execution 
on Fri Aug 03 19:42:23 2007

Identifying host WINXP-PROSP0
 . Can't find a TCP open port in the container. A small Syn Scan will be launched
 . Found TCP open port 445.
 . Can't find a TCP closed port in the container. Using port 58070
 . Can't find a UDP closed port in the container. Using port 43388
T1(ACK=S++%Resp=Y%DF=Y%FLAGS=AS%W=64240%OPTIONS=MNWNNT)
T2(ACK=S%Resp=Y%DF=N%FLAGS=AR%W=0%OPTIONS=)
T3(ACK=S++%Resp=Y%DF=Y%FLAGS=AS%W=64240%OPTIONS=MNWNNT)
T4(ACK=O%Resp=Y%DF=N%FLAGS=R%W=0%OPTIONS=)
T5(ACK=S++%Resp=Y%DF=N%FLAGS=AR%W=0%OPTIONS=)
T6(ACK=O%Resp=Y%DF=N%FLAGS=R%W=0%OPTIONS=)
T7(ACK=S++%Resp=Y%DF=N%FLAGS=AR%W=0%OPTIONS=)
PU(RIPTL=328%RIPCK=E%DF=N%TOS=0%Resp=Y%IPLEN=56%DAT=E%RID=E%UCK=E%ULEN=308)
TSEQ(IPID=I%TS=0%SI=8691%CLASS=RI%GCD=1)

Neural Networks Output (close to 1 is better) for host : WINXP-PROSP0

Relevant analysis
Relevant: 0.99994168672

Operating System analysis
Linux: -1.0
Solaris: -0.999999999957
OpenBSD: -1.0
FreeBSD: -0.99999999989
NetBSD: -0.999999999997
Windows: 1.0

Setting (guessing) architecture: i386
Setting OS to windows unknown
--
Module finished execution after 4 secs.

Module "OS Detect by DCE-RPC Endpoint Mapper" (v44816) started execution 
on Fri Aug 03 19:42:41 2007

Neural Network Output (close to 1 is better): 
2003: -0.976865900508
Editions:
	Enterprise Edition: -0.974276697062
	Web Edition: -0.997169839239
	Standard Edition: -0.989473487426
Service Packs:
	0: -0.976764400241
XP: 0.982025690672
Editions:
	Home: -0.782377835425
	Professional: 0.781603957055
Service Packs:
	1: -0.644117279137
	0: 0.643715919863
	2: -0.994853092045
2000: -0.986282244753
Editions:
	Professional: -0.981895215216
	Advanced Server: -0.986226167626
	Server: -0.999999583734
Service Packs:
	1: -0.999999996199
	0: -0.999994137454
	3: -0.996464013302
	2: -0.999998618603
	4: -0.999999997293
Vista: -0.999511745009
Editions:
	Enterprise Edition: -0.996903133104
	Home Premium: -0.998846009352
	Ultimate: -0.986315824624
	Business: -0.99972788018
	Home Basic: -0.997213542381
Service Packs:
	0: -0.999448076308
NT4: -0.995288740453
Editions:
	Enterprise Server: -0.992361551034
	Server: -0.99911291821
Service Packs:
	6a: -0.999991345275
	6: -0.984204011638
Setting OS to Windows XP Professional sp0
 . Setting (guessing) architecture: i386
--
Module finished execution after 2 secs.
\end{verbatim}

\subsection{Windows XP Professional SP 1}
\begin{verbatim}
Module "Neural Nmap OS Stack Fingerprinting" (v46560) started execution 
on Fri Aug 03 19:44:16 2007

Identifying host XPPROSP1SRC
 . Can't find a TCP open port in the container. A small Syn Scan will be launched
 . Found TCP open port 445.
 . Can't find a TCP closed port in the container. Using port 57802
 . Can't find a UDP closed port in the container. Using port 52521
T1(ACK=S++%Resp=Y%DF=Y%FLAGS=AS%W=64240%OPTIONS=MNWNNT)
T2(ACK=S%Resp=Y%DF=N%FLAGS=AR%W=0%OPTIONS=)
T3(ACK=S++%Resp=Y%DF=Y%FLAGS=AS%W=64240%OPTIONS=MNWNNT)
T4(ACK=O%Resp=Y%DF=N%FLAGS=R%W=0%OPTIONS=)
T5(ACK=S++%Resp=Y%DF=N%FLAGS=AR%W=0%OPTIONS=)
T6(ACK=O%Resp=Y%DF=N%FLAGS=R%W=0%OPTIONS=)
T7(ACK=S++%Resp=Y%DF=N%FLAGS=AR%W=0%OPTIONS=)
PU(RIPTL=32%RIPCK=E%DF=N%TOS=0%Resp=Y%IPLEN=56%DAT=E%RID=F%UCK=F%ULEN=12)
TSEQ(IPID=I%TS=0%SI=12356%CLASS=RI%GCD=1)

Neural Networks Output (close to 1 is better) for host : XPPROSP1SRC

Relevant analysis
Relevant: 0.999514119904

Operating System analysis
Linux: -1.0
Solaris: -0.999999999999
OpenBSD: -1.0
FreeBSD: -0.999803238507
NetBSD: -1.0
Windows: 0.999999999905

Setting (guessing) architecture: i386
Setting OS to windows unknown
--
Module finished execution after 3 secs.

Module "OS Detect by DCE-RPC Endpoint Mapper" (v44816) started execution 
on Fri Aug 03 19:44:43 2007

Neural Network Output (close to 1 is better): 
2003: -0.976866466777
Editions:
	Enterprise Edition: -0.974276445356
	Web Edition: -0.99716998999
	Standard Edition: -0.989472456135
Service Packs:
	0: -0.976764815683
XP: 0.982034078113
Editions:
	Home: -0.782390137405
	Professional: 0.781719584116
Service Packs:
	1: -0.643998315244
	0: 0.643722354671
	2: -0.994851901817
2000: -0.986282559979
Editions:
	Professional: -0.981890047816
	Advanced Server: -0.986236157306
	Server: -0.999999583371
Service Packs:
	1: -0.999999996202
	0: -0.999994129597
	3: -0.996462111441
	2: -0.999998617521
	4: -0.999999997295
Vista: -0.99951162021
Editions:
	Enterprise Edition: -0.996902670015
	Home Premium: -0.998846031131
	Ultimate: -0.986313621237
	Business: -0.999727882217
	Home Basic: -0.997213666256
Service Packs:
	0: -0.999447945135
NT4: -0.995291443616
Editions:
	Enterprise Server: -0.992364695536
	Server: -0.999113116513
Service Packs:
	6a: -0.999991346474
	6: -0.984212570005
Setting OS to Windows XP Professional sp0
 . Setting (guessing) architecture: i386
--
Module finished execution after 1 secs.
\end{verbatim}

\subsection{RedHat 7.2 - Linux kernel 2.4.7}
\begin{verbatim}
Module "Neural Nmap OS Stack Fingerprinting" (v46560) started execution 
on Fri Aug 03 19:54:13 2007

Identifying host 
 . Can't find a TCP open port in the container. A small Syn Scan will be launched
 . Found TCP open port 443.
 . Can't find a TCP closed port in the container. Using port 40552
 . Can't find a UDP closed port in the container. Using port 44615
The avg TCP TS HZ is: 92.713932
T1(ACK=S++%Resp=Y%DF=Y%FLAGS=AS%W=5792%OPTIONS=MNNTNW)
T2(Resp=N)
T3(ACK=S++%Resp=Y%DF=Y%FLAGS=AS%W=5792%OPTIONS=MNNTNW)
T4(ACK=O%Resp=Y%DF=Y%FLAGS=R%W=0%OPTIONS=)
T5(ACK=S++%Resp=Y%DF=Y%FLAGS=AR%W=0%OPTIONS=)
T6(ACK=O%Resp=Y%DF=Y%FLAGS=R%W=0%OPTIONS=)
T7(ACK=S++%Resp=Y%DF=Y%FLAGS=AR%W=0%OPTIONS=)
PU(RIPTL=328%RIPCK=E%DF=N%TOS=192%Resp=Y%IPLEN=356%DAT=E%RID=E%UCK=E%ULEN=308)
TSEQ(IPID=Z%TS=100HZ%SI=4531809%CLASS=RI%GCD=1)

Neural Networks Output (close to 1 is better) for host : 

Relevant analysis
Relevant: 1.0

Operating System analysis
Linux: 1.0
Solaris: -1.0
OpenBSD: -0.999999999942
FreeBSD: -1.0
NetBSD: -1.0
Windows: -1.0

Linux version analysis
2.0: -1.0
2.2: -0.811819709928
1: -1.0
2.1: -0.999999999999
2.3: -1.0
2.4: -0.572290330796
2.5: 0.963623880474
2.6: -0.999641979997

Setting (guessing) architecture: i386
Setting OS to linux 2.5
--
Module finished execution after 7 secs.
\end{verbatim}

\subsection{RedHat 7.3 - Linux kernel 2.4.18}
\begin{verbatim}
Module "Neural Nmap OS Stack Fingerprinting" (v46560) started execution 
on Fri Aug 03 19:57:01 2007

Identifying host REDHAT73
 . Can't find a TCP open port in the container. A small Syn Scan will be launched
 . Found TCP open port 25.
 . Can't find a TCP closed port in the container. Using port 55639
 . Can't find a UDP closed port in the container. Using port 45636
The avg TCP TS HZ is: 107.324767
T1(ACK=S++%Resp=Y%DF=Y%FLAGS=AS%W=5792%OPTIONS=MNNTNW)
T2(Resp=N)
T3(ACK=S++%Resp=Y%DF=Y%FLAGS=AS%W=5792%OPTIONS=MNNTNW)
T4(ACK=O%Resp=Y%DF=Y%FLAGS=R%W=0%OPTIONS=)
T5(ACK=S++%Resp=Y%DF=Y%FLAGS=AR%W=0%OPTIONS=)
T6(ACK=O%Resp=Y%DF=Y%FLAGS=R%W=0%OPTIONS=)
T7(ACK=S++%Resp=Y%DF=Y%FLAGS=AR%W=0%OPTIONS=)
PU(RIPTL=328%RIPCK=E%DF=N%TOS=192%Resp=Y%IPLEN=356%DAT=E%RID=E%UCK=E%ULEN=308)
TSEQ(IPID=Z%TS=100HZ%SI=6116470%CLASS=RI%GCD=1)

Neural Networks Output (close to 1 is better) for host : REDHAT73

Relevant analysis
Relevant: 1.0

Operating System analysis
Linux: 1.0
Solaris: -1.0
OpenBSD: -0.999999999942
FreeBSD: -1.0
NetBSD: -1.0
Windows: -1.0

Linux version analysis
2.0: -1.0
2.2: -0.81261567247
1: -1.0
2.1: -0.999999999999
2.3: -1.0
2.4: -0.571271875895
2.5: 0.963580536888
2.6: -0.999639959408

Setting (guessing) architecture: i386
Setting OS to linux 2.5
--
Module finished execution after 6 secs.
\end{verbatim}

\subsection{RedHat 8 - Linux kernel 2.4.18}
\begin{verbatim}
Module "Neural Nmap OS Stack Fingerprinting" (v46560) started execution 
on Fri Aug 03 19:59:16 2007

Identifying host 
 . Can't find a TCP open port in the container. A small Syn Scan will be launched
 . Found TCP open port 443.
 . Can't find a TCP closed port in the container. Using port 52267
 . Can't find a UDP closed port in the container. Using port 43185
The avg TCP TS HZ is: 495.076523
T1(ACK=S++%Resp=Y%DF=Y%FLAGS=AS%W=5792%OPTIONS=MNNTNW)
T2(Resp=N)
T3(ACK=S++%Resp=Y%DF=Y%FLAGS=AS%W=5792%OPTIONS=MNNTNW)
T4(ACK=O%Resp=Y%DF=Y%FLAGS=R%W=0%OPTIONS=)
T5(ACK=S++%Resp=Y%DF=Y%FLAGS=AR%W=0%OPTIONS=)
T6(ACK=O%Resp=Y%DF=Y%FLAGS=R%W=0%OPTIONS=)
T7(ACK=S++%Resp=Y%DF=Y%FLAGS=AR%W=0%OPTIONS=)
PU(RIPTL=328%RIPCK=E%DF=N%TOS=192%Resp=Y%IPLEN=356%DAT=E%RID=E%UCK=E%ULEN=308)
TSEQ(IPID=Z%SI=3448260%CLASS=RI%GCD=1)

Neural Networks Output (close to 1 is better) for host : 

Relevant analysis
Relevant: 1.0

Operating System analysis
Linux: 1.0
Solaris: -0.999999999995
OpenBSD: -0.999999999952
FreeBSD: -1.0
NetBSD: -1.0
Windows: -1.0

Linux version analysis
2.0: -1.0
2.2: 0.99985697843
1: -1.0
2.1: -0.992948310554
2.3: -0.999999999961
2.4: -0.996308385368
2.5: -0.969276727872
2.6: 0.997298116001

Setting (guessing) architecture: i386
Setting OS to linux 2.6
--
Module finished execution after 7 secs.
\end{verbatim}

\subsection{RedHat 9 - Linux kernel 2.4.20}
\begin{verbatim}
Module "Neural Nmap OS Stack Fingerprinting" (v46560) started execution 
on Fri Aug 03 20:00:58 2007

Identifying host 
 . Can't find a TCP open port in the container. A small Syn Scan will be launched
 . Found TCP open port 443.
 . Can't find a TCP closed port in the container. Using port 47776
 . Can't find a UDP closed port in the container. Using port 41173
The avg TCP TS HZ is: 98.167136
T1(ACK=S++%Resp=Y%DF=Y%FLAGS=AS%W=5792%OPTIONS=MNNTNW)
T2(Resp=N)
T3(ACK=S++%Resp=Y%DF=Y%FLAGS=AS%W=5792%OPTIONS=MNNTNW)
T4(ACK=O%Resp=Y%DF=Y%FLAGS=R%W=0%OPTIONS=)
T5(ACK=S++%Resp=Y%DF=Y%FLAGS=AR%W=0%OPTIONS=)
T6(ACK=O%Resp=Y%DF=Y%FLAGS=R%W=0%OPTIONS=)
T7(ACK=S++%Resp=Y%DF=Y%FLAGS=AR%W=0%OPTIONS=)
PU(RIPTL=328%RIPCK=E%DF=N%TOS=192%Resp=Y%IPLEN=356%DAT=E%RID=E%UCK=E%ULEN=308)
TSEQ(IPID=Z%TS=100HZ%SI=1173253%CLASS=RI%GCD=1)

Neural Networks Output (close to 1 is better) for host : 

Relevant analysis
Relevant: 1.0

Operating System analysis
Linux: 1.0
Solaris: -1.0
OpenBSD: -0.999999999943
FreeBSD: -1.0
NetBSD: -1.0
Windows: -1.0

Linux version analysis
2.0: -1.0
2.2: -0.810122652594
1: -1.0
2.1: -0.999999999999
2.3: -1.0
2.4: -0.574443724934
2.5: 0.963715560002
2.6: -0.999646226043

Setting (guessing) architecture: i386
Setting OS to linux 2.5
--
Module finished execution after 6 secs.
\end{verbatim}

\subsection{OpenBSD 3.1}
\begin{verbatim}
Module "Neural Nmap OS Stack Fingerprinting" (v46560) started execution 
on Fri Aug 03 20:24:13 2007

Identifying host 
 . Can't find a TCP open port in the container. A small Syn Scan will be launched
 . Found TCP open port 111.
 . Can't find a TCP closed port in the container. Using port 45877
 . Can't find a UDP closed port in the container. Using port 50092
The avg TCP TS HZ is: 1.818184
T1(ACK=S++%Resp=Y%DF=Y%FLAGS=AS%W=16445%OPTIONS=MNWNNT)
T2(Resp=N)
T3(ACK=S++%Resp=Y%DF=Y%FLAGS=AS%W=16445%OPTIONS=MNWNNT)
T4(ACK=O%Resp=Y%DF=Y%FLAGS=R%W=16384%OPTIONS=)
T5(ACK=S++%Resp=Y%DF=Y%FLAGS=AR%W=0%OPTIONS=)
T6(ACK=O%Resp=Y%DF=Y%FLAGS=R%W=0%OPTIONS=)
T7(ACK=S%Resp=Y%DF=Y%FLAGS=AR%W=0%OPTIONS=)
PU(RIPTL=308%RIPCK=F%DF=N%TOS=0%Resp=Y%IPLEN=56%DAT=E%RID=E%UCK=E%ULEN=308)
TSEQ(IPID=RD%CLASS=TR%TS=2HZ)

Neural Networks Output (close to 1 is better) for host : 

Relevant analysis
Relevant: 0.999739390711

Operating System analysis
Linux: -1.0
Solaris: -0.999999999984
OpenBSD: 1.0
FreeBSD: -0.999999999987
NetBSD: -1.0
Windows: -1.0

OpenBSD version analysis
2.7: -0.999999981748
2: -0.999959190418
3: 0.996988038738

Setting (guessing) architecture: i386
Setting OS to openbsd 3
--
Module finished execution after 7 secs.
\end{verbatim}

\subsection{OpenBSD 3.3}
\begin{verbatim}
Module "Neural Nmap OS Stack Fingerprinting" (v46560) started execution 
on Fri Aug 03 20:25:28 2007

Identifying host 
 . Can't find a TCP open port in the container. A small Syn Scan will be launched
 . Found TCP open port 23.
 . Can't find a TCP closed port in the container. Using port 42982
 . Can't find a UDP closed port in the container. Using port 58442
The avg TCP TS HZ is: 1.834863
T1(ACK=S++%Resp=Y%DF=Y%FLAGS=AS%W=16445%OPTIONS=MNWNNT)
T2(Resp=N)
T3(ACK=S++%Resp=Y%DF=Y%FLAGS=AS%W=16445%OPTIONS=MNWNNT)
T4(ACK=O%Resp=Y%DF=Y%FLAGS=R%W=16384%OPTIONS=)
T5(ACK=S++%Resp=Y%DF=Y%FLAGS=AR%W=0%OPTIONS=)
T6(ACK=O%Resp=Y%DF=Y%FLAGS=R%W=0%OPTIONS=)
T7(ACK=S%Resp=Y%DF=Y%FLAGS=AR%W=0%OPTIONS=)
PU(RIPTL=308%RIPCK=F%DF=N%TOS=0%Resp=Y%IPLEN=56%DAT=E%RID=E%UCK=E%ULEN=308)
TSEQ(IPID=RD%CLASS=TR%TS=2HZ)

Neural Networks Output (close to 1 is better) for host : 

Relevant analysis
Relevant: 0.999739390711

Operating System analysis
Linux: -1.0
Solaris: -0.999999999984
OpenBSD: 1.0
FreeBSD: -0.999999999987
NetBSD: -1.0
Windows: -1.0

OpenBSD version analysis
2.7: -0.999999981748
2: -0.999959190418
3: 0.996988038738

Setting (guessing) architecture: i386
Setting OS to openbsd 3
--
Module finished execution after 7 secs.
\end{verbatim}

\subsection{Mandrake 7.2 - Linux kernel 2.2.17}
\begin{verbatim}
Module "Neural Nmap OS Stack Fingerprinting" (v46560) started execution 
on Fri Aug 03 20:30:07 2007

Identifying host 
 . Can't find a TCP open port in the container. A small Syn Scan will be launched
 . Found TCP open port 23.
 . Can't find a TCP closed port in the container. Using port 51249
 . Can't find a UDP closed port in the container. Using port 44016
T1(ACK=S++%Resp=Y%DF=Y%FLAGS=AS%W=32595%OPTIONS=MENW)
T2(Resp=N)
T3(ACK=S++%Resp=Y%DF=Y%FLAGS=AS%W=32595%OPTIONS=MENW)
T4(ACK=O%Resp=Y%DF=N%FLAGS=R%W=0%OPTIONS=)
T5(ACK=S++%Resp=Y%DF=N%FLAGS=AR%W=0%OPTIONS=)
T6(ACK=O%Resp=Y%DF=N%FLAGS=R%W=0%OPTIONS=)
T7(ACK=S%Resp=Y%DF=N%FLAGS=AR%W=0%OPTIONS=)
PU(RIPTL=328%RIPCK=E%DF=N%TOS=192%Resp=Y%IPLEN=356%DAT=E%RID=E%UCK=E%ULEN=308)
TSEQ(IPID=I%TS=U%SI=4295012%CLASS=RI%GCD=1)

Neural Networks Output (close to 1 is better) for host : 

Relevant analysis
Relevant: 1.0

Operating System analysis
Linux: 1.0
Solaris: -0.999999862027
OpenBSD: -0.999999999999
FreeBSD: -1.0
NetBSD: -1.0
Windows: -1.0

Linux version analysis
2.0: -1.0
2.2: 1.0
1: -1.0
2.1: 0.999905561341
2.3: -0.999999999991
2.4: -1.0
2.5: -1.0
2.6: -0.999999999929

Setting (guessing) architecture: i386
Setting OS to linux 2.1
--
Module finished execution after 6 secs.
\end{verbatim}

\subsection{Mandrake 10 - Linux kernel 2.6.3}
\begin{verbatim}
Module "Neural Nmap OS Stack Fingerprinting" (v46560) started execution 
on Fri Aug 03 20:34:10 2007

Identifying host 
 . Can't find a TCP open port in the container. A small Syn Scan will be launched
 . Found TCP open port 445.
 . Can't find a TCP closed port in the container. Using port 44906
 . Can't find a UDP closed port in the container. Using port 45558
The avg TCP TS HZ is: 994.954627
T1(ACK=S++%Resp=Y%DF=Y%FLAGS=AS%W=5792%OPTIONS=MNNTNW)
T2(Resp=N)
T3(ACK=S++%Resp=Y%DF=Y%FLAGS=AS%W=5792%OPTIONS=MNNTNW)
T4(ACK=O%Resp=Y%DF=Y%FLAGS=R%W=0%OPTIONS=)
T5(ACK=S++%Resp=Y%DF=Y%FLAGS=AR%W=0%OPTIONS=)
T6(ACK=O%Resp=Y%DF=Y%FLAGS=R%W=0%OPTIONS=)
T7(ACK=S++%Resp=Y%DF=Y%FLAGS=AR%W=0%OPTIONS=)
PU(RIPTL=328%RIPCK=E%DF=N%TOS=192%Resp=Y%IPLEN=356%DAT=E%RID=E%UCK=E%ULEN=308)
TSEQ(IPID=Z%TS=1000HZ%SI=2506961%CLASS=RI%GCD=1)

Neural Networks Output (close to 1 is better) for host : 

Relevant analysis
Relevant: 1.0

Operating System analysis
Linux: 1.0
Solaris: -1.0
OpenBSD: -0.999999999946
FreeBSD: -1.0
NetBSD: -1.0
Windows: -1.0

Linux version analysis
2.0: -1.0
2.2: 0.826111486106
1: -1.0
2.1: -0.999999421132
2.3: -1.0
2.4: -0.999999697669
2.5: -0.999999999837
2.6: 0.999999986751

Setting (guessing) architecture: i386
Setting OS to linux 2.6
--
Module finished execution after 6 secs.
\end{verbatim}

\subsection{Solaris 9}
\begin{verbatim}
Module "Neural Nmap OS Stack Fingerprinting" (v46560) started execution 
on Fri Aug 03 20:40:18 2007

Identifying host
 . Can't find a TCP open port in the container. A small Syn Scan will be launched
 . Found TCP open port 25.
 . Can't find a TCP closed port in the container. Using port 57302
 . Can't find a UDP closed port in the container. Using port 44869
The avg TCP TS HZ is: 99.785162
T1(ACK=S++%Resp=Y%DF=Y%FLAGS=AS%W=49335%OPTIONS=NNTMNW)
T2(Resp=N)
T3(Resp=N)
T4(ACK=O%Resp=Y%DF=Y%FLAGS=R%W=0%OPTIONS=)
T5(ACK=S++%Resp=Y%DF=Y%FLAGS=AR%W=0%OPTIONS=)
T6(ACK=O%Resp=Y%DF=Y%FLAGS=R%W=0%OPTIONS=)
T7(Resp=N)
PU(RIPTL=328%RIPCK=E%DF=Y%TOS=0%Resp=Y%IPLEN=112%DAT=E%RID=E%UCK=E%ULEN=308)
TSEQ(IPID=I%TS=100HZ%SI=28457%CLASS=RI%GCD=1)

Neural Networks Output (close to 1 is better) for host :

Relevant analysis
Relevant: 1.0

Operating System analysis
Linux: -0.999999999933
Solaris: 0.999999492289
OpenBSD: -1.0
FreeBSD: -0.999999999998
NetBSD: -1.0
Windows: -0.999999999916

Solaris version analysis
8: -0.999676619492
9: 0.999966693126
7: -0.999999983392
2.X: -0.999999989425
Other Solaris: -0.999998804773

Setting (guessing) architecture: SPARC_v8
Setting OS to solaris 9
--
Module finished execution after 6 secs.
\end{verbatim}

\section{Comparación con los métodos clásicos}

La tabla siguiente muestra una comparación entre el módulo de Nmap que usa
redes neuronales y el módulo Nmap ``clásico" que usa un algoritmo de puntajes.
En el caso de sistemas Windows, los respectivos módulos DCE-RPC
se usaron para refinar la detección de la versión y edición.
La detección de SO usando redes neuronales es más precisa: se obtiene más concordancias
de versión y edición y se obtiene menos errores.

Los Cuadros \ref{comparacion-windows} y \ref{comparacion-otros}
muestran una comparación entre los resultados obtenidos con los módulos clásicos
y los módulos basados en redes neuronales, para un conjunto de máquinas de nuestro laboratorio.
Las máquinas utilizadas para estas pruebas son distintas de las máquinas usadas 
para el entrenamiento de las redes neuronales.

\begin{table}
\center
\footnotesize
\begin{tabular} {| l | l | l | }
\hline
Sistema operativo & Resultado con  & Resultado con  \\
& Métodos clásicos & Redes neuronales \\
\hline
Windows NT 4 Server SP1		& Compaq T1010 Thin Client  & Windows NT4 Server SP6a \\
& Windows CE 2.12 & \\
\hline
Windows NT 4 Server SP3		& Compaq T1010 Thin Client & Windows NT4 Server SP6a \\
& Windows CE 2.12 & \\
\hline
Windows 2000 Adv. Server SP2	& Windows 2000 & Windows 2000 Adv. Server SP0 \\
\hline
Windows 2000 Adv. Server SP3	& Windows 2000 & Windows 2000 Adv. Server \\
\hline
Windows 2000 Adv. Server SP4	& Windows 2000 & Windows 2000 Adv. Server \\
\hline
Windows 2000 Professional SP0		& Windows 2000 & Windows 2000 Professional \\
\hline
Windows 2000 Professional SP1		& Windows 2000 & Windows 2000 Professional \\
\hline
Windows 2000 Professional SP2		& Windows 2000 & Windows 2000 Professional \\
\hline
Windows 2000 Professional SP3		& Windows 2000 & Windows 2000 Professional \\
\hline
Windows 2000 Server SP0			& Windows 2000 & Windows 2000 Server \\
\hline
Windows 2000 Server SP1			& Windows 2000 & Windows 2000 Server SP1 \\
\hline
Windows 2000 Server SP2			& Windows 2000 & Windows 2000 Server  \\
\hline
Windows 2000 Server SP3			& Windows 2000 & Windows 2000 Server SP3 \\
\hline
Windows 2000 Server SP4			& Windows 2000 & Windows 2000 Server SP4 \\
\hline
Windows XP Professional SP0	& Cannot differentiate between   	& Windows XP Home SP1	\\
& Windows XP and 2003 & \\
\hline
Windows XP Professional SP1	& Cannot differentiate between & Windows XP Home SP1	\\
& Windows XP and 2003		& \\
\hline
Windows XP Professional SP0	& Cannot differentiate between & Windows XP Home SP1	\\
& Windows XP and 2003  	& \\
\hline
Windows XP Professional SP1	& Cannot differentiate between & Windows XP unknown SP1	\\
& Windows XP and 2003  	& \\
\hline
Windows 2003 Server Entreprise	& Cannot differentiate between & Windows 2003 unknown SP0 \\
& Windows XP and 2003  	& \\
\hline
Windows XP Pro SP0	& Microsoft Windows Millennium, & Windows 2000 unknown SP4 \\
&  Windows 2000 Professional or  & \\
& Advanced Server, or Windows XP & \\
\hline
Windows 2003 Server	&Cannot differentiate between & Windows XP unknown SP1 \\
& Windows XP and 2003  	& \\
\hline
Windows XP Home SP2	& Cannot differentiate between & no match	\\
& Windows XP and 2003  	& \\
\hline
Windows XP Home SP0	& Cannot differentiate between & Windows XP unknown SP1 \\
& Windows XP and 2003  	& \\
\hline
Windows XP Home SP1	& Cannot differentiate between & Windows XP Home SP1 \\
& Windows XP and 2003  	& \\
\hline
Windows XP Professional SP0		& Apple Mac OS X Server 10.2.8 & Windows XP Professional SP1 \\
\hline
\end{tabular}
\caption{Comparación entre los métodos clásicos y los métodos 
basados en redes neuronales (máquinas Windows)}
\label{comparacion-windows}
\end{table}

\begin{table}
\center
\footnotesize
\begin{tabular} {| l | l | l | }
\hline
Sistema operativo & Resultado con  & Resultado con  \\
& Métodos clásicos & Redes neuronales \\
\hline
OpenBSD 3.0			& Foundry FastIron Edge  & OpenBSD 3 \\
& Switch (load balancer) 2402 & \\
\hline
OpenBSD 3.1			& Foundry FastIron Edge  & OpenBSD 3 \\
& Switch (load balancer) 2402 & \\
\hline
OpenBSD 3.2				&OpenBSD 3.0 or 3.3 & OpenBSD 3 \\
\hline
OpenBSD 3.3			& Foundry FastIron Edge  & OpenBSD 3 \\
& Switch (load balancer) 2402 & \\
\hline
OpenBSD 3.5			& Foundry FastIron Edge & OpenBSD 3 \\
& Switch (load balancer) 2402 & \\
\hline
Solaris 9				& 2Wire Home Portal 100  & Solaris 9 \\
& residential gateway, v.3.1.0 & \\
\hline
Solaris 10					&Sun Solaris 9 or 10 & Solaris 9 \\
\hline
Solaris 9					&Sun Solaris 9 or 10 & Solaris 9 \\
\hline
Debian 2.2 (Linux kernel 2.2)	& Linux 2.1 & Linux 2.2 \\
\hline
Debian 3 (Linux kernel 2.2)		&Linux 2.1 & Linux 2.2 \\
\hline
Fedora 3 (Linux kernel 2.6)		& Linux 2.6.10 & Linux 2.6 \\
\hline
Fedora 4 (Linux kernel 2.6)		& Linux 2.6.10 & Linux 2.6 \\
\hline
RedHat 6.2 (Linux kernel 2.2)	& Linux 2.1 & Linux 2.2 \\
\hline
RedHat 6.2 (Linux kernel 2.2)	& Linux 2.1 & Linux 2.2 \\
\hline
RedHat 7.2 (Linux kernel 2.4)		&Linux 2.6 & Linux 2.5 \\
\hline
RedHat 8 (Linux kernel 2.4) 	& Linux 2.4 & Linux 2.5 \\
\hline
RedHat 9 (Linux kernel 2.4)		& Linux 2.4 & Linux 2.5 \\
\hline
Mandrake 7.2 (Linux kernel 2.2)	& Linux 2.2.12 - 2.2.25 & Linux 2.1 \\
\hline
Mandrake 7.2 (Linux kernel 2.2)	& ZyXel Prestige Broadband router & Linux 2.1 \\
\hline
Mandrake 8.2 (Linux kernel 2.4)	& Linux 2.4.0 - 2.5.20 & Linux 2.6 \\
\hline
Mandrake 9.2 (Linux kernel 2.4)	&Linux 2.4 & Linux 2.5 \\
\hline
Mandrake 10 (Linux kernel 2.6)	& Linux 2.4.7 - 2.6.11 & Linux 2.6 \\
\hline
Suse 9.0 (Linux kernel 2.4)		&Linux 2.6.10 & Linux 2.6 \\
\hline
\end{tabular}
\caption{Comparación entre los métodos clásicos y 
los métodos basados en redes neuronales (cont.)}
\label{comparacion-otros}
\end{table}

La tabla siguiente resume los resultados obtenidos en las máquinas Windows.
Se considera la concordancia de versión y edición (no de service pack).
El módulo con redes neuronales muestra mejores resultados, dado
que logra reconocer la versión y la edición de Windows, mientras que el módulo
solo llega a reconocer las diferentes versiones.
\begin{center}
\begin{tabular} {| l | c | c | }
\hline
Resultado & Módulo clásico & Módulo con \\
 & & redes neuronales \\
\hline
Concordancia de versión y edición & 0 & 16 \\
Concordancia de versión  		& 12 & 6 \\
Concuerda únicamente la familia  	& 12  &  2 \\
Errores 				& 1 & 0 \\
Sin respuesta 			& 0 & 1 \\
\hline
\end{tabular}
\end{center}

La tabla siguiente resume los resultados obtenidos para las máquinas de 
otras familias (no Windows). Nuevamente, los módulos basados
en redes neuronales muestran un mejor desempeño,
en particular no cometieron errores de clasificación (asignar la máquina
a una familia de operativos incorrecta).
\begin{center}
\begin{tabular} {| l | c | c | }
\hline
Resultado & Módulo clásico & Módulo con \\
 & & redes neuronales \\
\hline
Concuerda familia y versión 	& 10 & 14 \\
Concuerda únicamente la familia   & 7  &  9 \\
Errores 				& 6 & 0 \\
Sin respuesta 			& 0 & 0 \\
\hline
\end{tabular}
\end{center}

\section{Conclusión}

Una de las principales limitaciones de las técnicas clásicas de detección de Sistemas Operativos
es el análisis realizado sobre la información recolectada por los tests, basado en alguna variación
del algoritmo de ``best fit" (encontrar el punto más cercano de acuerdo con una distancia de Hamming).

Hemos visto como generar y colectar la información a analizar, y como extraer parte de la estructura 
de los datos de entrada.
La idea principal de nuestro enfoque, que motiva la decisión de usar redes neuronales, 
es de dividir el análisis en varios pasos jerárquicos,
y de reducir la cantidad de dimensiones de entrada.
Los resultados experimentales (de laboratorio) muestran que este enfoque
da un método de reconocimiento de SO más fiable.

Además, la reducción de la matriz de correlación y el análisis en componentes principales
dan un método sistemático para analizar las respuestas de una máquina a los estímulos que le mandamos.
Como resultado, podemos identificar los elementos clave de los tests de Nmap,
por ejemplo los campos que dan información acerca de las distintas versiones de OpenBSD.
Otra aplicación de este análisis sería de optimizar los tests de Nmap para generar menos tráfico.
Otra aplicación más ambiciosa sería de crear una base de datos con respuestas de una población representativa
a una gran batería de tests
(combinaciones de distintos tipos de paquetes, puertos y flags).
Las mismas técnicas de análisis permitirían encontrar en esta gran base de datos
los tests más discriminativos para el reconocimiento de SO.

\section{Posibles extensiones}

El análisis que proponemos también se puede aplicar a otros métodos de detección:
\begin{enumerate}
\item{Xprobe2, de Ofir Arkin, Fyodor Yarochkin \& Meder Kydyraliev, 
que basa la detección en tests 
ICMP (Internet Control Message Protocol), 
SMB (Server Message Block) y 
SNMP (Simple Network Management Protocol).\\
Ver {\tt http://xprobe.sourceforge.net}.
}

\item{Passive OS Identification (p0f) de Michal Zalewski,
método que tiene la ventaja de no generar ningún tráfico adicional.
P0f trata de identificar el sistema operativo de
máquinas que se conectan a la nuestra (modo SYN), 
máquinas a las cuales nos conectamos  (modo SYN+ACK),
máquinas a las cuales no podemos conectarnos (modo RST+)
y máquinas cuyas comunicaciones podemos observar (con un sniffer).
Es un desafío interesante, dado que el análisis involucra una cantidad de información considerable
(todo el tráfico sniffeado), y probablemente requiere métodos más dinámicos y evolutivos.\\
Ver {\tt http://lcamtuf.coredump.cx/p0f.shtml}.
}

\item{Detección de SO basada en información del portmapper SUN RPC,
que permite distinguir sistemas Sun, Linux y otras versiones de System V.}

\item{Recolección de información para penetration test de client side,
en particular para detectar versiones de aplicaciones.
Por ejemplo para distinguir Mail User Agents (MUA) como Outlook o Thunderbird,
extrayendo información de los encabezados de mail.}
\item{Otra idea para investigar es de agregar ruido y filtrado de firewall a los datos estudiados.
Un firewall (o pared cortafuegos) es un elemento de hardware o software utilizado en una red de 
computadoras para controlar las comunicaciones, permitiéndolas o prohibiéndolas
según las políticas de red que se hayan definido. 
El modo de funcionamiento de los firewalls está especificado en el RFC 2979 \cite{rfc2979}.
Al filtrar los paquetes que el atacante manda a una máquina, y los paquetes que esta genera en respuesta
en esos estímulos, un firewall modifica considerablemente el comportamiento de esa máquina observable por el atacante.
Entrenar redes neuronales con datos filtrados por diferentes firewalls
permitiría la detección de un firewall, de identificar diferentes firewalls y de hacer test
más robustos.}

\end{enumerate}


\end{document}